\documentclass{tdp}

\usepackage[utf8]{inputenc} 

\usepackage{graphicx}
\usepackage{amsthm,amsmath,amssymb,amsfonts,bbm}
\usepackage{xspace}
\usepackage{color}
\usepackage{cleveref}
\crefformat{footnote}{#2\footnotemark[#1]#3}
\usepackage[table]{xcolor}
\usepackage{hhline}
\usepackage{multirow}
\usepackage{array}
\usepackage{balance}
\usepackage{rotating}
\newcolumntype{L}[1]{>{\raggedright\let\newline\\\arraybackslash\hspace{0pt}}p{#1}} 
\newcolumntype{C}[1]{>{\centering\let\newline\\\arraybackslash\hspace{0pt}}p{#1}}
\newcolumntype{R}[1]{>{\raggedleft\let\newline\\\arraybackslash\hspace{0pt}}p{#1}}

\newcommand{\eg}{\textit{e.g.},\xspace}
\newcommand{\ie}{\textit{i.e.},\xspace}
\newcommand{\etal}{\textit{et al.}\xspace}
\newcommand{\data}{trajectory micro-data\xspace}
\newcommand{\Data}{Trajectory micro-data\xspace}

\begin{document}

\title{Privacy in trajectory micro-data\\publishing: a survey}

\author{Marco~Fiore$^{*}$, Panagiota~Katsikouli$^{**}$, Elli~Zavou$^{***}$, Mathieu~Cunche$^{***}$, Fran\c{c}oise~Fessant$^{****}$, Dominique~Le~Hello$^{****}$, Ulrich~Matchi~Aivodji$^{*****}$, Baptiste~Olivier$^{****}$, Tony~Quertier$^{****}$, Razvan~Stanica$^{***}$}
\address{$^{*}$IMDEA Networks Institute, Spain.\\
        $^{**}$Technical University of Denmark, Lyngby, Denmark.\\
		$^{***}$University of Lyon, Inria, INSA-Lyon, CITI, France.\\
		$^{****}$ Orange Labs,France.\\
		$^{*****}$Universite du Quebec \`a Montreal, Canada.\\
		E-mails: {\small \tt{marco.fiore@ieiit.cnr.it}}, {\small \tt{panka@dtu.dk}}, {\small \tt{elli.zavou@inria.fr}},\\ {\small \tt{mathieu.cunche@insa-lyon.fr}}, {\small \tt{francoise.fessant@orange.com}},\\ {\small \tt{dominique.le.hello@orange.com}}, {\small \tt{aivodji.ulrich@courrier.uqam.ca}},\\{\small \tt{baptiste.olivier@orange.com}}, {\small \tt{tony.quertier@orange.com}} ,\\{\small \tt{razvan.stanica@insa-lyon.fr}}
}

\maketitle

\begin{abstract} 
%
%
We survey the literature on the privacy of \data, \ie spatiotemporal information about the mobility of individuals, whose collection is becoming increasingly simple and frequent thanks to emerging information and communication technologies. The focus of our review is
on privacy-preserving data publishing (PPDP), \ie the publication of databases of \data that preserve the privacy of the monitored individuals.
We classify and present the literature of attacks against \data, as well as solutions proposed to date for protecting databases from such attacks.
This paper serves as an introductory reading on a critical subject in an era of growing awareness about privacy risks connected to digital services, and provides insights into open problems and future directions for research.
\end{abstract}


\section{Introduction}

Our lives are increasingly entangled with ubiquitous communication technologies.
Calling someone on a mobile phone, tweeting about an event, browsing the
World Wide Web, using a car navigation system, or paying with a credit card
are a few examples of situations that create a seamless trail of \emph{digital breadcrumbs}
about our daily activities. These actions are easily recorded and persistently stored
into databases. Today, the pervasiveness of mobile communication technologies allows
tracking millions of users simultaneously, leading to the collection of vast
amounts of personal mobility data, which are then mined for many and varied
purposes, such as location-based marketing, targeted advertising, behavioural
profiling, transportation analysis, liability attribution, or security
enforcement -- just to cite a few relevant applications.

The galloping pace of innovations in this field, along with the increasing trend
of digitalization of our lives, suggests that what we are experiencing nowadays
is just the tip of the iceberg. In fact, services based on personal data records
promise to be life-changers for the newer generations, with a clear trend of
innovation happening in the data domain, where social networks (and alike) collect
and exploit users' information more and more~\cite{segal17}.

\begin{figure*}
\centering
\includegraphics[width=0.8\columnwidth]{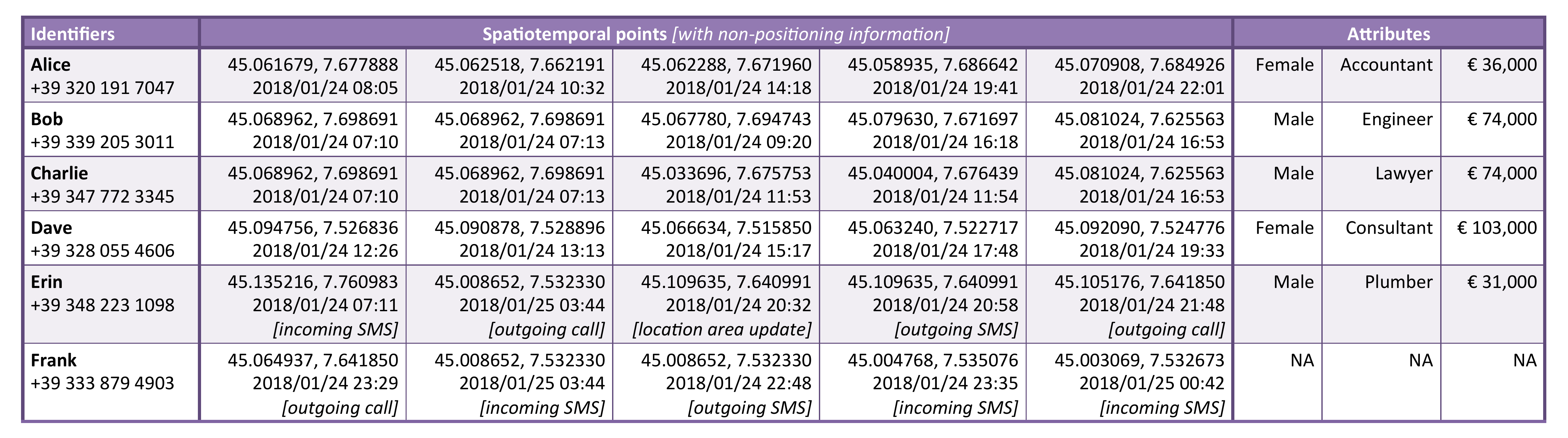}
\vspace*{-8pt}
\caption{Example of database of \data. Each record is composed of an identifier
(left), a spatiotemporal trajectory (middle), and additional attributes (right).
In this specific example, the person's name and phone address are the identifiers,
and spatiotemporal points in the trajectory are GPS locations augmented with
non-positioning information, within brackets, about their mobile communication
activity. Attributes consist of gender, employment and revenue.}
\vspace*{-8pt}
\label{fig:data}
\end{figure*}

A common trait to most of these emerging technologies is that they often build and
rely on databases that compose of or include \textit{\data}.
As the term indicates, these are micro-data, \ie information about single
individuals, that describe their \textit{spatiotemporal trajectories}, \ie sequences
of geographical positions of the monitored individuals over time.
Figure~\ref{fig:data} shows a toy example of a typical \data database: each record
corresponds to one person, and contains an identifier as well as a set of geo-referenced
and time-stamped elements, or \textit{spatiotemporal points}. Depending on the nature of
the database, the elements can also include non-positioning (\eg numerical or categorical)
information associated to each spatiotemporal point. Also, the database can present
additional fields that map to attributes beyond the spatiotemporal trajectory.

The definition of \data database above encompasses positioning
information gathered in a variety of ways, via different platforms and technologies.
For the sake of clarity, we illustrate below five prominent examples of \data sources. 
\begin{itemize}
\item Location-based services (LBS) are implemented as applications running on
mobile devices (\eg smartphones or tablets), which upload user position data as
required for service operation. Many extremely popular applications, such as Google
Maps, FourSquare, Twitter, Instagram, or Pokemon Go, fall in this category, and
relentlessly capture \data of individuals.
\item Cellular network operators deploy passive monitoring systems in their networks
to collect data about their subscribers' activity, for purposes including billing,
traffic engineering or added-value service development. Such data include
time-stamped user locations (\eg the location of the antenna which the user device is
associated to, or a triangulated point from signal strength indicators). For instance,
research-favoured call detail records (CDR) allow tracking mobile subscribers
every time their devices interact with the network. 
\item Mobile devices equipped with Wi-Fi interfaces typically broadcast probe messages
to discover nearby Access Points (APs). By letting APs (or sniffers, \ie dedicated
devices that passively monitor probe messages) record the unique Medium Access
Control (MAC) address of the devices emitting such probes, the Wi-Fi access provider
can track users within coverage of the Wi-Fi network. In presence of large deployments,
\eg covering municipalities or urban transportation infrastructures, mobile devices can
be potentially followed across a vast portion of their movements.
\item Modern car navigation systems have Internet connection capability, thanks to an
embedded mobile network interface. This setup allows notifying drivers about road traffic
conditions in real time, but also to collect fine-grained positioning data while the
vehicle engine is on. Such data are used by navigation system providers to determine
the congestion level of roads, and by insurance companies to determine liability in
case of accidents or to profile driving styles and associated risk levels.
\item Electronic payments are replacing cash in everyday's shopping. The resulting
transactions are easily linked to the address of the retailer who accepted the payment,
which allows companies in the banking sector to monitor the movements of their customers
as they use their debit or credit cards.
\end{itemize}
In all examples above, new and pervasive technologies allow the collection
of \data at very large scales, \ie enable tracking thousands to millions of
users at once.

It is precisely the possibility of knowing the movement patterns of large populations at
an individual level that paves the road to a wide range of applications for databases
of \data.
Such applications encompass a variety of contexts, including intelligent transportation,
assisted-life services, city planning, location-based marketing, data-driven
decision-making, or infrastructure optimization; as a recent and compelling example,
targeted containment measures enacted by many countries during the COVID-19 outbreak
in early 2020 have relied on contact information extracted from \data collected by
companies in the telecommunication sector~\cite{latif2020leveraging}.
In the light of these considerations, network and service providers are obviously eager
to exploit \data in new ways by continuously developing original dedicated
platforms~\cite{dans17,fluxvision}, and financial forecasts expect \data-driven services
to grow into a new multibillion-dollar market in the coming years~\cite{bi15,dornan15}.
Ultimately, the emergence of such a large and innovative technological and business
ecosystem creates a whole new need for gathering, storing, provisioning, circulating
and trading databases of \data.

\subsection{Privacy of \data}

Privacy is an obvious major concern at all stages of \data manipulation. This
consideration holds no matter whether the ultimate aim of the data processing is the
discovery of new knowledge or the monetization of embedded information.
As a matter of fact, owing to the nature of \data, incorrect stewardship can easily reveal
sensitive personal information about the users. Examples include iOS devices storing
their own spatiotemporal trajectories in unencrypted format and transmitting them to
Apple~\cite{chen11a,chen11b}, US mobile carriers selling real-time personal \data
to third party service providers~\cite{neustrom17}, or demand-side platforms for
targeted advertising in mobile phone apps paving the way to uncontrolled collection
of personal \data~\cite{vines17}. This is also generating a growing concern in the
general public, as awareness is raising about the privacy risks associated with spatiotemporal tracking~\cite{nytimes18}, and about how such personal information is shared in the data market~\cite{vallina17}.

Situations such as those mentioned above call for \textit{privacy-preserving
data publishing} (PPDP) of \data databases in all contexts where this kind of
data is stored or shared. PPDP recommends that databases should
be transformed prior to publication in potentially hostile environments, so as
to grant that the published data remains useful while individual privacy is
preserved~\cite{fung10}.

\begin{figure*}
\centering
\includegraphics[width=0.85\columnwidth]{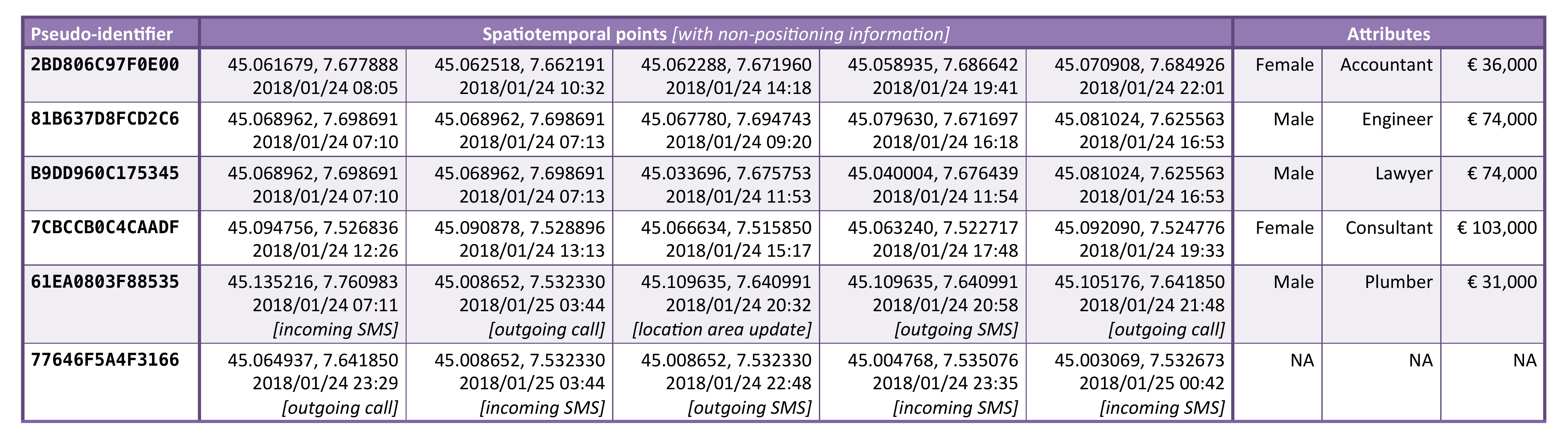}
\vspace*{-8pt}
\caption{Example of pseudonymisation of the database in Figure~\ref{fig:data}.
Identifiers (left) are replaced with values that cannot be linked with the
person's identity. The spatiotemporal trajectory and additional attributes
remain unchanged.}
\vspace*{-8pt}
\label{fig:pseudonymisation}
\end{figure*}

The common practice adopted by data collectors and data owners in order to protect the
privacy of the individuals they monitor is \emph{pseudonymisation}, also referred to
as \emph{de-personification}. This straightforward approach consists in removing all
personal identifiers (\eg information that is directly linked to the person's identity,
such as name, telephone number, precise address, plate number, etc.), and replacing them
with some \emph{pseudorandom identifier}; the latter can be a keyed hash of the
original personal identifiers, or simply a random number that is uniquely associated to
the \data of an actual individual. Figure~\ref{fig:pseudonymisation} provides an
example of pseudonymisation, for the database in Figure~\ref{fig:data}.

Unfortunately, pseudonymisation only provides a very mild level of protection. A number
of experiments, performed in recent times and using large-scale real-world datasets, have
repeatedly demonstrated the significant risks associated to pseudonymised \data.
In particular, naive cross-correlation of pseudonymised data with named side information
(obtained from, \eg public-access social network data) leads to \emph{re-identification},
\ie disclosure of the identities of users with high probability, making pseudonymisation
basically useless.
We will discuss in detail the investigations leading to such conclusions in Section\,\ref{sec:att}.
What is relevant here is that, in the light of these findings and quite unsurprisingly,
data controllers have nowadays become extremely cautious
in opening access to pseudononymised \data. A prominent example is that of TfL, the transport
regulator in London, UK, which recently ran a pilot experiment by tracking passengers
in the London Underground network via Wi-Fi probes broadcasted by mobile devices.
TfL later rejected a Freedom of Information (FOI) request to release the pseudonymised
dataset, exactly because of the potential re-identification risks of the data~\cite{lomas17}.

With such a growth of concerns about risks associated with uncontrolled gathering and
mining of \data, regulatory bodies have been working on new legal frameworks dedicated
to personal data protection. A leading act in this sense is the General Data Protection
Regulation (GDPR)~\cite{gdpr}, 
which became effective on May 2018 and applies to all European Union citizens.
The GDPR enforces that data controllers shall adopt the best measures for data protection
by design and by default. Such measures include pseudonymisation, as it can reduce the
risks for the data subjects concerned and help controllers and processors to meet their
data-protection obligations. However, the GDPR makes it very clear that pseudonymisation
alone is an insufficient privacy measure when it comes to PPDP. Indeed, the regulation
decrees that pseudonymised data has still to be treated as personal data, which must be
securely stored and cannot be circulated freely.
Instead, the GDPR lays down that a more open publication of data is allowed upon
\textit{anonymization}, a process which ensures that the data cannot be any longer linked to an identified or identifiable natural person or data subject. According to the GDPR, anonymized data is not personal anymore, hence is not concerned by the privacy-protection rules it defines.

Legislations such as the GDPR are thus an important part of the solution, as they make
procedures ensuring a correct data processing mandatory. However, they must be complemented
by sound technical solutions that implement the invoked ``best measures'' and achieve the
privacy goals set by PPDP. In the specific case of \data, developing anonymization
algorithms that provably prevent any re-identification or personal information inference
from the original spatiotemporal points is extremely challenging. As it is often the case
in presence of difficult tasks, the problem has drawn a substantial effort by the research
community: a plethora of scientific papers have appeared over the past decade, aiming both at
unveiling privacy risks connected with \data, and at proposing solutions to
cope with such risks. However, such a large body of works targets heterogeneous types of \data,
considers a variety of attacker models, relies on different privacy criteria, and uses disparate
data transformation techniques. This substantial diversity makes the literature tangled and
complicated to approach, raising questions about where the current state of the art actually stands.

\subsection{Objective, positioning and structure of the survey}

This survey serves as a comprehensive introduction to the domain of privacy of \data for PPDP.
It summarizes almost two decades of research, providing a review of a large number
of 
works that cover all aspects of the problem. These include the assessment of
privacy risks in \data, the definition of attacks realizing such risks, and the proposition
of solutions that protect user privacy from the aforementioned attacks.

Our survey joins a rather small family of reviews that previously explored similar
domains.
The early works by Decker~\cite{decker08} and Chow and Mokbel~\cite{chow11} review privacy
in LBS: as we will explain in Section\,\ref{sub:intro-remarks}, this is an orthogonal
problem with respect to that of PPDP of \data.
Garfinkel~\cite{garfinkel15} provides a general overview of personal information de-identification
across a wide range of database types, including geographic and map ones. The portion of the
work dedicated to \data is necessarily limited in such a holistic document, and does not cover
the subject in depth.
Haris \etal~\cite{haris14} discuss privacy leakages, associated risks and potential remedies
in the broad context of mobile computing. Their review of the literature has a very wide breadth,
which is fully orthogonal to ours: indeed, their work targets applications and services rather than data. 
Christin~\cite{christin16} focuses on mobile participatory sensing, discussing privacy threats
relevant to the different phases of the sensing process. Mobile participatory sensing can generate
in some cases \data, hence some reviewed works are also covered in our document. However, the
overlap is marginal, and the overall context, discussion of challenges and conclusions by
Christin~\cite{christin16} are related to data and processes of a different nature from
those of interest to us.
The overlap is also minimal with the short review by Al-Azizy \etal~\cite{al-azizy16}, who review
the literature on data de-anonymisation: on the one hand, they consider any class of data instead
of the more specific \data; on the other hand, they focus on de-anonymisation, whereas we cover all
aspects of the problem, from risk assessment to solutions for data protection.
Similar considerations hold for the recent book edited by Gkoulalas-Divanis and Bettini~\cite{gkoulalas18}, which provides a comprehensive overview of current challenges and solutions in mobile data privacy; the book covers a wide range of subjects, from data collection to management and analysis, but does not include a dedicated discussion on mobile data PPDP.
The work that is the closest in spirit to ours is that by Bonchi \etal~\cite{bonchi11}, who
discuss a selected set of seven papers that propose techniques to anonymise \data: this is a
small subset of the studies we survey, which cover a much larger literature on both attack and
protection techniques.

As a result, none of the existing surveys provides a literature overview that comprehensively
addresses \data privacy. This paper aims at closing this gap -- a significant one in the light of
the rapid emergence of real-world services that heavily rely on \data.
The document is structured into two main Sections, respectively dedicated to \textit{attacks}
against \data, and \textit{anonymisation} of \data.
The former, in Section\,\ref{sec:att}, reviews the body of works that assess the privacy
risks associated with \data, by devising, implementing and evaluating attacks that
allow re-identifying users in a \data database.
The latter, in Section\,\ref{sec:anonym}, surveys countermeasures proposed to protect \data
from the aforementioned privacy threats.
The contents of such Sections allows us to draw considerations, as well as present open issues
and research opportunities in Section\,\ref{sec:disc}.

\subsection{Remarks}
\label{sub:intro-remarks}

Before proceeding further, the following three important remarks are in order.

First, as anticipated above, our focus is on privacy-preserving publication (PPDP) of \data,
which is an entirely different problem than privacy in LBS. Indeed, the two scenarios entail
non-comparable system models. In the case of \data publishing, databases of millions of
records are mined offline, and the challenge is ensuring that their circulation does not
pose a threat to user privacy, but retains data utility.
In the case of LBS, single (geo-referenced and time-stamped) queries generated by
mobile devices must be processed in real-time, and the objective is \textit{location privacy},
\ie ensuring that such a process preserves users' privacy by preventing the service provider
from locating users.
These considerations make PPDP of large-scale datasets the relevant problem in the
context of \data, while LBS are more concerned with the real-time anonymization of small
sets of spatiotemporal points; ultimately, this difference entails attacker models and anonymization
techniques that are very diverse for the two scenarios.
Indeed, Xiao and Xiong~\cite{xiao15} and Bindschaedler \etal~\cite{bindschaedler16} have shown that individual spatiotemporal points anonymized via solutions for location privacy are still vulnerable to attacks when their time-ordered sequence is considered, \ie when they are treated as a spatiotemporal trajectory.
As our focus is on trajectory releasing, the vast body of literature on privacy of LBS is
out of the scope of the present document; we refer readers with an interest on privacy
in LBS to the dedicated surveys by Decker~\cite{decker08}, Chow and Mokbel~\cite{chow11}, and Bettini~\cite{bettini18}.

Second, consistently with the scope of the survey, we are interested in attacks and
anonymization solutions that target \data. Therefore, we do not consider in our review
attacks against metadata that contain spatiotemporal information but where such information
is not factually exploited, as in the case of the mobile phone call graphs considered by
Sharad and Danezis~\cite{sharad13}, or of the mobile subscriber communication history studied by
Mayer \etal~\cite{mayer16}.
Similarly, we target individual trajectories intended as sequences of spatiotemporal points, while we do not consider aggregate forms of such data, like people counts or density. Hence, we do not review attacks against aggregate data from trajectories, such as those designed, \eg by Xu \etal~\cite{xu17}; and, we do not delve in the details of privacy-preserving techniques for aggregate statistics, such as those proposed by Liu \etal~\cite{Liu2016}, or those briefly mentioned at the beginning of Section~\ref{sub:anon-uninf}.

Third, in the remainder of the document we will not make any distinction between
original and pseudonymised data, and just refer to both as \data. This is because
pseudonymisation does not provide any significant layer of privacy protection, and
the vast majority of studies assume that all their input data are already pseudonymised.



\section{Attacks on \data}
\label{sec:att}


The first part of our survey is dedicated to the body of works on attacks against \data.
The objective of these studies is to propose techniques that enable the re-identification
or inference of personal data from datasets of \data, implicitly revealing privacy risks
associated to the inconsiderate publication of this class of databases.
To structure our discussion, we present in Section\,\ref{sub:att-tax} an original taxonomy of
attack strategies against \data.
Then, in Sections\,\ref{sub:link-pos} to \ref{sub:link-aux}, we review relevant studies,
separating them into classes based on the proposed taxonomy.

\subsection{A taxonomy of attacks}
\label{sub:att-tax}

Attack models are typically defined by a precise objective, and by the background
knowledge that the adversary can exploit towards attaining such an objective. The
taxonomy we adopt builds on three orthogonal dimensions that fully capture these
features: \textit{(i)} the objective of the attack, presented in Section\,\ref{sub:att-tax-obj};
\textit{(ii)} the format of the adversary's knowledge and \textit{(iii)} its origin, 
which are introduced in Section\,\ref{sub:att-tax-format} and Section\,\ref{sub:att-tax-origin},
respectively.
The classification of the literature that results from the intertwining of these three
dimensions is finally presented and discussed in Section\,\ref{sub:att-tax-lit}.

\subsubsection{Attack objective}
\label{sub:att-tax-obj}

While the overall aim of an adversary remains that of re-identifying individuals
in the \data, or more generally acquiring sensitive information about them from the
\data, different approaches can be leveraged to those purposes. Each approach
translates into a specialized \textbf{\textit{attack objective}} (denoted by \textbf{O}
in the rest of the paper), which is the first dimension of our taxonomy. The attack
objective is in fact a rather standard way to categorize privacy threats on generic
micro-data datasets, and we borrow from the classic codification by Fung \etal~\cite{fung10}
to organize the literature along this dimension.

\begin{figure*}
\centering
\includegraphics[width=0.85\columnwidth]{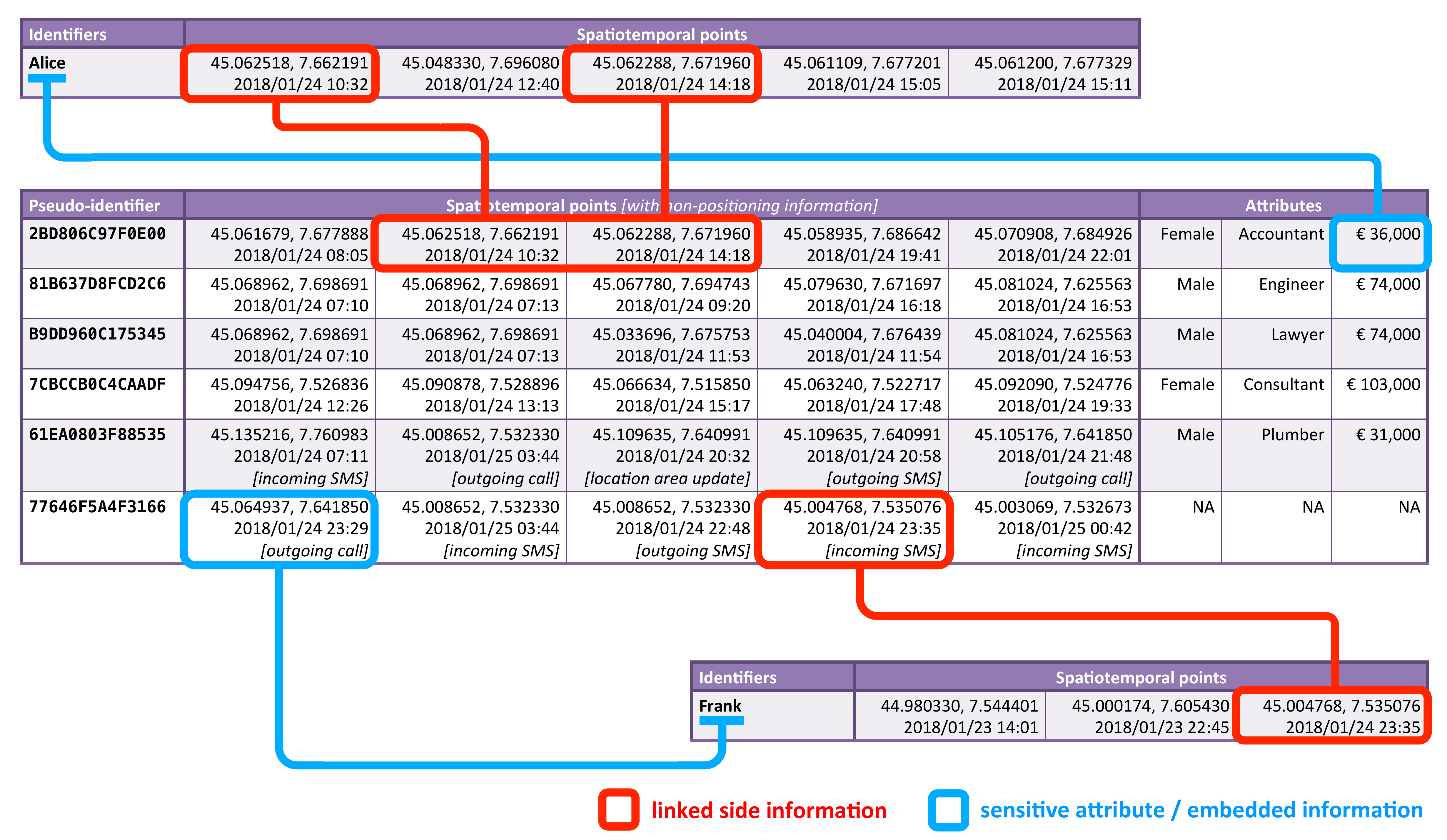}
\vspace*{-8pt}
\caption{Examples of record linkage attacks against the pseudonymised database in
Figure~\ref{fig:pseudonymisation}.
The side information (top and bottom) contains personal identifiers (the identities of
\textit{Alice} and \textit{Frank}, respectively) and some spatiotemporal points.
Record linkage exploits matches in the trajectory data (\eg identical GPS locations
in the side information and target database, in red) to associate the identifiers to one
record. It links the person to sensitive data (in blue), be those
attributes (\eg \textit{Alice}'s salary), or information embedded in the spatiotemporal
data itself (\eg \textit{Frank} visiting a gentlemen club at $45.064937,7.641850$).}
\vspace*{-8pt}
\label{fig:record-linkage}
\end{figure*}

The vast majority of attacks against \data investigated in
the literature belong to the category of \textbf{\textit{record linkage attacks}}
(\textbf{O.1}) as defined by Fung \etal~\cite{fung10}, which are often simply referred
to as \textit{linkage attacks}.
As illustrated in the toy example in Figure\,\ref{fig:record-linkage},
record linkage attacks aim at mapping records in the \textit{target} \data
with \textit{side information} owned by the adversary. The side information
must include personal identifiers, without which re-identification is not
possible, as well as some (possibly limited) data about the mobility of (a
subset of) users in the target database. These can
be collected in a variety of ways: examples include directly observing
target individuals (\eg by physically meeting or following them and recording
their movements), mining suitable open data (\eg via crawling of geo-referenced
social network metadata), or gaining access to samples of the actual \data
(\eg by leveraging a security breach).

A successful attack allows associating an identity to records in the target
database. Establishing such a link represents a privacy breach when the
records of the target database contain \textit{sensitive attributes}.
In the context of \data, there are two different situations where this
happens, both exemplified in Figure\,\ref{fig:record-linkage}.
\begin{itemize}
\item The typical assumption made in most of the works we will review is
that the database of \data also includes additional, separated sensitive
attributes. For instance, each record could include the spatiotemporal
points as well as personal data about the individual such as gender,
age, address, employment, or accounting information.
In this case, mobility data allows linking personal identities with the
non-positioning sensitive data, as in \textit{Alice}'s case in
Figure\,\ref{fig:record-linkage}.
\item A second, subtler perspective is that \data embed information that
is potentially sensitive per-se.
Gaining access to a large amount of timestamped locations visited by an
individual may allow an adversary understand where the individual lives, where
she works, and which kind of Points of Interest (PoIs) she visits.
The latter can reveal important locations (\eg home and workplace addresses),
commuting patterns (\eg periodic times of visits to public
transportation hubs), religious and political views (\eg regular visits
to places of worship or party meetings), or health conditions (\eg frequent
visits to healthcare structures), just to mention a few relevant
examples.
Extracting PoIs from \data is in fact a fairly easy task. The vast literature
on knowledge discovery from spatiotempral trajectories proposes a variety of
techniques to infer home, work, and other relevant locations from this kind
of data~\cite{naboulsi16}.
This would be the case of the record linkage against \textit{Frank}'s \data
in Figure\,\ref{fig:record-linkage}.
\end{itemize}

\begin{figure*}
\centering
\includegraphics[width=0.85\columnwidth]{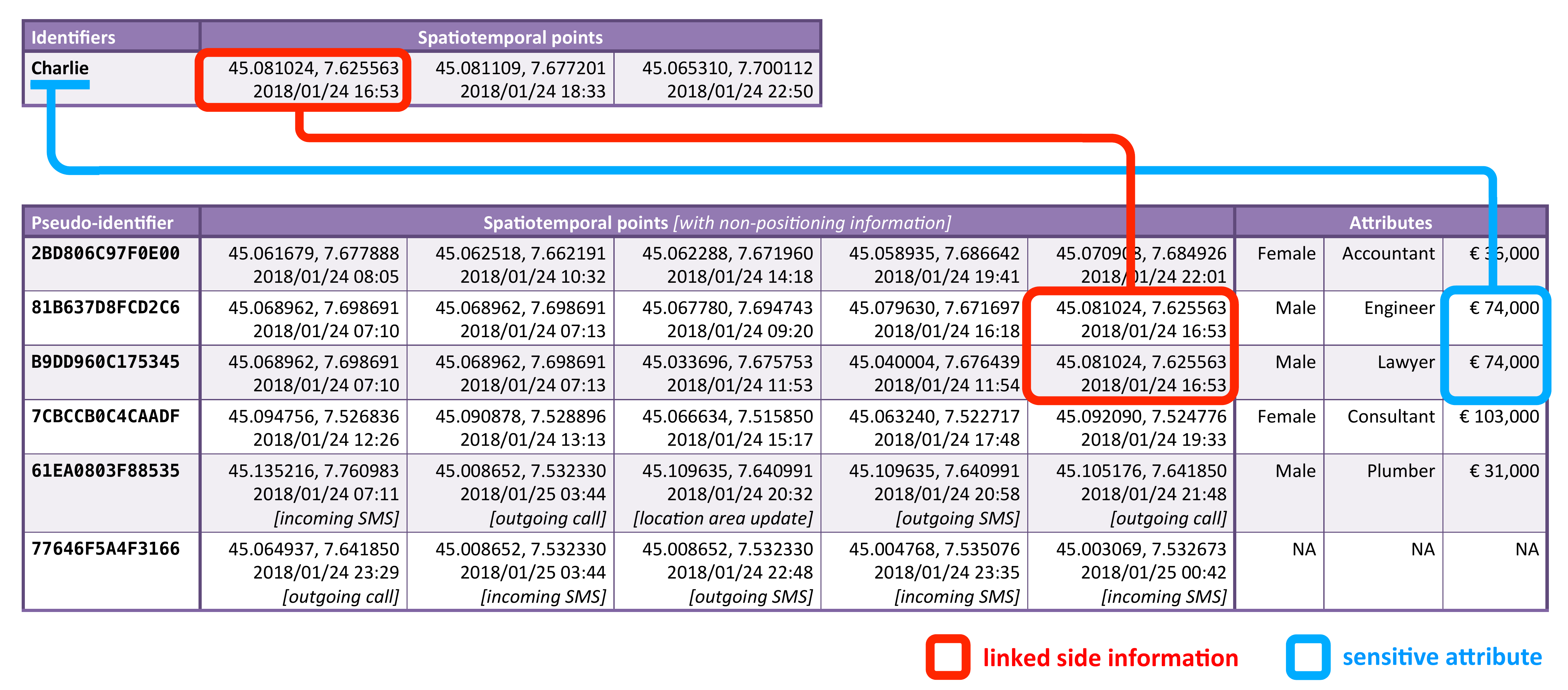}
\vspace*{-12pt}
\caption{Example of attribute linkage attacks against the pseudonymised database in
Figure~\ref{fig:pseudonymisation}.
The side information does not allow linking personal identifiers to one specific record,
since two different records contain the single matching spatiotemporal sample.
However, the two records yield the same revenue value, hence the attacker can still
link the identity of \textit{Charlie} to the sensitive information of a $74,000$-Euro yearly income.}
\vspace*{-8pt}
\label{fig:attribute-linkage}
\end{figure*}

\begin{figure*}
\centering
\includegraphics[width=0.85\columnwidth]{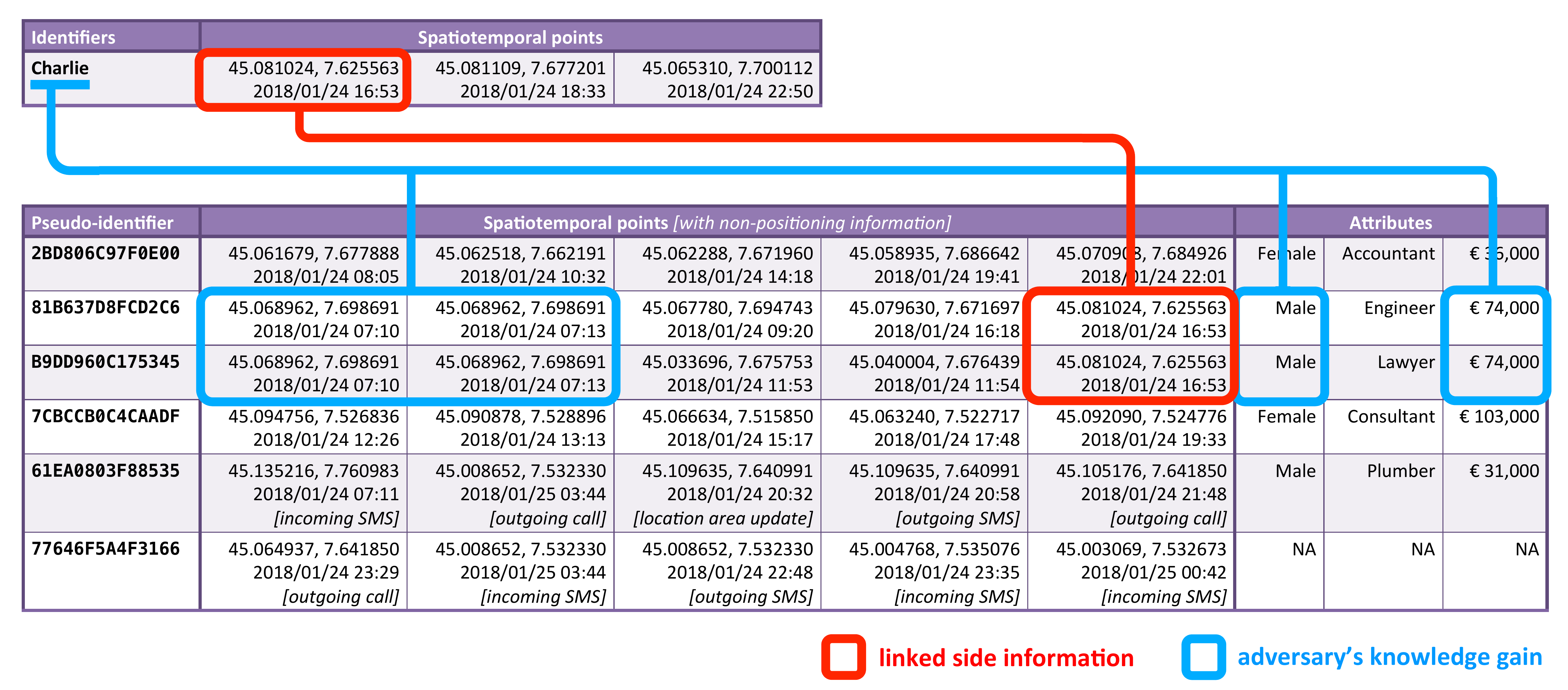}
\vspace*{-14pt}
\caption{Example of probabilistic attacks against the pseudonymised database in
Figure~\ref{fig:pseudonymisation}.
The side information allows the adversary to increase their knowledge, by discovering
two new spatiotemporal samples, as well as the gender and revenue of the target individual,
\textit{Charlie}.
All such additional information is considered sensitive in probabilistic attacks.}
\vspace*{-4pt}
\label{fig:probabilistic}
\end{figure*}

Classes of attack beyond record linkage have been rarely investigated in
works related to \data. Notable exceptions have explored two additional attack
categories.

The first is that of \textbf{\textit{attribute linkage attacks}} (\textbf{O.2}) in the terminology
by Fung \etal~\cite{fung10}, which are also known as \textit{homogeneity attacks}.
In this case, the objective of the adversary is to link its side information
with sensitive attributes (rather than specific records) in the target \data.
The typical scenario envisioned for attribute linkage is one were the attacker's
side information maps to multiple records (hence preventing record linkage), but
all such records share the same sensitive attributes (which are thus re-indentified
by attribute linkage).
Figure\,\ref{fig:attribute-linkage} depicts a toy example of database configuration
that is prone to a homogeneity attack on \data.
The privacy risks associated with a successful attribute linkage attack are
the same as for record linkage: indeed, record linkage yields a privacy breach
because the adversary can infer the sensitive information within a record,
and not the specific record itself. Yet, homogeneity attacks can be successful in
cases where record attacks are not possible, and thus pose a greater risk to privacy.

The other category of threats considered in the literature is that of
\textbf{\textit{probabilistic attacks}} (\textbf{O.3}) in the categorization
by Fung \etal~\cite{fung10},
which are also referred to as \textit{inference attacks} in the literature.
The goal of a probabilistic attack is increasing the adversary's knowledge
by accessing the target database. This is equivalent to generalizing the
notion of sensitive attribute to any information contained in a record:
in the context of \data, learning any additional, non-negligible portion of
the mobility of a user beyond the original side information already makes
the attack successful.
Figure\,\ref{fig:probabilistic} shows an example of probabilistic attack on \data.
Preventing probabilistic attacks is more challenging than countering
record or attribute linkage, since the adversary's goal is much broader.

\subsubsection{Format of the side information}
\label{sub:att-tax-format}

To achieve any of the attack objectives above, an adversary must leverage
some background knowledge. This side information can have different formats,
which drive the implementation of the attack. Thus, the second and third dimensions
considered in our taxonomy relate to the nature of the background data
of the adversary.
Specifically, we tell apart the two distinguishing characteristics of the side
information: \textit{(i)} its format, \ie the actual content of
the side information, discussed next; and, \textit{(ii)} its source, \ie how the
side information is gathered, which is instead presented in Section\,\ref{sub:att-tax-origin}.

\begin{figure*}
\centering
\includegraphics[width=0.85\columnwidth]{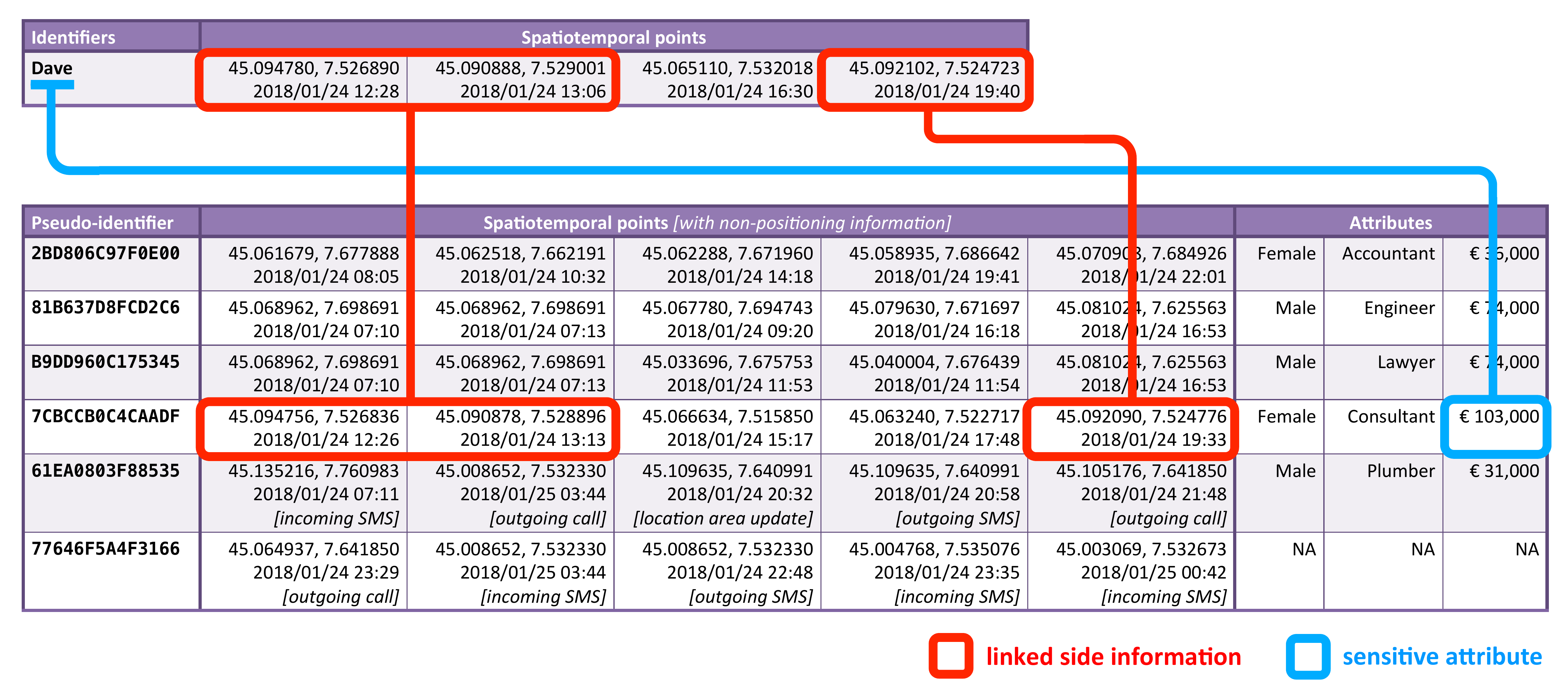}
\vspace*{-14pt}
\caption{Example of record linkage attack against the pseudonymised database in
Figure~\ref{fig:pseudonymisation}.
The side information includes a set of spatiotemporal points that does not
perfectly match any of those in the database. The adversary can still map
his knowledge to the most similar trajectory: multiple points in his possession
are very close to those in one specific record, which allows linking \textit{Dave}'s
identity to the sensitive revenue attribute.}
\vspace*{-12pt}
\label{fig:diverse-sampling}
\end{figure*}

Concerning the \textbf{\textit{format of side information}} (\textbf{F}),
we identify three classes from studies on \data privacy.
In the baseline case, the side information has the exact same format
of the mobility data contained in the target database, \ie a sequence of
\textbf{\textit{spatiotemporal points}} (\textbf{F.1}). Two situations can
occur under this format, as follows.
\begin{itemize}
\item In a simpler case, the spatiotemporal points in the side information are
a \textbf{\textit{subset}} (\textbf{F.1a}) of those contained in records of the
target \data. This is the situation portrayed in the previous examples in
Figures\,\ref{fig:record-linkage} through~\ref{fig:probabilistic}.
It assumes that the adversary gathers side information using the exact same
tracking technology employed to build the target database. This can be regarded
as a best-case scenario for the attacker, which is however unlikely to occur in
practical settings.
\item In many practical scenarios, the adversary does not possess a perfect
subset of the \data. Instead, the spatiotemporal points in the side information
are obtained from a different source than that generating the \data, hence they
represent a \textbf{\textit{diverse sampling}} (\textbf{F.1b}) of the underlying
spatiotemporal trajectories of the users.
Figure\,\ref{fig:diverse-sampling}
exemplifies this situation.
\end{itemize}

\begin{figure*}
\centering
\includegraphics[width=0.85\columnwidth]{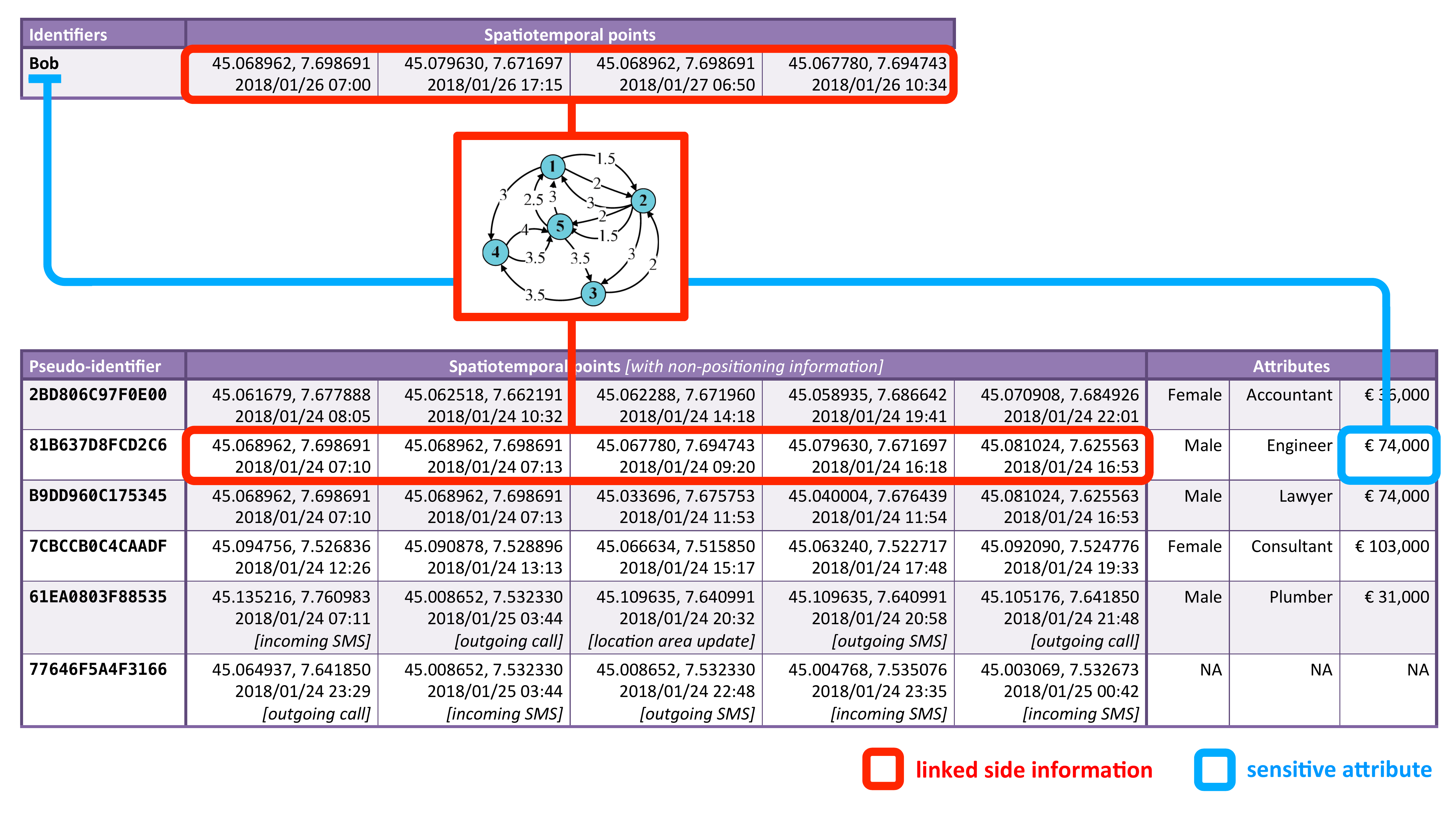}
\vspace*{-12pt}
\caption{Example of record linkage attack against the pseudonymised database in
Figure~\ref{fig:pseudonymisation}.
The side information includes a set of spatiotemporal points that does not
perfectly match any of those in the database. The adversary can build a profile
of the victim (\eg a probabilistic mobility model of transitions among locations)
from their knowledge,
and match this to similar profiles of all records in the database. A very similar
profile is identified in the second record, allowing linking \textit{Bob}'s
identity to the sensitive revenue attribute.}
\vspace*{-8pt}
\label{fig:profile}
\end{figure*}

A second type of side information format is represented by indirect knowledge
inferred from \data. The adversary does not have access to precise spatiotemporal
points of his victim's \data, but knows instead some high-level \textbf{\textit{profiles}}
(\textbf{F.2}) that characterize the movements of the target individuals.
The notion of profile is general and can accommodate a wide range of scenarios,
among which we identify the following four prominent situations:
\begin{itemize}
\item \textbf{\textit{mobility models}} (\textbf{F.2a}) are mathematical representations that summarize
the complete movement behavior of the target individuals, as illustrated by Figure~\ref{fig:profile};
\item \textbf{\textit{important locations}} (\textbf{F.2b}) are places frequently visited by or
especially significant for the target individuals;
\item \textbf{\textit{mobility features}} (\textbf{F.2c}) are specific properties found in the movement
patterns of the target individuals;
\item \textbf{\textit{social graphs}} (\textbf{F.2d}) are structures that describe the social relationships
of the target individuals with other users whose trajectories are also present in the database.
\end{itemize}

\begin{figure*}
\centering
\includegraphics[width=0.85\columnwidth]{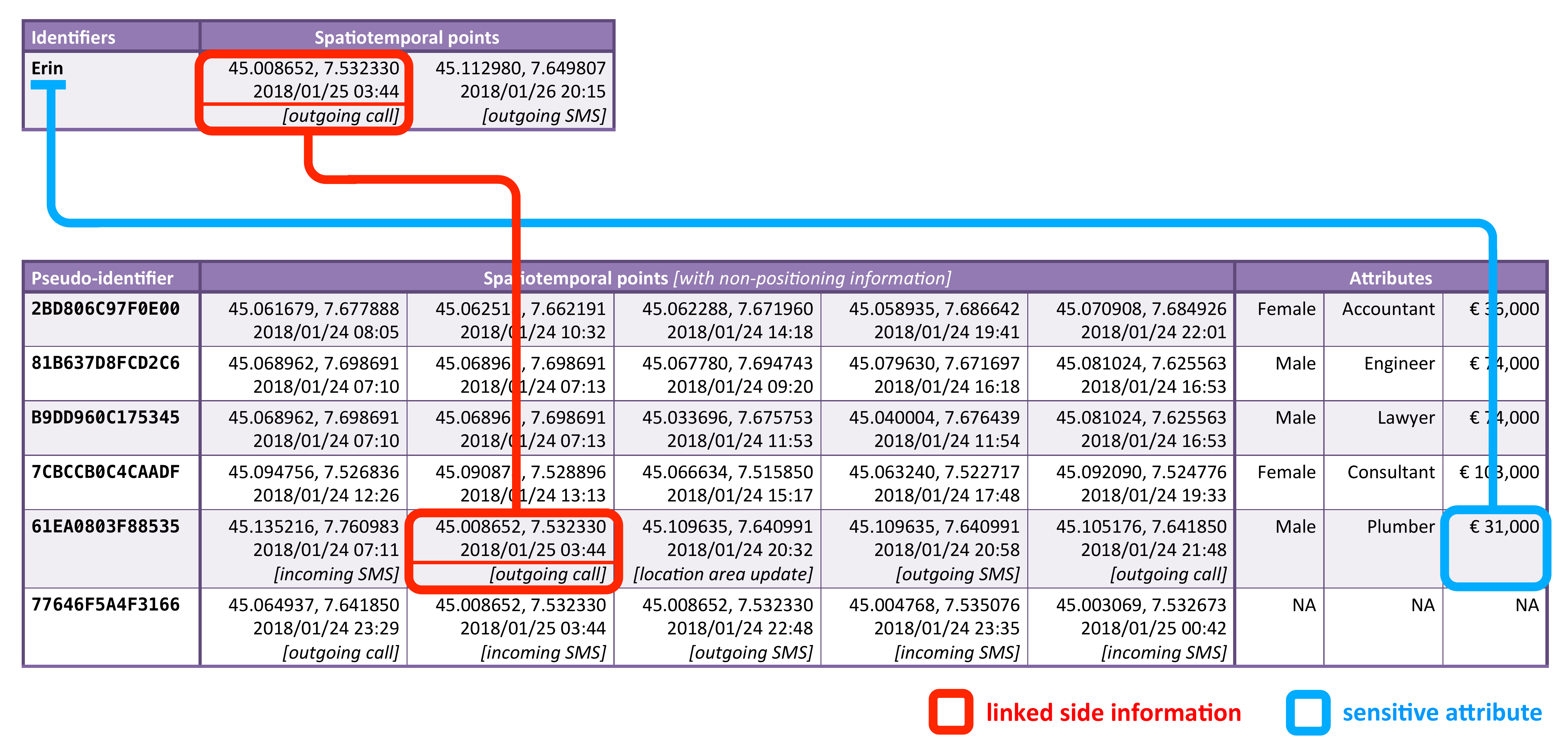}
\vspace*{-12pt}
\caption{Example of record linkage attack against the pseudonymised database in
Figure~\ref{fig:pseudonymisation}.
The side information includes a set of spatiotemporal points, each associated with
auxiliary data, within brackets, on mobile communication activities.
While the positioning data alone matches two records with different sensitive
attributes, the additional knowledge provided by the auxiliary data allows the
adversary to tell apart the records, and link \textit{Erin}'s identity to the
sensitive revenue attribute.}
\vspace*{-8pt}
\label{fig:auxiliary}
\end{figure*}

The third format of side information is a combination of \data and some
\textbf{\textit{auxiliary data}} (\textbf{F.3}) that is not related to the
mobility of the target individuals.
We can see this format as an augmented version of the two above, thanks to the
addition of the auxiliary data.
Under the assumption that records on the target database also contain fields related
to the auxiliary data, such reinforced side information grants an additional degree
of freedom to carry out attacks. An intuitive example is provided in Figure\,\ref{fig:auxiliary}.

\subsubsection{Origin of the side information}
\label{sub:att-tax-origin}

The third dimension we consider in our taxonomy of attacks on \data is that
of the \textbf{\textit{source of side information}} (\textbf{S}) owned by the adversary.
Side information can be retrieved from sources belonging to one of two categories, as follows.
\begin{itemize}
\item The vast majority of works in the literature directly extract the side information
from the target \data. We refer to this approach as \textbf{\textit{intra-record}}
(\textbf{S.1}), since the
source of the side information is the data contained in the database records themselves.
There exist two subcategories of intra-record sources. So-called
\textbf{\textit{intra-record subsampling}} (\textbf{S.1a}) leaves the original \data
unmodified once the side information is extracted: therefore, the side information is
necessarily present in the target database.
Instead, \textbf{\textit{intra-record training}} (\textbf{S.1b}) removes from the
target \data the side information, separating the original database into
training (used as the side information) and test (regarded as the target \data).
\item A more realistic approach, which we name \textbf{\textit{cross-database}}
(\textbf{S.2}), consists in considering a side information source that is entirely different
from the target \data: indeed, in practical cases, an adversary would derive his
background knowledge from direct observations of his victims' movements, or from
external datasets that are fully disjoint from the target one.
However, a cross-database approach requires suitable side-information databases
that contain mobility data for (a subset of) the users in the target \data, and
that are collected via a different technology. Acquiring such databases can be
complicated, which is why only a small number of works in the literature adopted
this strategy to date.
\end{itemize}



\begin{table*}
\renewcommand{\arraystretch}{1.3}
\caption{Classification of the literature of attacks on \data based on
our proposed taxonomy. The three dimensions of the taxonomy are highlighted
in bold. The leftmost column separates rows according to different classes
of attacker objective (\textbf{O}) as per Section\,\ref{sub:att-tax-obj}.
The subsequent two columns refine rows according to different categories and
subcategories of side information format (\textbf{F}) as per
Section\,\ref{sub:att-tax-format}. The last three columns distinguish types and
subtypes of side information source (\textbf{S}) as per Section\,\ref{sub:att-tax-origin}.
Grayed-out cells denote unfeasible combinations of side information
format and source.\vspace*{8pt}}
\label{tab:attacks}
\centering
\scriptsize
\setlength{\tabcolsep}{1.5pt}
\begin{tabular}[c]{r @{}p{3pt}@{} l @{}p{3pt}@{} l l @{}p{3pt}@{} l l l}
	& &
	\multicolumn{1}{c}{} & &
	\multicolumn{2}{c}{} & &
	\multicolumn{3}{l}{\textbf{Side information source (\textbf{S})\vspace*{2pt}}} \\
\hhline{~~~~~~~---}
	& &
	\multicolumn{1}{l}{} & &
	\multicolumn{2}{l}{} & &
	\multicolumn{2}{|L{158pt}|}{Intra-record \newline \vspace*{2pt}\textbf{S.1}} &
	\multicolumn{1}{L{75pt}|}{Cross-database \newline \vspace*{2pt}\textbf{S.2}} \\
\hhline{~~~~~~~--~}
	& &
	\multicolumn{1}{l}{} & &
	\multicolumn{2}{p{77pt}}{\textbf{Side information \newline format} (\textbf{F})} & &
	\multicolumn{1}{|L{70pt}|}{Subsampling \newline \vspace*{2pt}\textbf{S.1a}} &
	\multicolumn{1}{L{78pt}|}{Training \newline \vspace*{2pt}\textbf{S.1b}} &
	\multicolumn{1}{L{75pt}|}{} \\
\hhline{~~~~~~~---}\\[-5pt]
\hhline{~~-~--~---}
	\parbox[t]{7pt}{\multirow{9}{*}{\rotatebox[origin=c]{90}{\textbf{Attacker objective} (\textbf{O})}}} & &
	\multicolumn{1}{|L{52pt}|}{Record linkage \newline \vspace*{2pt}\textbf{O.1}} & &
	\multicolumn{1}{|L{25pt}|}{Points \newline \vspace*{2pt}\textbf{F.1}} &
	\multicolumn{1}{L{53pt}|}{Subset \newline \vspace*{2pt}\textbf{F.1a}} & &
	\multicolumn{1}{|L{70pt}|}{Bettini \etal~\cite{bettini05} \newline
                             De Montjoye \etal~\cite{de-montjoye13} \newline
														 Rossi \etal~\cite{rossi15} \newline
	                           Sapiezynski \etal~\cite{sapiezynski15} \newline
	                           De Montjoye \etal~\cite{de-montjoye15}} &
	\multicolumn{1}{l|}{\cellcolor{gray!12}} &
	\multicolumn{1}{l|}{\cellcolor{gray!12}} \\
\hhline{~~~~~-~---}
	& &
	\multicolumn{1}{|l|}{} & &
	\multicolumn{1}{|l|}{} &
	\multicolumn{1}{L{50pt}|}{Diverse sampling \newline \vspace*{2pt}\textbf{F.1b}} & &
	\multicolumn{1}{|L{70pt}|}{Ma \etal~\cite{ma13} \newline
	                           Rossi \etal~\cite{rossi15}} &
	\multicolumn{1}{L{78pt}|}{Rossi and Musolesi~\cite{rossi14}} &
	\multicolumn{1}{L{75pt}|}{Tockar~\cite{tockar14} \newline
	                          Cecaj \etal~\cite{cecaj14,cecaj16} \newline
	                          Kondor \etal~\cite{kondor17} \newline
	                          Riederer \etal~\cite{riederer16} \newline
	                          Wang \etal~\cite{wang18}} \\
\hhline{~~~~--~---}
	& &
	\multicolumn{1}{|l|}{} & &
	\multicolumn{1}{|L{25pt}|}{Profile \newline \vspace*{2pt}\textbf{F.2}} & 
	\multicolumn{1}{L{53pt}|}{Mobility model \newline \vspace*{2pt}\textbf{F.2a}} & &
	\multicolumn{1}{|l|}{} &
	\multicolumn{1}{L{75pt}|}{De Mulder \etal~\cite{de-mulder08} \newline
	                          Shokri \etal~\cite{shokri11} \newline
	                          Gambs \etal~\cite{gambs14} \newline
	                          Murakami \etal~\cite{murakami17}} &
	\multicolumn{1}{l|}{} \\
\hhline{~~~~~-~---}
	& &
	\multicolumn{1}{|l|}{} & &
	\multicolumn{1}{|l|}{} & 
	\multicolumn{1}{L{53pt}|}{Important locations \newline \vspace*{2pt}\textbf{F.2b}} & &
	\multicolumn{1}{|L{70pt}|}{Freudiger \etal~\cite{freudiger12} \newline
	                           Zang and Bolot~\cite{zang11}} &
	\multicolumn{1}{L{78pt}|}{Unnikrishnan and Naini~\cite{unnikrishnan13} \newline
		                        Naini \etal~\cite{naini16} \newline
	                          Rossi and Musolesi~\cite{rossi14}} &
	\multicolumn{1}{L{75pt}|}{Krumm~\cite{krumm07} \newline
	                          Goga~\cite{goga13}} \\
\hhline{~~~~~-~---}
	& &
	\multicolumn{1}{|l|}{} & &
	\multicolumn{1}{|l|}{} & 
	\multicolumn{1}{L{53pt}|}{Mobility features \newline \vspace*{2pt}\textbf{F.2c}} & &
	\multicolumn{1}{|L{70pt}|}{Rossi \etal~\cite{rossi15} \newline
	                           Zan \etal~\cite{zan13}} &
	\multicolumn{1}{l|}{} &
	\multicolumn{1}{l|}{} \\
\hhline{~~~~~-~---}
	& &
	\multicolumn{1}{|l|}{} & &
	\multicolumn{1}{|l|}{} & 
	\multicolumn{1}{L{53pt}|}{Social graph \newline \vspace*{2pt}\textbf{F.2d}} & &
	\multicolumn{1}{|l|}{} &
	\multicolumn{1}{l|}{} &
	\multicolumn{1}{L{75pt}|}{Srivatsa and Hicks~\cite{srivatsa12} \newline
	                          Ji \etal~\cite{ji14,ji16}} \\
\hhline{~~~~--~---}
	& &
	\multicolumn{1}{|l|}{} & &
	\multicolumn{2}{|L{77pt}|}{Auxiliary \newline \vspace*{2pt}\textbf{F.3}} & &
	\multicolumn{1}{|L{70pt}|}{Zang and Bolot~\cite{zang11} \newline
	                           De Montjoye \etal~\cite{de-montjoye15}} &
	\multicolumn{1}{l|}{} &
	\multicolumn{1}{L{75pt}|}{Goga~\cite{goga13}} \\
\hhline{~~-~--~---}
	& &
	\multicolumn{1}{|L{52pt}|}{Attribute linkage \newline \vspace*{2pt}\textbf{O.2}} & &
	\multicolumn{1}{|L{25pt}|}{Points \newline \vspace*{2pt}\textbf{F.1}} &
	\multicolumn{1}{L{53pt}|}{Subset \newline \vspace*{2pt}\textbf{F.1a}} & &
	\multicolumn{1}{|L{70pt}|}{Sui \etal~\cite{sui16}} &
	\multicolumn{1}{l|}{\cellcolor{gray!12}} &
	\multicolumn{1}{l|}{\cellcolor{gray!12}} \\
\hhline{~~-~--~---}
	& &
	\multicolumn{1}{|L{52pt}|}{Probabilistic \newline \vspace*{2pt}\textbf{O.3}} & &
	\multicolumn{1}{|L{25pt}|}{Points \newline \vspace*{2pt}\textbf{F.1}} &
	\multicolumn{1}{L{53pt}|}{Subset \newline \vspace*{2pt}\textbf{F.1a}} & &
	\multicolumn{1}{|L{70pt}|}{Gramaglia \etal~\cite{gramaglia17}} &
	\multicolumn{1}{l|}{\cellcolor{gray!12}} &
	\multicolumn{1}{l|}{\cellcolor{gray!12}} \\
\hhline{~~-~--~---}
\end{tabular}
\vspace*{-8pt}
\end{table*}

\subsubsection{Literature classification}
\label{sub:att-tax-lit}

We can now classify the existing works in the literature based on the three-dimensional
taxonomy proposed in the Sections above.
Table\,\ref{tab:attacks} summarizes how attacks against \data proposed in the literature are
positioned according to our taxonomy.
We note that record linkage attacks (\textbf{O.1}) have been thoroughly investigated, whereas
very little attention has been paid to other types of attacks (\textbf{O.2}, \textbf{O.3}).
This also results in that a variety of side information formats have been considered for
record linkage; instead, attribute linkage and probabilistic attacks have been only evaluated
with baseline format (\textbf{F.1a}).

The intersections of attacker objective and side information format (on rows) with the side
information source (on columns) also deserve attention. An important remark is that a single
type of source can generate multiple formats of side information. For instance, let us
look at the case of intra-record subsampling (\textbf{S.1a}): it results in a spatiotemporal
subset format (\textbf{F.1a}), if the points extracted from the target \data are used
as they are by the adversary; it can be cast to a spatiotemporal diverse sampling format
(\textbf{F.1b}), if the extracted points are perturbed in time and space; or, it can lead to
any profile formats (\textbf{F.2}), if the extracted points are post-processed to infer,
\eg important locations or specific mobility features.
Conversely, intra-record training (\textbf{S.1a}) and cross-database (\textbf{S.2}) sources
feature inherently diverse samplings with respect to the target data, hence cannot generate
a spatiotemporal subset format (\textbf{F.1a}), and the corresponding table cells are grayed out.
However, these sources can still produce all other side information formats.

Interestingly, our taxonomy highlights how some patterns are more frequent than others.
As an example, side information in the format of a mobility model (\textbf{F.2a}) typically
requires that the model is trained and tested on different datasets, making an intra-record
training (\textbf{S.1b}) the most appropriate type of source. Or, important locations (\textbf{F.2b})
are by far the most popular type of profile considered in attacks against \data, and have been
tested with all kinds of sources of side information. Also, there exist
substantial gaps in the literature when it comes to practical attacks that leverage a
cross-database source (\textbf{S.2}) and exploit profiles in the form of either mobility
models (\textbf{F.2a}) or features (\textbf{F.2c}).

Overall, Table\,\ref{tab:attacks} offers an outlook on well-explored as well as less investigated
attack surfaces against \data. It also motivates us to opt for a structure of the next
Sections that follows the rows of the table: indeed, it is along the objective dimension that
we identify most of the diversity among the reviewed studies; moreover, the format of the background
knowledge is what really guides the design of the attack strategy, whereas the side information
source is easily shaped into different formats, and does not allow for a rigorous classification.
Therefore, in the remainder of this Section we first present works that investigate record
linkage attacks based on different side information formats, and then discuss studies on attribute
linkage attacks and probabilistic attacks.

\subsection{Record linkage via subset of \data}
\label{sub:link-pos}

Record linkage attacks are straightforward when the side information database stores
a subset of the same spatiotemporal points present in the \data, as in class
\textbf{O.1}/\textbf{F.1a} in Table\,\ref{tab:attacks}. As shown in
Figure\,\ref{fig:record-linkage}, it is sufficient that the adversary looks for
records in the \data that include the points he owns.

The key question here is: how much such side information is required to perform a successful
linkage? The answer roots in the concept of \textit{unicity}, which is a measure of the
diversity that characterizes the movement patterns of different individuals. The higher
the unicity of \data, the higher the actual privacy risk connected with them: if the
monitored users have very diverse spatiotemporal trajectories, an adversary will likely
find a single \data record matching the side information points he owns.

In a pioneering work, Bettini \etal~\cite{bettini05} first hint at unicity in \data by introducing the notion of a location-based quasi-identifier (LBQID). The formal definition%
\footnote{The definition is inspired by the literature on privacy in traditional
relational databases. There, each record is associated with some data values, and a set of
quasi-identifiers: for instance, in a renowned study Sweeney~\cite{sweeney00} analyses a
database of all US citizens, where their individual healthcare records (\ie the actual data,
including admit/discharge dates, diagnoses, cures, and charges) are stored along with three
quasi-identifiers (\ie date of birth, gender, and ZIP code). The author finds that 87\% of
individuals have a unique combination of the three quasi-identifiers above, which clearly
raises questions on the privacy of the dataset.
In databases of \data, there is no clear distinction between quasi-identifiers and the actual
data, as any portion of the spatiotemporal trajectory itself can become a quasi-identifier,
depending on the positioning side information possessed by an adversary.  The formal definition
of LBQID captures this condition.}
of LBQID is \emph{``a spatio-temporal pattern specified by a sequence of spatio-temporal
constraints each one defining an area and a time span, and by a recurrence formula''}.
In other words, a LBQID is a sequence of spatiotemporal areas (which could specialize to
points at high granularity) that define a mobility pattern, and some associated frequency
of occurrence (which could specialize to just one occurrence).
The intuition of Bettini \etal~\cite{bettini05} is that it is possible to define LBQIDs
that require a minimum amount of knowledge about the spatiotemporal trajectory of a user,
and yet allow uniquely pinpointing his \data among those of a large population. In this
sense, an LBQID becomes a spatiotemporal pattern that is unique to one individual in a
database of \data.


\begin{figure}
\begin{minipage}[t]{0.3\textwidth}\vspace{0pt}
	\includegraphics[width=\textwidth]{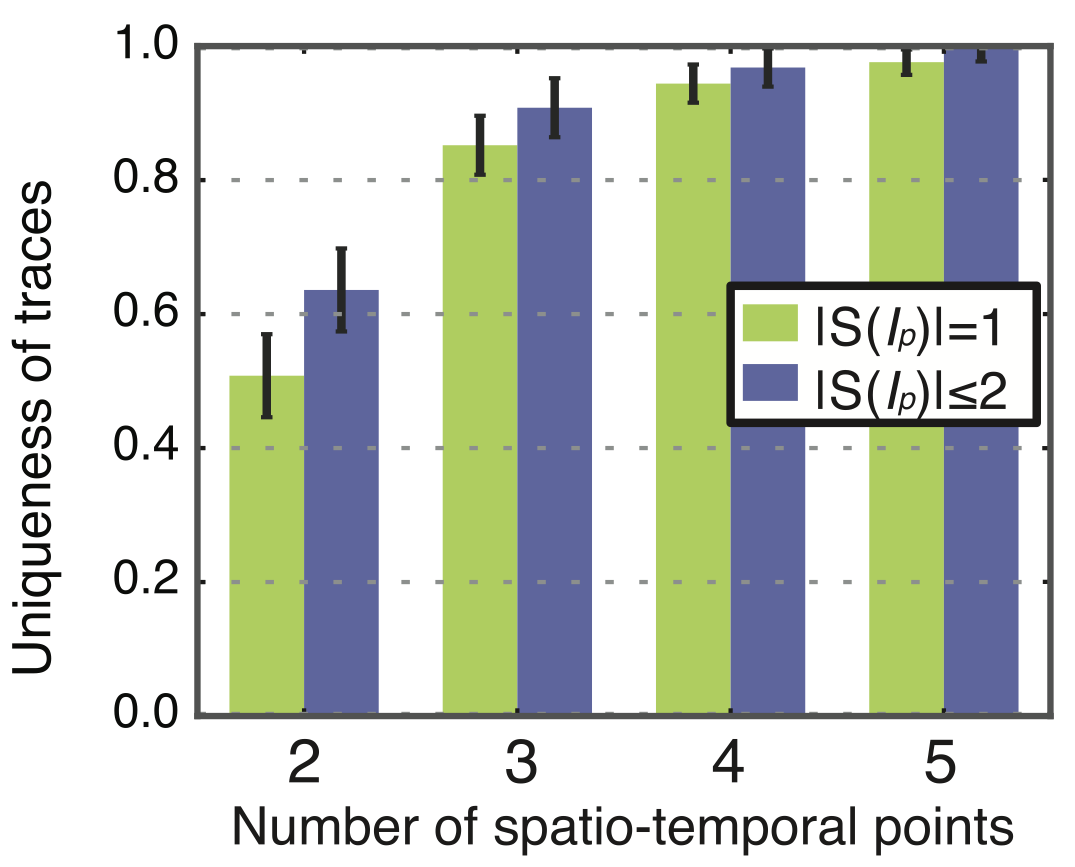}
\end{minipage}\hfill
\begin{minipage}[t]{0.68\textwidth}\vspace{-8pt}
	\caption{The unicity of trajectories with respect to the number of random spatiotemporal
points known to the adversary, denoted by $I_p$. The green bars represent the fraction
of unique trajectories, while the blue bars show the fraction of cases where the random
points identify one or two trajectories. Knowledge of four random points pinpoints 95\% of the
trajectories in a 1.5-million-record database.
Reproduced from De Montjoye \etal~\cite{de-montjoye13}.
%
}
	\label{fig:de-montjoye13_fig2b}
\end{minipage}
\vspace*{-8pt}
\end{figure}

The demonstration of such a conjecture is provided by the seminal work of
De Montjoye \etal~\cite{de-montjoye13}, who show that minimal LBQIDs, \ie very
little positioning side information, is sufficient to carry out successful linkage
attacks against \data.
Specifically, the authors prove that knowledge of a few
random points in the \data of a user allow pinning him down almost certainly, even
within a very large population%
\footnote{The dataset is composed of CDRs of 1.5 million users, collected by a
network operator during 15 months.\label{fn:de-montjoye13}}.
For instance, an adversary having observed the whereabouts of a target individual at
two random moments (whose corresponding spatiotemporal points are present in the target
database) during a whole year has a 50\% probability of recognizing his target
in a dataset of millions; the percentage grows to 95\% if as little as four random points
are known to the attacker. Figure\,\ref{fig:de-montjoye13_fig2b} shows the exact dynamics
of unicity versus the number of haphazard spatiotemporal points in the side information,
in the considered scenario.

The works above analyse mobile phone trajectories, but unicity is a general
characteristic of \data, no matter their original source. This is confirmed by
subsequent works that test the effectiveness of linkage attacks against \data
collected in a variety of ways.
Rossi \etal~\cite{rossi15} investigate unicity in GPS traces%
\footnote{The study leverages GPS \data from CabSpotting~\cite{epfl-mobility}
CenceMe~\cite{cenceme} and GeoLife~\cite{geolife}. The three datsets cover 536, 20 and
182 users, respectively, for several weeks and with diverse sampling frequencies.
A subset of the original data, of variable size, is used as side information; the
complement is used as target data.\label{fn:rossi15}},
and show that the high spatial accuracy (in the order of meters) of this kind of \data
exacerbates the phenomenon. Indeed, 100\% of the users in the datasets are pinpointed with
just two random spatiotemporal points.
Sapiezynski \etal~\cite{sapiezynski15} employ Wi-Fi \data%
\footnote{The data consists of Wi-Fi scans of 63 users, obtained by storing the list
of APs that respond to probe messages broadcasted by the users' mobile devices. As
AP locations are easily obtained from services such as Google Geolocation API, timestamped
Wi-Fi scans are a form of client-centric \data. The authors also use GPS traces of the
same users as ground truth.}
and demonstrate that knowing as little as 0.1\% of the APs seen by a user (\ie around 20
APs on average) allows tracking down a typical user during 90\% of her movements.
De Montjoye \etal~\cite{de-montjoye15} confirm that unicity persists also in the
case of \data from digital payments%
\footnote{The work used 3 months of credit card transactions of 1.1 million
users in 10,000 geo-referenced shops.\label{fn:de-montjoye15}}.
In this context, 90\% of
users are unique by assuming knowledge of four random points in their \data.
Interestingly, the authors also show that unicity varies with gender (women being 1.2
times more prone to unicity than men) and income level (high-income users being
1.7 times more prone to unicity than low-income ones): the authors speculate that
the reason is the location entropy, \ie the variety of different shops visited,
which is higher in women and high-income individuals.

\subsection{\hspace*{-0.7pt}Record\hspace*{-0.7pt} linkage\hspace*{-0.7pt} via\hspace*{-0.7pt} diverse\hspace*{-0.7pt} sampling\hspace*{-0.7pt} of\hspace*{-0.7pt} \data}
\label{sub:link-div}



As mentioned in Section\,\ref{sub:att-tax}, in practical cases the background knowledge
of the adversary is hardly a subset of the target \data. In a more credible scenario, the attacker
collects side information from a source that has the same spatiotemporal granularity of the
target database, but a different sampling of the actual trajectory of the target individuals.
This maps to class \textbf{O.1}/\textbf{F.1b} in Table\,\ref{tab:attacks}.

As illustrated in Figure\,\ref{fig:diverse-sampling}, record linkage attacks must reconciliate
spatiotemporal trajectories that are not one a subset of the other, which clearly makes the
problem more challenging. A first set
of strategies is proposed in a seminal work by Ma \etal~\cite{ma13}, who devise four
different estimators to measure the similarity of mobility traces in \data and
diversely sampled side information%
\footnote{Three pseudonymised GPS traces are considered in the study: 536 cabs in
San Francisco from CabSpotting~\cite{epfl-mobility}, 2,348 buses in Shanghai 
and 4,438 cabs in Shanghai. 
The side information is generated by adding noise to randomly sampled portions of the \data. \label{fn:ma13}}.
These are \textit{(i)} a Maximum Likelihood Estimator based on Euclidean distances,
\textit{(ii)} a Minimum Square Approach computing the negative sum of the squared
absolute value of the trajectory difference, \textit{(iii)} a Basic Approach that 
assumes Gaussian-noisy side information and defines the identity between two spatiotemporal
points based on the standard deviation of such noise, and \textit{(iv)} a Weighted
Exponential Approach that generalizes the previous technique to any noise distribution.
The methods achieve a linkage success rate of 50\%, when the adversary owns as little
as 10 observations of the original trajectory. The result then confirms the high privacy
risks associated to \data, even when the side information does not precisely match the spatiotemporal points in the target database.

A similar study is carried out by Rossi and Musolesi~\cite{rossi14} in the case of
location-based social networks (LBSNs), where localised and timestamped check-ins of
individual users result in \data%
\footnote{The data is crawled from three different LBSNs, \ie Brightkite and Gowalla~\cite{cho11},
and Foursquare~\cite{gao12}, and covers users in three US cities, \ie San Francisco,
New York and Los Angeles, for several months. The location information is approximated
by the check-in venue in the first two datasets, while it corresponds to the actual
user position in the third one. In each LBSN, a subset of the data is used as side
information, and the rest as the target \data.\label{fn:rossi14}}.
The authors show that a classifier based on a modified Hausdorff distance between
trajectories allows linking between 30\% and 60\% of the users in the \data with
10 LBSN check-ins.
Further evaluations of this approach with other datasets\cref{fn:rossi15}
presented by Rossi \etal~\cite{rossi15} show success probabilities over 90\%.

While the previous studies approximate diversely sampled side information by perturbing
or splitting the \data, further experiments consider the more realistic case
where \data and side information come from two sources that are actually different.
That boundary is first crossed by Tockar~\cite{tockar14}, who carries out a linkage
attack against \data of cabs in New York City, US%
\footnote{The data comprises pickup and drop-off times and locations, fare and tip
amounts of every yellow cab ride in New York City, US, in 2013. The dataset, released
by the New York City Taxi and Limousine Commission under Freedom of Information Law
(FOIL), is pseudonymised.}.
Tockar~\cite{tockar14} proves that the unicity of taxi trips makes them easily linked
to other databases that contain information about taxi rides of specific individuals.
To this end, he gathers an ad-hoc database by browsing gossip blogs and collecting
where and when celebrities used yellow cabs in the NYC area in 2013. Linkage of
spatiotemporal points allows the author to re-identify VIP passengers in the \data,
and hence the origin or destination of their trips, as well as the associated
tip amount. The latter are treated as the sensitive information, unveiling frequently
visited locations of celebrities, as well as their (lack of) generosity in tipping.
Although these pieces of information are deemed ``relatively benign'' by the author,
the trial represents a first clear example of actual privacy breach through a linkage
attack on \data.

Cecaj \etal~\cite{cecaj14,cecaj16} employ a real-world dataset of \data, and use
geo-tagged social network metadata as side information%
\footnote{The \data consist of pseudonymised CDRs of 2 million mobile subscribers, while
the side information are timestamped and geo-referenced posts of 700 usernames crawled
from Flickr and Twitter.}.
By applying a simple statistical learning approach based on matching and mismatching of
spatiotemporal points in the trajectories, they can link tens of social network usernames
to specific \data records. The result is in fact merely probabilistic, as it is not based
on actual ground truth (\ie the identity -- as social network username -- of users in the
\data): instead, a maximum a-posteriori estimation is used to compute the match probability.
A similar evaluation approach is considered by Kondor \etal~\cite{kondor17}, who investigate
the \textit{matchability} of large-scale datasets of \data%
\footnote{They use CDRs of 2.8 million mobile subscribers in Singapore, and smart card
bus/train transportation data of 3.3 million users collected by the Singapore Land
Transportation Authority (LTA). The two datasets are collected in the same time period.}.
To this end, they define space and time thresholds tailored towards the characteristics of
urban movements, and identify matching points (which are within the aforementioned
thresholds) and alibis (which are within the threshold in time, but not in space) in
trajectory pairs across the two datasets. Then, each trajectory in one dataset is linked
to that with the highest number of matching points and no alibis on the other dataset.
According to the results, the authors expect a successful match for around 8\% of users
in one-week datasets, and for about 33\% in one-month datasets. Such percentages grow
to 15\% and 60\% respectively, when focusing on very active, regular users only.

A more elaborate approach and dependable evaluation are proposed by Riederer \etal~\cite{riederer16},
who design a dedicated algorithm for linking \data and social network metadata.
The algorithm starts by computing a score for each pair of users across the two databases,
representing the likelihood of the user pair being actually the same person; it then
maximizes the overall score via a bipartite matching. The algorithm is proven to be
theoretically correct under the assumption that visits to a specific location during
a certain period follow a Poisson distribution and are independent of other visits.
Tests with real-world datasets%
\footnote{Three different pairs of \data databases are employed, \ie 862 users in two databases
crawled from Foursquare and Twitter, 1717	users in two databases crawled from Instagram
and Twitter, and 452 users in CDR and credit card record databases. In each scenario,
all databases are pseudonymised, but the ground-truth mapping of users between the two
databases is known.}
show that the algorithm outperforms approaches based on sparsity, frequency of visit
and density, reaching up to 0.95 precision and 0.7 recall in the best case.

The first test at scale is that recently performed by Wang \etal~\cite{wang18}. They
leverage an impressive collection of large-scale real-world datasets%
\footnote{Three types of datasets are considered in the study: mobile network CDRs of
more than 2 million subscribers, which is the target \data; GPS data of 56,000
Weibo social network users; check-in locations of 10,000 Weibo and 45,000 Dianping
application users.
All datasets are collected during the same week, and the side information datasets
only contain users who are also present in the target CDR database. User pseudonyms
are consistent across the datasets, which allows validate the results of record
linkage attacks on ground truth.}
to carry out a comparative analysis of record linkage attacks proposed in the literature,
including those by Ma \etal~\cite{ma13}, Rossi and Musolesi~\cite{rossi14},
Cecaj \etal~\cite{cecaj14,cecaj16}, and Riederer \etal~\cite{riederer16}.
All these strategies achieve hit precision sensibly higher than zero, hence they
can successfully link users across databases. However, and quite interestingly, the
results show that the performance of these strategies in presence of large-scale
real-world datasets are reduced with respect to those reported in the original papers,
with hit precision well below 20\% even when the side information comprises tens
of spatiotemporal points.
The authors' explanation is that each attack only addresses a subset of the
issues emerging in practical settings, which mostly stem from spatiotemporal
mismatches between target and side information data, and from database sparsity.
They then propose an attack technique that leverages a probabilistic representation
of the spatiotemporal mismatch and uses a simple Markov model to estimate missing
spatiotemporal points. The approach achieves a substantial gain in linking records,
with a maximum hit ratio of 40\%.




\subsection{Record linkage via mobility models}

A different attack surface for linkage is represented by indirect knowledge inferred from \data.
Let us imagine an adversary who does not have access to precise spatiotemporal points of his
target's \data, but knows instead some profiles that characterize the movements of his target.
Such an attacker could then extract the same profiles for all records in the \data, and try
to link a specific record to his side information.
Figure~\ref{fig:profile} illustrates this concept.

Mobility models are the first type of profile that has been considered in the literature,
as per class \textbf{O.1}/\textbf{F.2a} in Table\,\ref{tab:attacks}.
The strong \textit{regularity} that is known to characterize human movements~\cite{gonzalez08}
allows constructing simple models of individual movements that approximate well the
actual mobility patterns, and that thus represent a valuable side information.
The early work by De Mulder \etal~\cite{de-mulder08} is especially influential in this
sense. The authors assume that an adversary owns a Markovian model of his target's mobility;
the model describes memoryless transition probabilities among visited locations.
The authors then propose two matching strategies: the first builds a Markovian model
from each record in the target \data, and then compares the Markovian models directly;
the second calculates the probability that the specific sequence of locations in each
record of the \data is generated by the side information Markovian model.
Evaluations with real-world data%
\footnote{The study employs a processed version of the Reality Mining dataset~\cite{eagle06}.
Individual spatiotemporal trajectories of 100 volunteers are recovered from the entry
and exit time of each user at GSM cells. Side information is constructed from two months
of mobility, whereas another two months are used as target \data.}
show that the linkage is successful in 80\% of cases. Although the scale of the experiment
is small, the result still demonstrates that a risk exists. 

Markovian representations of mobility have been then considered as the side information
by several follow-up studies.
Shokri \etal~\cite{shokri11} assume an adversary who aims at linking transition probability
matrices to whole trajectories in a dataset of \data%
\footnote{The \data comprises of 20 mobility traces from the CabSpotting dataset~\cite{epfl-mobility},
where the sampling frequency is fixed at 5 minutes over 8 hours. Users move within the
San Francisco bay area, which is divided into 40 regions forming a 5 $\times$ 8 grid.
Side-information Markovian models are directly extracted from noisy subsets of the \data.}
via a maximum likelihood approach.
Gambs \etal~\cite{gambs14} consider a wide range of techniques to pair Markovian models
with trajectories in the target \data. Tests with heterogeneous datasets%
\footnote{The study employs GPS logs from five different scenarios, \ie Arum~\cite{killijian10},
GeoLife~\cite{geolife}, Nokia~\cite{nokia}, San Francisco cabs~\cite{epfl-mobility}, and Borlange~\cite{freudiger12}.
All datasets describe individual spatiotemporal trajectories, covering from a few users
to almost two hundred individuals. Side information Markov models are extracted from
a subset of each dataset, while the rest is considered as the \data for linkage.}
result in a success ratio of linkage attacks between 10\% and 50\%. Interestingly, the
authors show that the percentages grow proportionally with the sampling rate of
the \data.
Murakami \etal~\cite{murakami17} investigate the case where the side information
possessed by the attacker is limited and possibly collected from a non-identical set of
users. These conditions result in transition probability matrices of the Markovian models
that are sparse and erroneous.
In this scenario, previous techniques for linkage, such as the maximum likelihood used
by Shokri \etal~\cite{shokri11}, do not perform well. The authors then propose an attack
that \textit{(i)} reduces the problem dimension by means of group sparsity regularization
(\ie clustering) of the locations visited by users, and \textit{(ii)} estimates the
complete transition probability matrices via a dedicated factorization of the tensor
composed by all side-information (sparse) transition matrices.
Tests with real-world datasets%
\footnote{The authors employ data from 80 GeoLife users~\cite{geolife}, 250 Gowalla users~\cite{cho11},
and 400 Foursquare users~\cite{yang15}. All datasets cover tens of months of \data.
The side information is computed from 1.5\% to 9\% of the individual data, and
the remaining data is used as the target \data. \label{fn:murakami17}}
prove the effectiveness of the solution, which attains around 70\% AUC (Area Under the Curve)
of the ROC (Receiver Operating Characteristic) of the true positive rate against the false
positive rate. 

\subsection{Record linkage via important locations}


\begin{figure}
\begin{minipage}[t]{0.3\textwidth}\vspace{0pt}
	\includegraphics[width=\textwidth]{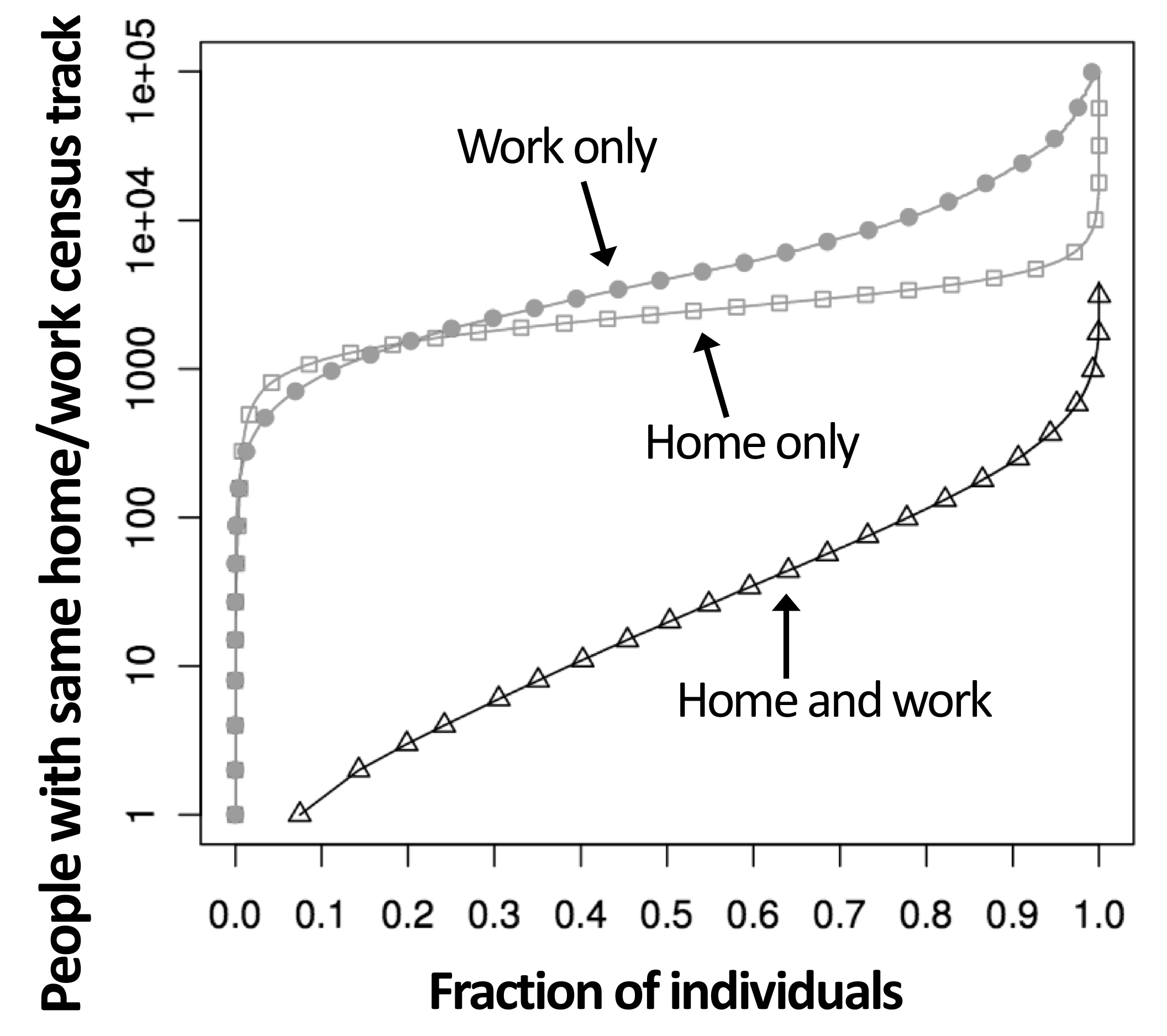}
\end{minipage}\hfill
\begin{minipage}[t]{0.68\textwidth}\vspace{-8pt}
	\caption{Fraction of population ($x$ axis) whose home (squares), work (circles) or
both home and work (triangles) locations (recorded at the census tract level) are shared with at most
$y$ other individuals.
While the set of people sharing the same home or work places are typically large (in the order
of several hundreds to tens of thousands), more than 5\% of workers do not share their combination of home
and work locations with any other individual in the US.
Adapted from Golle and Partridge~\cite{golle09}.}
	\label{fig:golle09_fig1a}
\end{minipage}
\vspace*{-8pt}
\end{figure}

Another class of \data profile used as side information are locations that are
frequently visited by the users, which maps to class \textbf{O.1}/\textbf{F.2b}
in Table\,\ref{tab:attacks}.
The most intuitive example is that of home and work locations. Such an apparently
basic knowledge poses in fact a very severe risk to privacy: Golle and Partridge~\cite{golle09}
show that the home and work locations of over 100 million individuals in the US, collected by
the Longitudinal Employer-Household Dynamics (LEHD) program, suffer from unicity.
Namely, more than 5\% of the population shows unique home-work location pairs at
tract granularity.
Figure\,\ref{fig:golle09_fig1a} provides a complete view of the result.

Based on this observation, in his seminal work, Krumm~\cite{krumm07} leverages home location side information to run linkage attacks on \data%
\footnote{The analysis is based on two-week (or longer) GPS \data of 172 volunteers.
The side information is easily obtained from an online ``white pages'' service, which
provides an association between the name and home address of individuals.
The ground truth are the actual home and work locations of the volunteers, who
communicated them as part of the experiment.}.
The author proposes four simple heuristic algorithms to infer the latitude and longitude of
home locations of each pseudonymised user whose trajectories are stored in a given \data
record. He then performs a reverse lookup for the home location in the side
information: if a unique matching entry is present, the attack allows linking
the \data record to an identity.
In the considered scenario, the chance of success for an adversary is at least 5\%.


The study is extended to the case where the attacker has side information about both
home and work locations by Freudiger \etal~\cite{freudiger12}. Their linkage attack
infers home and work locations in \data by: \textit{(i)} clustering spatial points
that map to frequently visited locations, via a variant of $k$-means;
\textit{(ii)} tagging the most popular location overnight as home, and that
during working hours as work.
The unicity of home and work locations allows pinpointing users in real-world
datasets%
\footnote{The study uses data from two-year GPS mobility of 24 cars in Borlange, Sweden
and one-year GPS mobility of 143 users in Lausanne, Switzerland~\cite{nokia}.
Home and work side information is artificially obtained by selecting a limited subset
of points from each user trace, via different subsampling strategies, and then applying
the same heuristic described above.}
with probabilities from 10\% to 90\%.

\begin{figure}
\centering
\includegraphics[width=0.9\columnwidth]{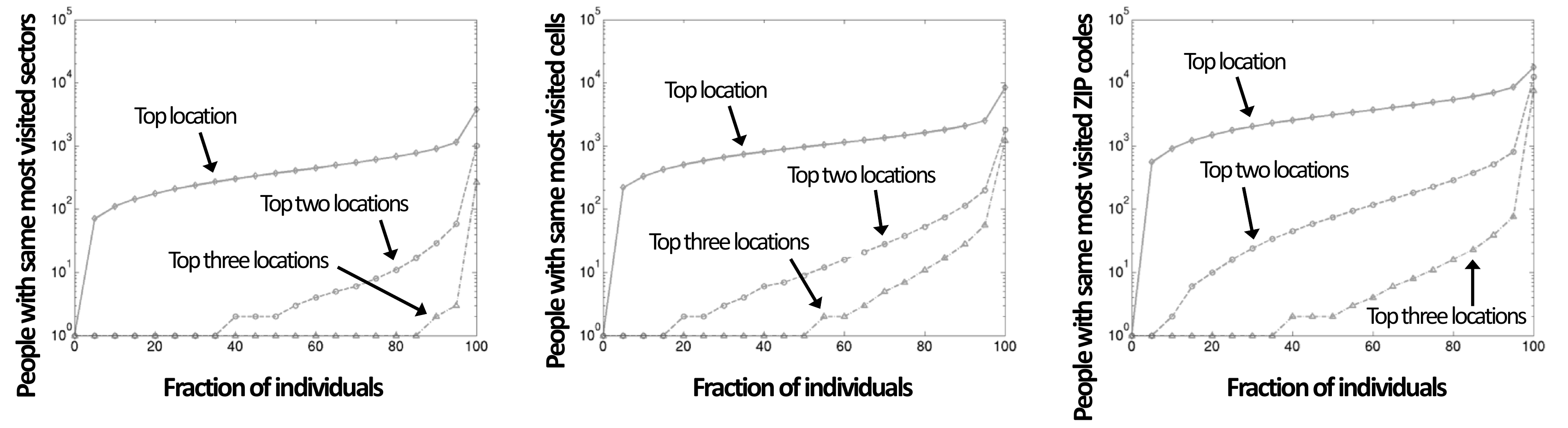}
\vspace*{-8pt}
\caption{Fraction of trajectories ($x$-axis) that share the top one, two or three most frequently
visited locations with at most $y$ other trajectories in the
target database. Plots refer to different spatial granularity levels with decreasing resolution:
US census sector (left), network cell (middle), and US ZIP code (right). While no trajectory is unique when
considering that only the top visited location is known to the attacker, all other situations
allow the adversary to perform successful record linkage on 5\% (top two locations known at a
ZIP code level) to 85\% (top three locations known at a sector level) of the trajectories of
20 million users.
Adapted from Zang and Bolot~\cite{zang11}.}
\vspace*{-8pt}
\label{fig:zang11_fig1abc}
\end{figure}

Zang and Bolot~\cite{zang11} consider a more generic notion of important locations, as
the top $n$ locations that are most frequently visited by a user, where $n$ is a small
number. They show that such side information is sufficient to distinguish a large
fraction of users among millions others%
\footnote{The study is performed on Call Detail Records (CDRs) of a nationwide US mobile
network operator, collected over a month and describing the spatiotemporal trajectories
of approximately 20 million subscribers. Side-information on important locations is
directly extracted from the same data used as target \data.\label{fn:zang11}}.
In their tests, 50\% of the individuals can be singled out by considering the top 3
mobile network cells they are observed at, and linkage is shown to be successful also
at different spatial granularity levels. Figure\,\ref{fig:zang11_fig1abc} portrays complete
results for $n\in[1,3]$.

Alternative versions of the \data profile of Zang and Bolot~\cite{zang11} are proposed in
subsequent works by Unnikrishnan and Naini~\cite{unnikrishnan13}, Naini \etal~\cite{naini16},
Goga \etal~\cite{goga13}, and Rossi and Musolesi~\cite{rossi14}.
In the first study, Unnikrishnan and Naini~\cite{unnikrishnan13} investigate the case where
the adversary's side information is in the form of histograms of the time spent by users
at different locations. Once comparable histograms are derived from the target \data,
linkage is formulated as a matching problem on a bipartite graph where vertices represent
records in the two datasets, and a maximum likelihood technique is used to solve it.
Their evaluation%
\footnote{The authors employ \data from Wi-Fi access to geo-referenced APs by over a thousand
students in the campus of EPFL, Lausanne, Switzerland. The data covers Mondays and Tuesdays in
two weeks. The histogram side information is inferred from days in the first week of the \data,
while days in the second week are used in their original format as the target dataset for linkage.}
results in a success probability of more than 50\% of the subject when using one day per
week, and of 70\% when considering two days.
A generalization of the study is proposed by Naini \etal~\cite{naini16}, using non-indentical
sets of individuals in the side information and target \data.

The approach is similar in the work by Goga \etal~\cite{goga13}, where the frequency histograms,
called location profiles, are computed at ZIP-code geographic granularity, and weighted so that
locations that are less common across all profiles but more representative of specific profiles
are valued more. The similarity between histograms is computed using a Cosine distance, as
other similarity metrics are shown to yield little difference in results. Interestingly, the
authors employ target and side information \data from actual different databases%
\footnote{The databases are crawled from Flickr, Twitter, and Yelp. Ground truth on the
identity of users appearing across databases is built by looking for accounts in the
three platforms that are associated to a same e-mail address, for all addresses mined
from a very large e-mail dataset. The final datasets include check-in data of 232,000 Twitter
users (used as the target \data), 22,000 Flickr users and 28,000 Yelp users (used as
side information).\label{fn:goga13}},
and show that their attack strategy is highly effective, linking records with 40\% to 80\% of
true positives and 1\% of false positives.
Finally, Rossi and Musolesi~\cite{rossi14} assume a slightly different form of side information,
\ie (time-dependent) distributions of visit frequencies at different locations. They leverage a
(time-dependent) multinomial na\"ive Bayes model to match the adversary knowledge to equivalent
distributions extracted from \data.
Experimental tests\cref{fn:rossi14} show that the attack has an accuracy between 40\% and 90\%,
depending on the dataset, when the adversary knowledge is built from 10 spatiotemporal points
from each record in the original \data.

\subsection{Record linkage via mobility features}

\begin{figure}
\centering
\includegraphics[width=0.75\columnwidth]{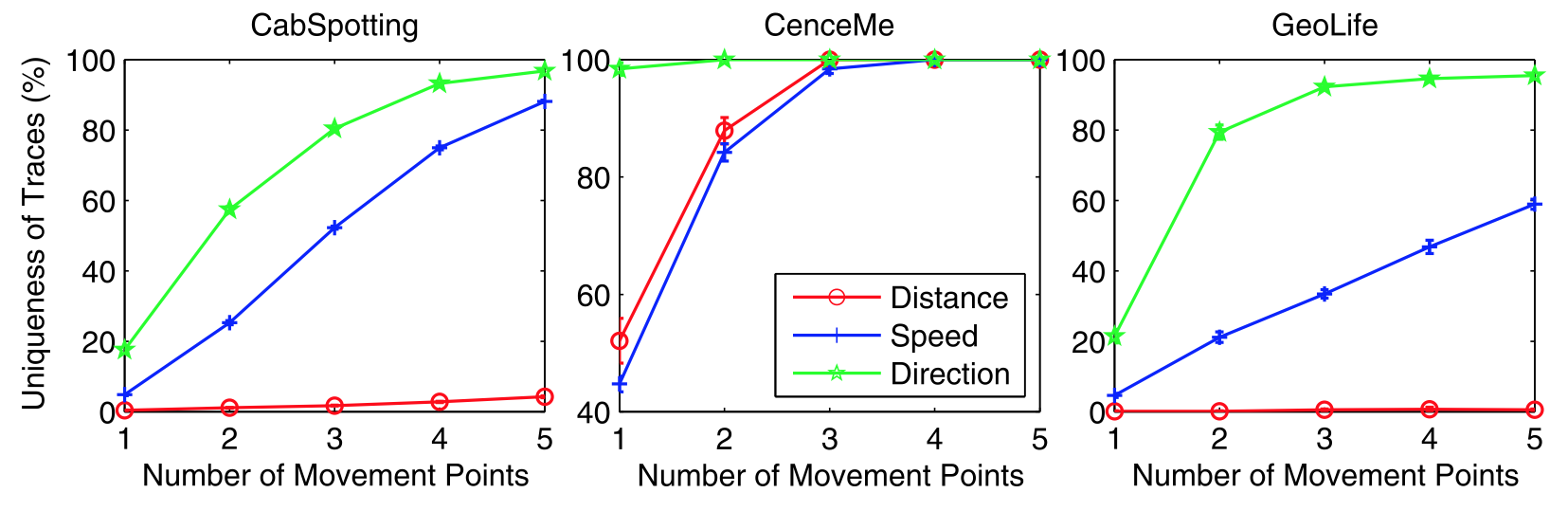}
\vspace*{-8pt}
\caption{Average percent unicity of trajectories in terms of their mean travelled distance (red circles), speed (blue crosses), or direction of movement (green stars) over time windows of 30 seconds. Unicity is portrayed versus the number of distance, speed or direction feature points in the adversary's side information. The three plots refer to different datasets: CabSpotting~\cite{epfl-mobility} (left), CenceMe~\cite{cenceme} (middle), and GeoLife~\cite{geolife} (right). Unicity typically rises well above 50\%
with just a few known feature points, and travel direction appears to be the most distinguishing feature.
Reproduced from Rossi \etal~\cite{rossi15}.
}
\vspace*{-8pt}
\label{fig:rossi15_fig2}
\end{figure}

A couple of works examine situations where the side information consists of specific
features that characterize the movements of individuals, hence falling in class
\textbf{O.1}/\textbf{F.2c} in Table\,\ref{tab:attacks}.
Rossi \etal~\cite{rossi15} assume that the adversary possesses knowledge of the travelled
distance, average speed, and heading direction collected during a number $n$ of different
(and typically small) time window.
This is equivalent to replacing spatiotemporal points with distance, speed or heading points
in the linkage attack. As all these features are easily inferred from \data records, they
can then be mapped to the side information.
The authors demonstrate that knowledge of these features poses an equivalent privacy risk
to that of regular spatiotemporal points. Tests with real-world \data\hspace*{-3pt}\cref{fn:rossi15}
demonstrate that unicity can reach values up to 95\%, although there is significant
variability across datasets and feature types.
Performance are summarized in Figure\,\ref{fig:rossi15_fig2}.

Another case where features are retained as side information is the peculiar scenario
envisioned by Zan \etal~\cite{zan13}, who focus on vehicular \data.
Their side information comprises features of the driving style
of each individual. An attack is proposed that builds on machine learning models
to classify vehicles in the \data based on type (car, truck, motorcycle).
The performance evaluation shows that this classification is already sufficient to
significantly reduce diversity in real-world fine-grained vehicular movement traces%
\footnote{Vehicular \data are collected by NGSIM on the US101 highway~\cite{ngsim},
during 10 minutes. Side information comprises profiles of speed, acceleration,
lane changing, and headway distance of each vehicle.},
hence making the unicity problem at least more severe.

\subsection{Record linkage via social graphs}
\label{sub:rl_social-graph}

In a distinctive study, Srivatsa and Hicks~\cite{srivatsa12} consider social graphs
as a profile side information, \ie class \textbf{O.1}/\textbf{F.2d} in Table\,\ref{tab:attacks}.
The assumption of the authors is that physical encounters are more
frequent among friends: hence, a contact graph derived from \data is tightly correlated to
a social graph such as that obtained by crawling friendships in social networks.
Under this condition, it is then possible for an adversary to link names in the social graph
with records in the \data, by extracting a contact graph from the latter.
Their proposed attack aims at linking nodes in the two graphs: it takes as an input the
pseudonymised contact graph, where a small number of nodes, called landmarks, are de-anonymized
either by leveraging centrality features or by exploiting leaked information. Starting from
landmarks, the method completes the mapping between graphs using distance vectors, spanning
tree matching or local subgraph features. The output mapping allows assigning social network
identifiers to the spatiotemporal trajectories in \data.
Evaluations with real-world datasets%
\footnote{The analysis leverages \data in the form of individual trajectories from geo-referenced
associations of mobile devices with Wi-Fi APs in the university campus of St Andrews, UK~\cite{standrews}.
The side information is derived from Facebook friendship relationships of the same set of student
volunteers in the \data. Ground-truth information, mapping Facebook identities to mobility traces,
is also provided by the experiment participants.
\label{fn:srivatsa12}}
show that the attack can achieve a high accuracy above 80\%; however, it is important to
note that the result holds when considering a relatively small user base (less than
$125$ individuals), and that the scalability of the attack to much larger datasets
is unclear.
Improvements to the approach have been more recently proposed by Ji \etal~\cite{ji14,ji16},
which however do not address the questions on scalability.

\subsection{Record linkage via auxiliary side information}
\label{sub:link-aux}


Auxiliary data that is unrelated to the movements of the target individual may also be exploited during attacks on \data. The nature of the auxiliary information is potentially very diverse.

For instance, one could consider as auxiliary data the knowledge of whether the target person is actually present in the database under attack; in fact, this piece of information is typically taken for granted in the works reviewed in the previous Sections, hence all previously discussed attacks implicitly leverage auxiliary data in a sense.
As another example, in the case where attacks occur against anonymized datasets, knowledge of the privacy-preserving transformation used to protect the \data can also be regarded as auxiliary information; this type of auxiliary data is instead never part of the adversary's side information in the attacks surveyed in the present Section, which are only run on pseudonimyzed datasets.

A couple of works in the literature show how auxiliary side information impacts
the unicity of \data in practical case studies, and fall in class \textbf{O.1}/\textbf{F.3}
in Table\,\ref{tab:attacks}.
Zang and Bolot~\cite{zang11} show that unicity in real-world large-scale
\data\cref{fn:zang11} is sensibly increased when the auxiliary data consists of
minimal social information about the target user. Namely, they assume that the
adversary also knows whether his target individual is especially social or not, \ie he calls
more than 20 unique persons in a month or otherwise.
This notion is also included as a flag field in each record of the target \data database.
Knowledge of this auxiliary one-bit piece of information permits unicity increase
by around 50\% on average in the million-strong dataset considered in the study.
De Montjoye \etal~\cite{de-montjoye15} examine instead the case where auxiliary
purchase cost data is associated to each spatiotemporal point in \data from credit
card transactions\cref{fn:de-montjoye15}. If the adversary happens to be able to
associate a purchase cost to each of the target's trajectory point he knows, 
his chances of success grow by 22\% on average.

The auxiliary data leveraged by Goga \etal~\cite{goga13} consists instead of
language and timing information, which is used to complement the important
locations that represent the adversary's baseline side information. Tests
with different real-world datasets\cref{fn:goga13} reveal that such an auxiliary
information allows for mild improvements of re-identification rates: in
particular, knowing the language of individuals does not help pinpointing
users in a more accurate way than just relying on important location profiles.

\subsection{Attribute linkage via subset of \data}
\label{sub:homo}

We now move to a different category of attacks, whose objective is not record
linkage but attribute linkage. These map to class \textbf{O.2}/\textbf{F.1a}
in Table\,\ref{tab:attacks}. In this case, the adversary exploits a weakness of
databases that is referred to as \textit{homogeneity}, and is subtler than unicity.
Hence, as mentioned in Section\,\ref{sub:att-tax-obj}, attribute linkage attacks
are also called homogeneity attacks in the literature.

In order to clarify the privacy risk associated with homogeneity, let us consider
a \data database where unicity is completely absent, and each record contains in
fact \data that cannot be told apart from those of a large number of other records.
In this case, any side information always matches many records, and a linkage with
the correct record has low chances of success.
However, recall that the ultimate goal of an adversary is the inference of 
the sensitive information within a record, and not of the record itself. And, the
fact that the \data in each record are not unique does not necessarily imply the
same for the sensitive part of the database.
Homogeneity is precisely the lack of diversity in the sensitive information across
the set of indistinguishable \data records. In other words, it can be understood
as an extension of unicity to the level of combined trajectory data and sensitive
attributes, rather than at the trajectory level only.

As depicted in Figure\,\ref{fig:attribute-linkage}, attribute linkage attacks exploit
homogeneity. The figure refers to the case where sensitive attributes are specific
attribute fields in each record, separated from the spatiotemporal points; yet,
attribute linkage can be also cast to the other perspective outlined in
Section\,\ref{sub:att-tax-obj}, where sensitive information is embedded in the
spatiotemporal trajectory of the user.

To date, there are no investigations of attribute linkage attacks against
\data databases that also contain separate fields with sensitive information.
Instead, a recent work by Sui \etal~\cite{sui16} explores the case where the sensitive
knowledge is embedded in the \data. More precisely, this study assumes that the two
most frequently locations visited by an individual are sensitive implicit attributes
in \data. Considering that the side information of an adversary consists of three random
spatiotemporal points, the authors show that real-world \data%
\footnote{The evaluation employs association logs of 150,000 users to 2,670 APs of
the Wi-Fi network of Tsinghua University, China. The authors assume the side information
available to the attacker to be formed by a subset of the target \data.}
is affected by significant homogeneity.
Namely, 35\% of records in their database are not unique, but 40\% of such non-unique
records are homogeneous.
An interesting corollary observation from the results in this study is that homogeneity
scales exponentially with the number of indistinguishable records in the considered scenario:
therefore, the problem of homogeneity may thus be inherently mitigated in fairly large sets
of records where a large number of records share similar, non-unique \data.

\subsection{Probabilistic attacks via subset of \data}

Probabilistic attacks are the third type of threat against \data that has been considered
in the literature, and correspond to class \textbf{O.3}/\textbf{F.1a} in Table\,\ref{tab:attacks}.
In this case, the adversary is successful if, upon accessing the database,
he increases his knowledge of the target individuals' trajectories by any non-negligible amount.
In a sense, probabilistic attacks can be understood as a generalization of attribute linkage:
while attribute linkage leverages the homegeneity of specific attributes (\ie those deemed to be
sensitive), a probabilistic attack can take advantage of the homogeneity of any field in the
database. Therefore, probabilistic attacks have a significantly broader scope than record or
linkage attacks. An illustration of this concept is in Figure\,\ref{fig:probabilistic}, where
a successful probabilistic attack is performed on a trajectory dataset where unicity and
homogeneity are removed.

The concept of probabilistic attacks against \data is evoked in a recent work by
Gramaglia \etal~\cite{gramaglia17}, under the simple assumption of a spatiotemporal
subset format obtained from intra-record subsampled data%
\footnote{The study leverages CDR datasets provided by mobile network operators
and describing spatiotemporal trajectories of subscribers in the
cities of Abidjan (29,191 individuals), Dakar (71,146 individuals) and Shenzhen (50,000 individuals),
and in the countries of Ivory Coast (82,728 individuals) and Senegal (286,926 individuals)
during a period of two weeks in each scenario.}.
In their work, the authors only introduce the threat from a conceptual standpoint, and do
not run actual attacks on real-world data. In conclusion, we currently lack
comprehensive, data-driven assessments of the effectiveness of probabilistic attacks in the
context of \data, which limits our current understanding of the actual risks entailed by this
kind of adversarial strategy.

\section{Anonymization of \data}
\label{sec:anonym}

Having reviewed the different classes of attacks on \data, in the second part of the survey we turn
our attention to the different techniques adopted in order to protect databases of \data from
the threaths in Section\,\ref{sec:att}. We first propose a taxonomy of the anonymization solutions
in the literature, in Section\,\ref{sub:anon-tax}; this allows us to structure the subsequent
detailed discussion of relevant works, in Sections\,\ref{sub:anon-mit} to \ref{sub:anon-uninf}.



%
%

\subsection{A taxonomy of anonymization techniques}
\label{sub:anon-tax}

Anonymization techniques are primarily characterized by the \emph{privacy principle} they seek to implement in the data. A privacy principle expresses some conditions on the maximum knowledge that can be gained by the attacker upon accessing the target database. Two major privacy principles have been considered in the literature, as far as the anonymization of \data is concerned.
\begin{itemize}
\item \emph{Indistinguishability} commends that each record in a database must not be distinguishable from a large enough group of other records in the same database, called \emph{anonymity set}. In the context of \data, the principle implicitly assumes that the adversary's knowledge is limited to some portion of the movement of the target individual. It
ensures that such an adversary will not be able to pinpoint a single trajectory in the target database; instead, the attacker will retrieve the whole batch of indistinguishable trajectories of all users in the anonymity set. Ultimately, when applied at the record level, the principle effectively removes unicity in \data. Extensions to the baseline principle can also tackle homogeneity, by applying the same principle at the attribute level, and granting that a sufficiently large set of trajectories share not only mobility information but also sensitive attributes. OVerall, indistinguishability is a sound countermeasure against record and attribute linkage attacks.
\item \emph{Uninformativeness} enforces that the difference between the knowledge of the adversary before and after accessing a database must be small. It is apparent that uninformativeness is a much more general principle that does not make any assumption on the adversary knowledge. This principle is suitable to address probabilistic attacks, hence providing much stronger privacy guarantees than indistinguishability.
\end{itemize}
In addition to the principles above, a substantial amount of works adopt less rigorous
privacy notions, and we group those under the following loose privacy principle.
\begin{itemize}
\item \emph{Mitigation} aims at reducing circumstantial privacy risks associated with the data,
without pursuing a well-defined privacy principle.
\end{itemize}

\begin{table*}
\caption{Classification of the techniques proposed in the literature to anonymize \data databases. Our taxonomy is outlined by the first two columns, that tell apart anonymization solutions based on the privacy principle they adopt, and on the privacy criterion used to implement the principle. The last three columns indicate which class of attack each solution aims at countering, based on the adversary's objective (\textbf{O}) outlined in Section\,\ref{sub:att-tax-obj}. Solutions based on the three privacy principles in the leftmost column are presented in Section~\ref{sub:anon-mit} (mitigation), Section~\ref{sub:anon-ind} (indistinguishability), and Section~\ref{sub:anon-uninf} (uninformativeness), respectively.\vspace*{8pt}}
\label{tab:anon-tax}
\scriptsize
\centering
\renewcommand{\arraystretch}{1.3}
\setlength{\tabcolsep}{1.5pt}
\begin{tabular}[c]{l l @{}p{3pt}@{} l l l}
	
	\multicolumn{2}{c}{} & &
	\multicolumn{3}{l}{\textbf{Attacker objective (\textbf{O}) \vspace*{2pt}}} \\
\hhline{~~~---}
	
	\multicolumn{1}{p{61pt}}{\textbf{Privacy \newline principle}} &
	\multicolumn{1}{p{78pt}}{\textbf{Privacy \newline criterion}} & &
	\multicolumn{1}{|L{96pt}|}{Record linkage \newline (\emph{Re-identification}) \newline \vspace*{2pt}\textbf{O.1}} &
	\multicolumn{1}{L{63pt}|}{Attribute linkage \newline (\emph{Homogeneity}) \newline \vspace*{2pt}\textbf{O.2}} &
	\multicolumn{1}{L{81pt}|}{Probabilitic \newline (\emph{Inference}) \newline \vspace*{2pt}\textbf{O.3}} \\
\hhline{~~~---}\\[-7pt]
\hhline{--~---}
	
\multicolumn{1}{|L{61pt}|}{Mitigation} &
	\multicolumn{1}{l|}{Obfuscation} & &
	\multicolumn{1}{|L{96pt}|}{Srivatsa \etal~\cite{srivatsa12}} &
	\multicolumn{1}{l|}{} &
	\multicolumn{1}{l|}{} \\
\hhline{~-~---}

\multicolumn{1}{|l|}{} &
	\multicolumn{1}{l|}{Cloaking} & &
	\multicolumn{1}{|L{96pt}|}{Hoh \etal~\cite{hoh06} \newline
							   Murakami \etal~\cite{murakami17} \newline
							   Ma \etal~\cite{ma13} \newline
							   Rossi \etal~\cite{rossi15}} &
	\multicolumn{1}{l|}{} &
	\multicolumn{1}{l|}{} \\
\hhline{~-~---}

\multicolumn{1}{|l|}{} &
	\multicolumn{1}{l|}{Segmentation} & &
	\multicolumn{1}{|L{96pt}|}{Song \etal~\cite{song14}} &
	\multicolumn{1}{l|}{} &
	\multicolumn{1}{l|}{} \\
\hhline{~-~---}

\multicolumn{1}{|l|}{} &
	\multicolumn{1}{l|}{Swapping} & &
	\multicolumn{1}{|L{96pt}|}{Salas \etal~\cite{salas18}} &
	\multicolumn{1}{l|}{} &
	\multicolumn{1}{l|}{} \\
\hhline{--~---}\\[-7pt]
\hhline{--~---}


\multicolumn{1}{|L{61pt}|}{Indistinguishability} &
	\multicolumn{1}{L{78pt}|}{$k$-anonymity via spatiotemporal generalization} & &
	\multicolumn{1}{|L{96pt}|}{Yavonoy \etal~\cite{yarovoy09} \newline
                             De Montjoye \etal~\cite{de-montjoye13} \newline
    	                       Zang \etal~\cite{zang11} \newline
	                           Gramaglia and Fiore~\cite{gramaglia15}} &
	\multicolumn{1}{l|}{} &
	\multicolumn{1}{l|}{} \\
\hhline{~-~---}

	\multicolumn{1}{|l|}{} &
	\multicolumn{1}{L{78pt}|}{$k$-anonymity via suppression} & &
	\multicolumn{1}{|L{96pt}|}{Terrovitis and \newline Mamoulis~\cite{terrovitis08}} &
	\multicolumn{1}{l|}{} &
	\multicolumn{1}{l|}{} \\
\hhline{~-~---}

	\multicolumn{1}{|l|}{} &
	\multicolumn{1}{L{78pt}|}{$k$-anonymity via generalization and suppression} & &
	\multicolumn{1}{|L{96pt}|}{Nergiz \etal~\cite{nergiz09} \newline
	                           Monreale \etal~\cite{monreale10} \newline
		                         Gramaglia and Fiore~\cite{gramaglia15} } &
	\multicolumn{1}{l|}{} &
	\multicolumn{1}{l|}{} \\
\hhline{~-~---}

	\multicolumn{1}{|l|}{} &
	\multicolumn{1}{L{78pt}|}{$k$-anonymity via micro-aggregation and suppression} & &
	\multicolumn{1}{|L{96pt}|}{Domingo-Ferrer and Trujilo-Rasua~\cite{domingo12} \newline
	                           Torres and Trujilo-Rasua~\cite{torres16} \newline
	                           Naini \etal~\cite{naini16}} &
	\multicolumn{1}{l|}{} &
	\multicolumn{1}{l|}{} \\
\hhline{~-~---}
	
	\multicolumn{1}{|l|}{} &
	\multicolumn{1}{L{78pt}|}{Generalized $k$-anonymity with spatial uncertainty} & &
	\multicolumn{1}{|L{96pt}|}{Abul \etal~\cite{abul08} \newline
	                           Abul \etal~\cite{abul10}	\newline
	                           Kopanaki \etal~\cite{kopanaki16}} &
	\multicolumn{1}{l|}{} &
	\multicolumn{1}{l|}{} \\
\hhline{~-~---}

	\multicolumn{1}{|l|}{} &
	\multicolumn{1}{|l|}{$l$-diversity, $t$-closeness} & &
	\multicolumn{1}{|L{96pt}|}{Tu \etal~\cite{tu2017beyond,tu18}} &
	\multicolumn{1}{l|}{Tu \etal~\cite{tu2017beyond,tu18}} &
	\multicolumn{1}{l|}{} \\
\hhline{--~---}\\[-7pt]
\hhline{--~---}

\multicolumn{1}{|L{61pt}|}{Uninformativeness} &
	\multicolumn{1}{l|}{$(\epsilon$,$\delta)$-differential privacy} & &
	\multicolumn{1}{|L{96pt}|}{} &
	\multicolumn{1}{l|}{} &
	\multicolumn{1}{l|}{Shao \etal~\cite{shao13}} \\
\hhline{~-~---}

\multicolumn{1}{|l|}{} &
	\multicolumn{1}{l|}{$\epsilon$-differential privacy} & &
	\multicolumn{1}{|L{96pt}|}{} &
	\multicolumn{1}{l|}{} &
	\multicolumn{1}{L{81pt}|}{Chen \etal~\cite{chen12} \newline
						Chen \etal~\cite{chen12_ccs} \newline
						Bonomi and Li~\cite{Bonomi13} \newline
						Qardaji \etal~\cite{qardaji13} \newline
						Zhang \etal~\cite{zhang16} \newline
						He \etal~\cite{he15} \newline
						Mir \etal~\cite{mir13} \newline
						Roy \etal~\cite{Roy2016} \newline
						Gursoy \etal~\cite{gursoy18}} \\
\hhline{~-~---}

\multicolumn{1}{|l|}{} &
	\multicolumn{1}{l|}{Plausible deniability} & &
	\multicolumn{1}{|L{96pt}|}{} &
	\multicolumn{1}{l|}{} &
	\multicolumn{1}{L{81pt}|}{Bindschaedler and Shokri~\cite{bindschaedler16} \newline
						Bindschaedler \etal~\cite{bindschaedler17}} \\
\hhline{~-~---}

\multicolumn{1}{|l|}{} &
	\multicolumn{1}{l|}{$k^{\tau,\epsilon}$-anonymity} & &
	\multicolumn{1}{|L{96pt}|}{Gramaglia \etal~\cite{gramaglia17}} &
	\multicolumn{1}{l|}{} &
	\multicolumn{1}{l|}{Gramaglia \etal~\cite{gramaglia17}} \\
\hhline{--~---}

\end{tabular}
\end{table*}

The privacy principles above constitute the first dimension of our taxonomy, reflected in the
first column in Table\,\ref{tab:anon-tax}.
Note, however, that privacy principles are abstract definitions. In order to be applied in practical
cases, they need to be specialized into \emph{privacy criteria} that define the exact requirements
that a database needs to meet in order to comply with the principle. Privacy criteria are categorized as per the second column of Table\,\ref{tab:anon-tax}. Next, we provide brief primers on each privacy principle, possibly introducing the key criterion used to implement it in the literature. Then, we discuss in detail mitigation techniques in Section~\ref{sub:anon-mit}, approaches based on indistinguishability in Section~\ref{sub:anon-ind}, and those aiming at uninformativeness in Section~\ref{sub:anon-uninf}.

\subsubsection{Mitigation of privacy risks} 
\label{sub:mix_def}

A large body of works does not target a rigorously defined privacy principle, rather aims at mitigating privacy risks in \data. Approaches in this category propose perturbation of the data records, reduction of the spatial or temporal resolution of the data, or arbitrary trimming of the trajectories. However, such strategies do not offer any formal guarantee in terms of privacy.
A number of variants of these simple models exist, and we survey them in Section\,\ref{sub:anon-mit}.

\subsubsection{Indistinguishability via $k$-anonymity.}
\label{sub:k-anon_def}

Indistinguishability is mainly implemented via \emph{$k$-anonymity}, a privacy criterion first
introduced by Sweeney~\cite{sweeney02} for relational micro-data that has also found wide application
with trajectories. The idea behind $k$-anonymity is that any subset of the spatiotemporal
points of each user in a \data database shall not be distinguishable from the spatiotemporal points
of at least $k-1$ other users in the same database. Formally:
\begin{definition}
Let $\mathcal{D}$ be a database of \data and $LBQID$ the location-based quasi-identifier associated
with it, 
and let $\mathcal{D}[LBQID]$ be the set of records returned by a query for $LBQID$ on $\mathcal{D}$.
Then, $\mathcal{D}$ satisfies $k$-anonymity if and only if the records in $\mathcal{D}[LBQID]$ are
at least $k$.
\end{definition}
The $k$-anonymity criterion can be implemented with many and varied techniques, from generalization to microaggregation, and has been also augmented in several ways. These variants map to different
rows under the indistinguishability privacy principle in Table\,\ref{tab:anon-tax}, and we will discuss them in detail in Section\,\ref{sub:anon-ind}.

\subsubsection{Uninformativeness via differential privacy}
\label{sub:dp_def}

Uninformativeness is typically achieved through \emph{differential privacy}, whose original definition
by Dwork~\cite{dwork06} imposes that the result of a query on a differentially private database must yield only a small variation depending on whether a specific record is present or not in the database. Formally, $\epsilon$-differential privacy, which is the standard form of differential privacy, is defined as follows:
\begin{definition}
A randomized algorithm $\mathcal{A}$ offers $\epsilon$-differential privacy if for all datasets $\mathcal{D}'$
and $\mathcal{D}''$ differing in one element ($|\mathcal{D}'-\mathcal{D}''|=1$), and for all subsets $S$ of
the output of $\mathcal{A}$, it holds
$Pr[\mathcal{A}(D') \in S] \leq e^\epsilon \times Pr[\mathcal{A}(D'') \in S]$.
\end{definition}
The concept of ``small'' difference between query results is embodied by the so-called budget parameter $\epsilon$, which regulates the amount of diversity in the query result allowed when removing a single individual from the database. Thus, differential privacy realizes uninformativeness by ensuring that the additional knowledge gained by the adversary when he accesses the database is bounded to $\epsilon$.

An important remark is that differential privacy, as defined above, is a condition on the algorithm used
to extract information from the database, and not on the database itself. Therefore, differential privacy
is not immediately related to PPDP of \data, rather to privacy-preserving data mining (PPDM). PPDM is a
completely different problem from PPDP, as it assumes that the exact operations that will be run on the
database are known a-priori, and can be included in the anonymization process.
However, as we will see, this criterion can be adapted to the case where databases of \data are to be
published for generic future mining purposes.

Moreover, differential privacy is not the only criterion that implements uninformativeness. Other models,
based on extensions of $k$-anonymity, have also been proposed. We will review all anonymization techniques
aiming at satisfying the uninformativeness principle for PPDP of \data in Section\,\ref{sub:anon-uninf}.

\subsubsection{Literature classification}

We classify the works in the literature according to the privacy criterion they implement, as listed in
the rows of Table\,\ref{tab:anon-tax}. However, we also complete our taxonomy with one additional dimension,
orthogonal to privacy principles and criteria, \ie the type of attack on \data that each anonymization technique
is intended to tackle. The columns of Table\,\ref{tab:anon-tax} tell apart different attacker objectives (\textbf{O}),
categorized according to the discussion in Section\,\ref{sub:att-tax-obj}. The labels within parenthesis report alternative
names for these attacks that are frequently used in the literature on anonymization: specifically,
record and attribute linkage are often referred to as re-identification and homogeneity attacks, respectively;
inference is the terminology typically employed to indicate probabilistic attacks.
	
The taxonomy in Table\,\ref{tab:anon-tax} allows us to catalogue works that propose anonymization techniques
for \data based on the combination of their underlying privacy principle/criterion and the type of attack they
are effective against.
We should note that some of the works discussed in this survey do not make their assumed attacker model explicit.
However, the proposed anonymization model description implies the type of attack they could be used against.
For example, in the cases of Torres and Trujillo-Rasua~\cite{torres16} and Kopanaki \etal~\cite{kopanaki16}, the proposed $k$-anonymity models naturally protect against record linkage attacks, even if not specified in the papers.

Although fairly sparse, the table highlights how the vast majority of the literature is focused on mitigating
or preventing record linkage on published databases of \data. Also, some expected correlations emerge:
indistinguishability is mostly suitable to counter record linkage, while uninformativeness tends to be used
to develop solutions against probabilistic attacks. Interestingly, mitigation techniques can only cope with
the simplest class of attack, \ie record linkage, due to their heuristic nature.
Variations of these baseline matches of criterion and attack are rare, and we will
detail them in our following discussion.

\subsection{Solutions providing mitigation of privacy risks}
\label{sub:anon-mit}

We start our review of solutions for the anonymization of \data by presenting techniques that do not implement any well-defined privacy principle, rather mitigate privacy risks without theoretical or provable guarantees. In the following, we tell apart such heuristic solutions based on the type of transformation they perform on the data.

\subsubsection{Obfuscation}

A very simple solution consists in distorting location data by adding noise to it. The \textit{value distortion} technique is originally introduced for privacy preserving data mining (PPDM) of location data by Agrawal and Srikant~\cite{agrawal00}, and later formalized as \textit{obfuscation} in LBS environments by Duckham and Kulik~\cite{duckham05}. Srivatsa and Hicks~\cite{srivatsa12} add different models of random noise to their social-graph representations of trajectories (see Section~\ref{sub:rl_social-graph}), and show\cref{fn:srivatsa12} that they can reduce the success of record linkage attacks in a substantial way only if the level of noise is high.

\subsubsection{Cloaking}

Another baseline strategy is to reduce the granularity of the trajectory data in space or time, which is often referred to as \textit{cloaking} as per the seminal work by Gruteser and Grunwald~\cite{gruteser03} in LBS systems. Hoh \etal~\cite{hoh06} show that increasing the sampling interval from one to four minutes (\ie only retaining every fourth sample) in their \data%
\footnote{Tests are conducted on real-world GPS traces of vehicles in the Detroit area, tracked during a week with a frequency of $1$ sample per minute.}
reduces home identification rates from $85$\% to $40$\%, although the risk is far from being removed.
Murakami \etal~\cite{murakami17} adopt a slightly different approach, and selectively remove a given fraction of points from the original trajectories, either randomly or so as to minimize the opportunities for linkage by an attacker (this second option assumes knowledge of the adversary's side information). The authors report that, by deleting up to $5$ points from all trajectories in their reference databases\cref{fn:murakami17}, the performance of record linkage attacks are halved, yet remain high in absolute terms.

Ma \etal~\cite{ma13} tamper instead with the spatial dimension of \data, and show that, in the case of their datasets\cref{fn:ma13}, reducing the geographical accuracy of the spatiotemporal points does not have a clear positive effect on unicity: the chances that a record linkage attack is successful stay above 50\% when the adversary knows as little as $8$ points of its target's mobility. In fact, in situations where the adversary knowledge is also inaccurate, a lower granularity may even lead to increased record linkage: the authors ascribe this effect to the fact that a coarser cell structure mitigates mistakes in the attacker's side information.
Similar conclusions are drawn by Rossi \etal~\cite{rossi15}, who reduce the accuracy of GPS data by truncating the longitude and latitude values to increasingly fewer decimal places: this effectively allows them to consider geographical resolutions that range from around $1$ m\textsuperscript{2} to over $10$ km\textsuperscript{2}. However, even at the lowest spatial granularity, $5$\% to $60$\% of users are still unique in the considered datasets\cref{fn:rossi15}.


\subsubsection{Segmentation}

A third straightforward technique is that suggested by Song \etal~\cite{song14}, \ie segmenting each trajectory and using a different pseudo-identifier for each segment. The rationale is that the unicity of a trajectory increases with its length, hence slicing each original trajectory into many output trajectories typically makes the latter less unique and easier to protect via anonymization. However, the authors show this simple technique cannot reduce unicity in a significant way: $80$\% of truncated trajectories in their dataset%
\footnote{Experiments are conducted on CDR-based trajectories of $1.14$ million users, tracked for one week, with a sampling interval of 15 minutes.}
are still unique even when they only span $6$ consecutive hours. Moreover, this approach risks to dramatically reduce the utility of the \data, preventing many analyses that require complete movement information about each user.

\subsubsection{Swapping}

A recent work by Salas \etal~\cite{salas18} proposes a model where portions of the trajectories are iteratively swapped among users, so that the output trajectories are in fact composed of segments belonging to multiple actual users. The technique, named SwapMob,
operates opportunistically on pairs of trajectories that come close enough to be swapped.
Tests with real-life data%
\footnote{Experiments are run on GPS trajectories of $10,357$ taxies in Beijing, China, during one week in February 2008. The database contains over $15$ million spatiotemporal points, with an average sampling interval of $177$ seconds and $623$ meters.}
demonstrate that SwapMob effectively dissociates the segments of trajectories from the subject that generated them, significantly reducing the space for record linkage. Yet, an adversary knowing $10$ spatiotemporal points is still able to link 42\% of the users, and learn more than 50\% of the original trajectories in 5\% of cases.
Also, it holds again the consideration that the output \data does not retain any utility for studies that require the possibility of following users for long, continued time intervals.

\subsection{Solutions providing indistinguishability}
\label{sub:anon-ind}

Indistinguishability is the first proper privacy criterion that we consider in our survey.
As already mentioned, $k$-anonymity is the de-facto standard privacy criterion for indistinguishability
in \data. $k$-anonymity is attained by transforming the spatiotemporal points of the trajectories in
the database, so that all points in every spatiotemporal trajectory are found in least $k-1$ other
trajectories. Different types of transformations can be applied to the spatiotemporal points, telling
apart the diverse methods to implement $k$-anonymity that are outlined by the first set of rows in
Table\,\ref{tab:anon-tax}, and that we will review in the rest of this Section.

\begin{sidewaystable}
\caption{Comparative roster of the main features of the techniques proposed to achieve $k$-anonymization in \data databases. The columns indicate: \textit{(i)--(ii)} the reference and acronym of the solution; \textit{(iii)--(iv)} the type of trajectories they operate with and the LBQID they assume; \textit{(v)--(vi)} the approach they adopt, including the distance metric between trajectories; \textit{(vii)--(viii)} typical performance figures, in terms of removed spatiotemporal points, and resulting data quality.\vspace*{8pt}}
\label{tab:anon-feat}
\scriptsize
\centering
\renewcommand{\arraystretch}{1.3}
\setlength{\tabcolsep}{1.5pt}
\begin{tabular}[c]{l l @{}p{3pt}@{} l l @{}p{3pt}@{} l l @{}p{3pt}@{} l l}
	
	\multicolumn{1}{p{62pt}}{\textbf{Reference}} &
	\multicolumn{1}{p{26pt}}{\textbf{Name}} & &
	\multicolumn{1}{p{54pt}}{\textbf{Trajectory}} &
	\multicolumn{1}{p{52pt}}{\textbf{LBQID}} & &
	\multicolumn{1}{p{58pt}}{\textbf{Approach}} &
	\multicolumn{1}{p{71pt}}{\textbf{Pairwise trajectory \newline distance metric}} & &
	\multicolumn{1}{p{41pt}}{\textbf{Suppressed \newline points}} &
	\multicolumn{1}{p{52.5pt}}{\textbf{Data quality \newline ($k=2$)}} \\
\hhline{~~~~~~~~~~~}

%

\hhline{--~--~--~--}
	\multicolumn{1}{|L{62pt}|}{Terrovitis and \newline Mamoulis~\cite{terrovitis08}} &
	\multicolumn{1}{|L{26pt}|}{--} & &
	\multicolumn{1}{|L{54pt}|}{Spatial (discrete)} &
	\multicolumn{1}{|L{52pt}|}{Subset of points} & &
	\multicolumn{1}{|L{58pt}|}{Suppression} &
	\multicolumn{1}{|L{71pt}|}{Euclidean distance} & &
	\multicolumn{1}{|L{41pt}|}{30-50\%} &
	\multicolumn{1}{|L{52.5pt}|}{--} \\

\hhline{--~--~--~--}
	\multicolumn{1}{|L{62pt}|}{Yarovoy \etal~\cite{yarovoy09}} &
	\multicolumn{1}{|L{26pt}|}{--} & &
	\multicolumn{1}{|L{54pt}|}{Spatial} &
	\multicolumn{1}{|L{52pt}|}{Subset of points} & &
	\multicolumn{1}{|L{58pt}|}{Generalization} &
	\multicolumn{1}{|L{71pt}|}{Hilbert distance} & &
	\multicolumn{1}{|L{41pt}|}{--} &
	\multicolumn{1}{|L{52.5pt}|}{7-62\% query \newline distortion} \\

\hhline{--~--~--~--}
	\multicolumn{1}{|L{62pt}|}{Nergiz \etal~\cite{nergiz09}} &
	\multicolumn{1}{|L{26pt}|}{--} & &
	\multicolumn{1}{|L{54pt}|}{Spatiotemporal} &
	\multicolumn{1}{|L{52pt}|}{Any} & &
	\multicolumn{1}{|L{58pt}|}{Generalization \newline \& suppression} &
	\multicolumn{1}{|L{71pt}|}{Log cost metric (LCM)} & &
	\multicolumn{1}{|L{41pt}|}{3-4\%} &
	\multicolumn{1}{|L{52.5pt}|}{50-90\% clustering \newline accuracy} \\

\hhline{--~--~--~--}
	\multicolumn{1}{|L{62pt}|}{Monreale \etal~\cite{monreale10}} &
	\multicolumn{1}{|L{26pt}|}{KAM} & &
	\multicolumn{1}{|L{54pt}|}{Spatial} &
	\multicolumn{1}{|L{52pt}|}{Any} & &
	\multicolumn{1}{|L{58pt}|}{Generalization \newline \& suppression} &
	\multicolumn{1}{|L{71pt}|}{Longest common \newline subsequence} & &
	\multicolumn{1}{|L{41pt}|}{--} &
	\multicolumn{1}{|L{52.5pt}|}{0.5-0.7 clustering \newline precision} \\

\hhline{--~--~--~--}
	\multicolumn{1}{|L{62pt}|}{\multirow{2}{*}[4pt]{Gramaglia \etal~\cite{gramaglia15}}} &
	\multicolumn{1}{|L{26pt}|}{\multirow{2}{*}[4pt]{GLOVE}} & &
	\multicolumn{1}{|L{54pt}|}{\multirow{2}{*}[4pt]{Spatiotemporal}} &
	\multicolumn{1}{|L{52pt}|}{\multirow{2}{*}[4pt]{Any}} & &
	\multicolumn{1}{|L{58pt}|}{\multirow{2}{*}{\shortstack[l]{Generalization \\ \& suppression}}} &
	\multicolumn{1}{|L{71pt}|}{\multirow{2}{*}{\shortstack[l]{Fingerprint \\ stretch effort}}} & &
	\multicolumn{1}{|L{41pt}|}{--} &
	\multicolumn{1}{|L{52.5pt}|}{1~km, 1~hour} \\

\hhline{~~~~~~~~~--}
	\multicolumn{1}{|L{62pt}|}{} &
	\multicolumn{1}{|L{26pt}|}{} & &
	\multicolumn{1}{|L{54pt}|}{} &
	\multicolumn{1}{|L{52pt}|}{} & &
	\multicolumn{1}{|L{58pt}|}{} &
	\multicolumn{1}{|L{71pt}|}{} & &
	\multicolumn{1}{|L{41pt}|}{5\%} &
	\multicolumn{1}{|L{52.5pt}|}{0.5~km, 40~min} \\

\hhline{--~--~--~--}
	\multicolumn{1}{|L{62pt}|}{Naini \etal~\cite{naini16}} &
	\multicolumn{1}{|L{26pt}|}{--} & &
	\multicolumn{1}{|L{54pt}|}{Spatiotemporal} &
	\multicolumn{1}{|L{52pt}|}{Location \newline histogram} & &
	\multicolumn{1}{|L{58pt}|}{Microaggregation} &
	\multicolumn{1}{|L{71pt}|}{Normalized \newline information loss} & &
	\multicolumn{1}{|L{41pt}|}{--} &
	\multicolumn{1}{|L{52.5pt}|}{--} \\

\hhline{--~--~--~--}
	\multicolumn{1}{|L{62pt}|}{Torres \etal~\cite{torres16}} &
	\multicolumn{1}{|L{26pt}|}{--} & &
	\multicolumn{1}{|L{54pt}|}{Spatiotemporal} &
	\multicolumn{1}{|L{52pt}|}{Any} & &
	\multicolumn{1}{|L{58pt}|}{Microaggregation} &
	\multicolumn{1}{|L{71pt}|}{Fr\'echet/Manhattan \newline coupling distance} & &
	\multicolumn{1}{|L{41pt}|}{29\%} &
	\multicolumn{1}{|L{52.5pt}|}{0.2-0.95 S/T \newline range query \newline distortion} \\

\hhline{--~--~--~--}
	\multicolumn{1}{|L{62pt}|}{Domingo \etal~\cite{domingo12}} &
	\multicolumn{1}{|L{26pt}|}{--} & &
	\multicolumn{1}{|L{54pt}|}{Spatiotemporal} &
	\multicolumn{1}{|L{52pt}|}{Any} & &
	\multicolumn{1}{|L{58pt}|}{Microaggregation \newline \& suppression} &
	\multicolumn{1}{|L{71pt}|}{Synchronized \newline trajectory distance} & &
	\multicolumn{1}{|L{41pt}|}{80\%} &
	\multicolumn{1}{|L{52.5pt}|}{2.4~km, 100~min} \\

\hhline{--~--~--~--}
	\multicolumn{1}{|L{62pt}|}{Abul \etal~\cite{abul08}} &
	\multicolumn{1}{|L{26pt}|}{NWA} & &
	\multicolumn{1}{|L{54pt}|}{Spatial} &
	\multicolumn{1}{|L{52pt}|}{Any} & &
	\multicolumn{1}{|L{58pt}|}{Microaggregation \newline \& suppression} &
	\multicolumn{1}{|L{71pt}|}{Euclidean distance} & &
	\multicolumn{1}{|L{41pt}|}{--} &
	\multicolumn{1}{|L{52.5pt}|}{Several km} \\

\hhline{--~--~--~--}
	\multicolumn{1}{|L{62pt}|}{Abul \etal~\cite{abul10}} &
	\multicolumn{1}{|L{26pt}|}{W4M} & &
	\multicolumn{1}{|L{54pt}|}{Spatiotemporal} &
	\multicolumn{1}{|L{52pt}|}{Any} & &
	\multicolumn{1}{|L{58pt}|}{Microaggregation \newline \& suppression} &
	\multicolumn{1}{|L{71pt}|}{EDR / LSTD} & &
	\multicolumn{1}{|L{41pt}|}{5-20\%} &
	\multicolumn{1}{|L{52.5pt}|}{Several km, several hours} \\

\hhline{--~--~--~--}
	\multicolumn{1}{|L{62pt}|}{Kopanaki \etal~\cite{kopanaki16}} &
	\multicolumn{1}{|L{26pt}|}{WCOP} & &
	\multicolumn{1}{|L{54pt}|}{Spatiotemporal} &
	\multicolumn{1}{|L{52pt}|}{Any} & &
	\multicolumn{1}{|L{58pt}|}{Microaggregation \newline \& suppression} &
	\multicolumn{1}{|L{71pt}|}{EDR / LSTD} & &
	\multicolumn{1}{|L{41pt}|}{--} &
	\multicolumn{1}{|L{52.5pt}|}{Several km, several hours} \\
\hhline{--~--~--~--}
\end{tabular}
\end{sidewaystable}

Table~\ref{tab:anon-feat} provides a summary of the main features of solutions proposed in the literature to implement $k$-anonymity in \data. It offers a quick outlook of the assumptions, approach and performance of each technique, and is thus a useful reference to start comparing different strategies for the $k$-anonymization of trajectory databases. Before delving into the details of these techniques, two important remarks about $k$-anonymity are in order, which apply beyond the context of \data.

First, the privacy level granted by $k$-anonymity is very much dependent on the value of $k$: in presence
of a $k$-anonymous database, the probability of re-identification under a random guess by the adversary
is $1/k$, hence $k$ is inversely proportional to the chance of success of a record linkage. Yet, there
is no clear consensus on which $k$ is safe enough, and the values adopted in the literature tend to be
application-dependent.
Also, it should be noted that attaining higher $k$ values typically reduces the utility of the \data,
as it requires distorting the spatiotemporal points when applying the transformation. Again, this creates
a trade-off that is not simply solved, and is highly use-case-dependent.

Second, $k$-anonymity has well-known and severe limitations. Basically, this privacy criterion offers
strong protection against record linkage attacks only; however, it does not remove privacy risks
associated to attribute linkage or any form of probabilistic attack. This has been repeatedly shown,
considering, \eg attacks aiming at localizing users, or at disclosing their presence, meetings and sensitive places~\cite{shokri11,machanavajjhala07,gramaglia17}.
The fact that $k$-anonymity has been at times misunderstood or oversold as a comprehensive solution for
PPDP of micro-data has led to diffused criticism on its use over the past years. Still, it remains a sensible
privacy criterion within its scope of application~\cite{fung10}. Moreover, as we will see in the following,
$k$-anonymity represents a basis on top of which more complex privacy-preserving solutions can be
developed, so as to counter complex attacks that go beyond record linkage.

\subsubsection{$k$-anonymity via spatiotemporal generalization}

{\em Spatiotemporal generalization} is the baseline method used to achieve $k$-anonymity in \data.
It reduces the spatial accuracy and temporal granularity of the spatiotemporal points in the
trajectories contained in the target database, so that all points of each trajectory are indistinguishable
from the points of $k-1$ other trajectories in the same database. 

\begin{figure*}
\centering
\includegraphics[width=0.31\columnwidth]{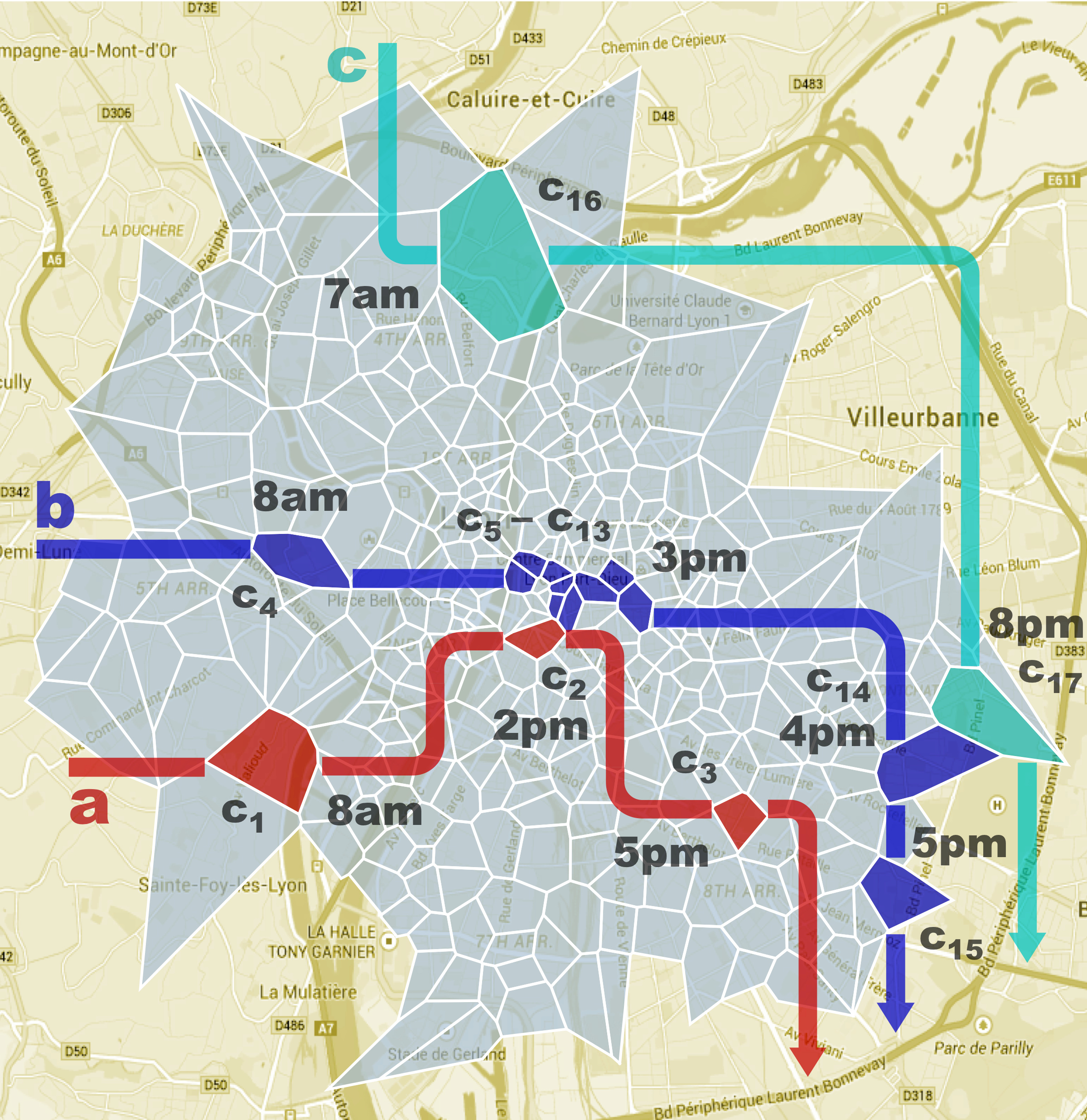}
\hspace*{5pt}
\includegraphics[width=0.31\columnwidth]{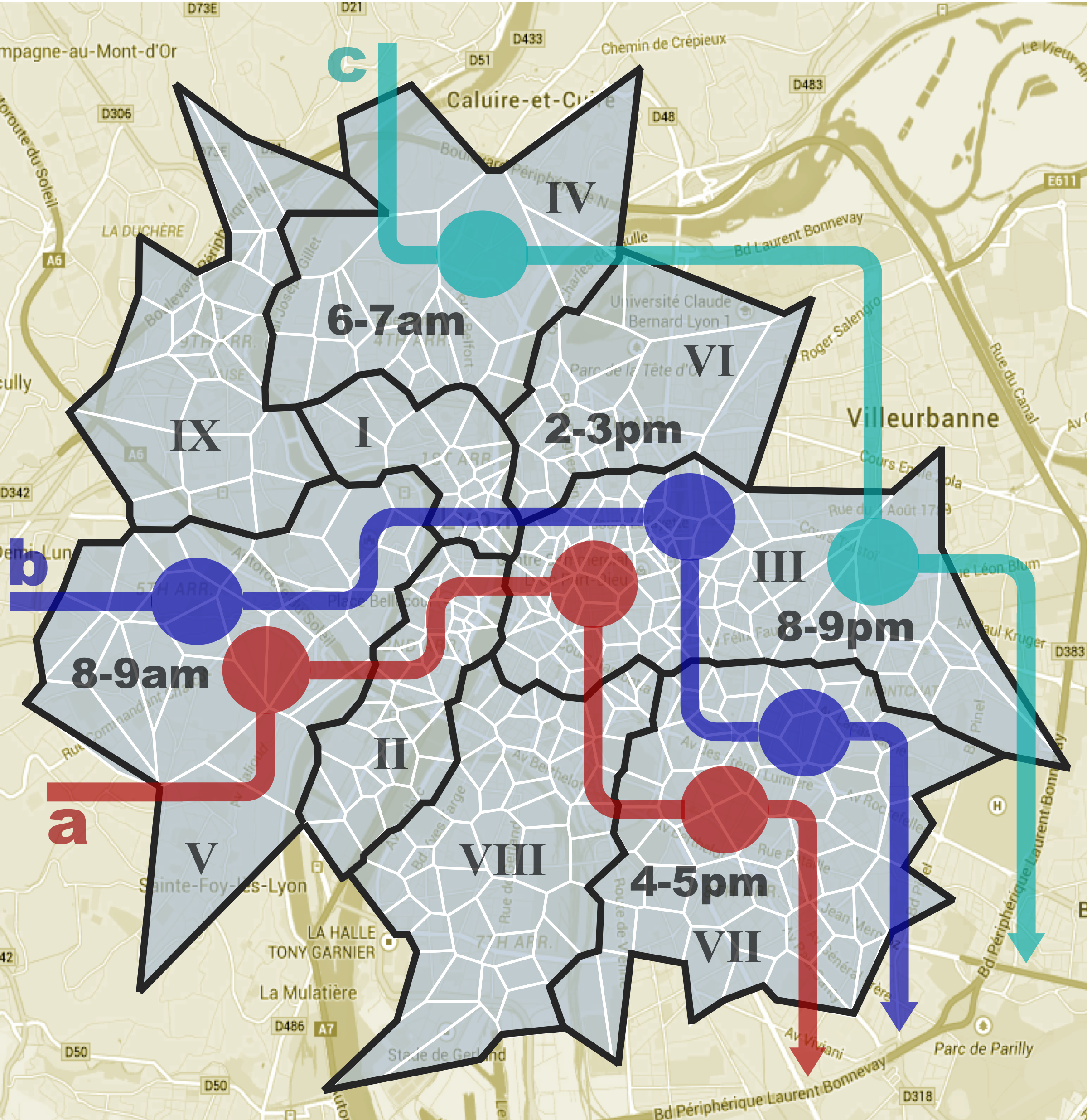}
\hspace*{5pt}
\includegraphics[width=0.31\columnwidth]{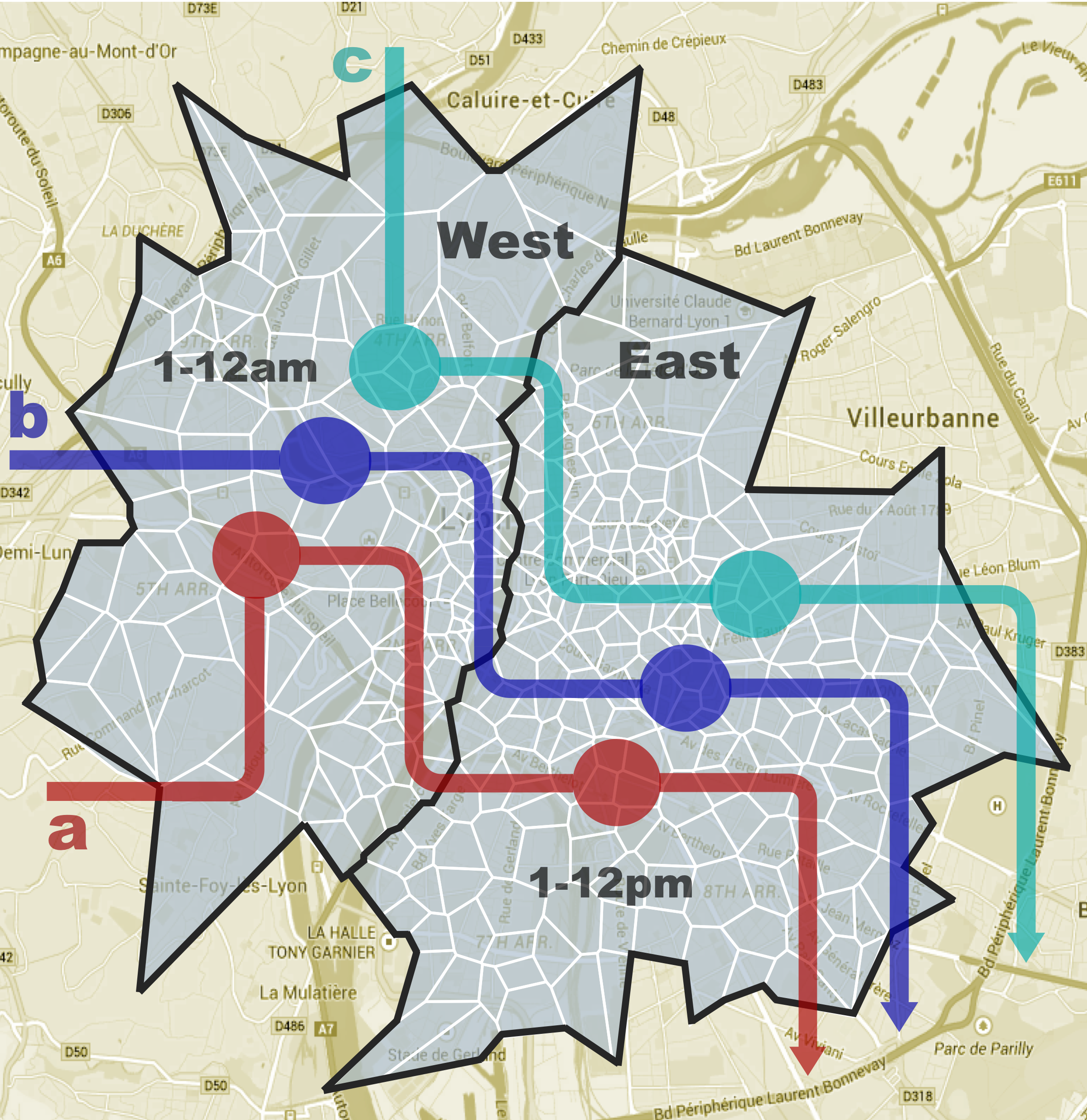}
\vspace*{-4pt}
\caption{Example of $k$-anonymity via spatiotemporal generalization. (a) Original database: user locations
are represented at cell level, and the temporal information has hourly precision. The \data of three users,
$a$, $b$, and $c$ are highlighted. (b) Spatiotemporal generalization: positions are recorded at the city
administrative zoning level, and the time granularity is reduced to two hours. The \data of users $a$ and
$b$ is now indistinguishable, and those two users are $2$-anonymized. (c) Increased generalization: locations are limited to the Eastern or Western half of the city, and time has 12-hour granularity.
All three users are indistinguishable and $3$-anonymized.
Reproduced in part from Gramaglia and Fiore~\cite{gramaglia15}, with permission.
}
\label{fig:kanon}
\vspace*{-8pt}
\end{figure*}

An illustrative example of spatiotemporal generalization of \data is shown in Figure\,\ref{fig:kanon}.
Despite its simplicity, the example highlights how achieving $k$-anonymity has a cost in terms of spatiotemporal
accuracy of the database: ensuring that all trajectories are $k$-anonymized may force reducing the spatial
and temporal accuracy of records up to the point where the \data becomes useless. Several works have quantified
such a trade-off between privacy and utility, when enforcing $k$-anonymity on \data.
Zang and Bolot~\cite{zang11} investigate how unicity decreases as the spatial granularity of a \data database is
lowered; this is equivalent to assessing which reduction of geographical accuracy is needed to attain $2$-anonymity,
\ie the minimum level of $k$-anonymity that removes unicity and grants indistinguishability. The study unveils
how unicity is very hard to eliminate from spatiotemporal trajectories: if the three most frequently visited
locations are known to the attacker, the only way to grant $2$-anonymity is to just publish user movements
among US States, even for very large databases\cref{fn:zang11}.

A more thorough evaluation is carried out by De Montjoye et al.~\cite{de-montjoye13}, who establish an empirical
relation between unicity and the spatiotemporal resolution of the \data dataset\cref{fn:de-montjoye13}. They
find that unicity decreases as a power law of both spatial and temporal granularity: this implies that the cost
in terms of data accuracy loss increases very quickly as a larger fraction of trajectories in a same dataset becomes
$2$-anonymized, and $2$-anonymizing the last percentile of trajectories may require a generalization that is orders of
magnitude stronger than that needed to $2$-anonymize the first percentile. Even worse, the power law exponent 
decreases linearly in the number of spatiotemporal points known to the attacker, and a few additional points make
$k$-anonymization much more expensive.

The reasons behind the high cost of $k$-anonymization in \data are studied by Gramaglia and Fiore~\cite{gramaglia15}.
They show that spatiotemporal trajectories in large databases%
\footnote{The authors employ two \data databases from two-week nationwide CDRs of 82,000 and 320,000 mobile network subscribers located in Ivory Coast and Senegal,
respectively, released in the context of the D4D Challenge~\cite{d4d}.\label{fn:gramaglia15}}
typically have a substantial fraction of points that are easily generalized into those of a single other trajectory
in the same database; however, they also have a small but non-negligible set of points that are very unique, and
hard to hide into another trajectory. Since $k$-anonymity is only achieved when all points are made indistinguishable, the cost of generalization is driven by such hard-to-hide points.
This study also investigates cases when $k>2$, demonstrating how higher $k$ values induce a superlinear growth in the utility loss.

Based on these observations, the authors propose GLOVE, an algorithm that achieves $k$-anonymity of \data
via spatiotemporal generalization, at a sensibly lower cost than the approaches by Zang and Bolot~\cite{zang11}
and De Montjoye et al.~\cite{de-montjoye13}. The key idea is to operate generalization on each spatiotemporal point
individually, instead of applying the same reduction of granularity to all points of all trajectories, as done
previously. Based on this intuition, the authors define a pairwise trajectory similarity metric named
\emph{fingerprint stretch effort}, which quantifies the loss of spatial and temporal granularity needed
to hide each sample of one trajectory into the closest sample of the other trajectory.
Then, a simple greedy clustering based on fingerprint stretch efforts lets GLOVE $2$-anonymize a complete \data
database with tens of thousands of records\cref{fn:gramaglia15} while retaining median resolutions of 1~km in
space and 1~hour in time.
Interestingly, performance tends to improve, \ie the data loss is reduced, as the database size grows.

All the above studies consider the $k$-anonymization of full-length spatiotemporal trajectories, \ie they assume
that all spatiotemporal points of each trajectory must be indistinguishable from the points of other $k-1$ trajectories
in the same database. Yarovoy et al.~\cite{yarovoy09} relax this challenging constraint, and study $k$-anonymity in a
setting where a known subset of the points of each trajectory is used as LBQID: therefore, only that subset needs to
be $k$-anonymized, for each trajectory. This significantly reduces the cost of generalization, since the LBQID
contains a number of points much smaller than that in the complete trajectory; however, it also introduces the
new problem of selecting the so-called anonymization group, \ie the set of $k-1$ records within which the LBQID
of each trajectory must be made indistinguishable. Indeed, a careless choice may lead to a successful record
linkage by an adversary with knowledge of the LBQIDs of multiple users.

The authors then propose algorithms that select anonymization groups so as to ensure proper $k$-anonymity in
this scenario. A first solution identifies sets of $k$ records based on a Hilbert distance measure, and
ensures that every trajectory in a group is generalized with respect to all LBQIDs of all other trajectories
in the same group. A second solution operates on a per-record basis rather than on a per-group basis: for each
trajectory $i$ in the database, it finds suitable trajectories $j$ to enforce symmetric $k$-anonymization of
the LBQID, \ie it generalizes the LBQID points of $i$ into those of $j$ and vice-versa.
Tests with databases featuring fixed temporal periodicity%
\footnote{The authors employ a real-life dataset includes GPS \data of cars in Milan, Italy. The data is
pre-processed to include one sample every 5 minutes, resulting in more than 45,000 trajectories and 2009
timestamps. A synthetic dataset is also used; it is created using Brinkhoff's generator~\cite{brinkhoff03} and includes 150,000
trajectories with 400 timestamps over the road-network of Oldenbur, Germany.}
show that, when LBQIDs include between 5\% and 50\% of the total spatiotemporal points of each trajectory,
the proposed schemes achieve $k$-anonymity, with $k$ from 2 to 32; however, they also induce spatial distortions
that cause 7\%--62\% of location-based queries to fail in the anonymized database.

\subsubsection{$k$-anonymity via suppression}

A different technique to achieve $k$-anonymity is \emph{suppression}, which removes spatiotemporal points from the original trajectories. Terrovitis and Mamoulis~\cite{terrovitis08} propose an algorithm that iteratively removes points from trajectories, simplifying the movement description until $k$-anonymity is satisfied. At each iteration, all points that break $k$-anonymity are identified, and the one entailing minimum Euclidean distortion is selected for removal. However, the simplicity of the solution entails strong assumptions on the \data format and attacker model in order to produce reasonable results: \emph{(i)} trajectories are purely spatial, \ie do not have a temporal dimension; \emph{(ii)} space is discretized in a finite number of locations; and, \emph{(iii)} adversaries are in a small number, and their exact knowledge is available and can be used as an input to the anonymization process. The latter point implies that the $k$-anonymization is limited to a very specific set of LBQIDs, \ie sequences of points. These aspects are reflected in the performance evaluation, carried out with synthetic data%
\footnote{The authors use 2,000 to 15,000 trajectories returned by the Brinkhoff's moving object generator~\cite{brinkhoff03} in Oldenburg, Germany.}, where 2 to 7 adversaries have side information (known to the anonymization algorithm) of all points in 1 out of 100 total locations.

\subsubsection{$k$-anonymity via generalization and suppression}


Generalization and suppression can in fact be used jointly. In the light of the analysis by Gramaglia and Fiore~\cite{gramaglia15}, suppression can be highly beneficial to $k$-anonymization: indeed, discarding the small fraction of unique points may take away a substantial portion of the diversity among trajectories, whose generalization then retains a higher accuracy level.

The first example of approach based on suppression is that by Nergiz \etal~\cite{nergiz09}. The solution is
close in spirit to GLOVE, as it also relies on per-spatiotemporal point generalization. However,
\textit{(i)} it enforces that no two points of one trajectory can be generalized with a single point 
of the other trajectory, which leads to suppression in presence of trajectories with a non-matching number
of points;
\textit{(ii)} it is based on a different pairwise trajectory similarity metric, named \emph{log cost metric},
which scales logarithmically the loss of spatial and temporal accuracy and accounts for suppressed points.
Evaluations with real-world and synthetic trajectory data%
\footnote{The real dataset includes 1,000 GPS trajectories of taxis in Milan, Italy, with a total of
98,544 samples, collected as part of the GeoPKDD project~\cite{giannotti09}. The synthetic dataset comprises
1,000 trajectories and 70,118 samples, which are obtained using Brinkhoff's moving object generator~\cite{brinkhoff03}.}
show that the proposed solution can achieve $2$-anonymization by suppressing 3-4\% of data, while $2$-anonymization has a much higher cost typically around 25\% of removed points. Although the authors do not report on exact error figures of the anonymized trajectories in space and time, they show that the results of one specific analysis, \ie clustering, are preserved with precision and recall in the range 50-90\%.

A different solution is proposed by Monreale \etal~\cite{monreale10}, which however only operates on
spatial trajectories that do not have time labels. Their strategy involves a first phase in which space
is discretized via a Voronoi tessellation: the seeds are obtained by clustering all spatiotemporal
points in the dataset in a way that a minimum number of trajectories is ensured to flow between any two
adjacent Voronoi cells. Trajectories are then all generalized in space according to the voronoi tessellation.
Two algorithms are proposed for the second phase, which actually implements $k$-anonymity.
KAM\_CUT is intended for dense datasets: it first creates an efficient tree structure of trajectories,
where common subtrajectories are the parent nodes to child nodes representing more complete (but diverse)
subtrajectories of the same users; it then traverses the tree by suppressing branches shared by less than $k$
trajectories.
KAM\_REC extends the above for sparse datasets: to this end, it tries to re-insert the pruned sub-trajectories
back in the tree, by finding their longest subsequence of points that maps to some popular sub-trajectory
either still in the tree, or shared by at least $k$ other trimmed sub-trajectories.
Experiments are run on measurement data%
\footnote{The dataset consists of 5,707 GPS trajectories of cars moving around Milan, collected by the
automotive service provider Octotelematics within the GeoPKKD project~\cite{giannotti09}. Note that the
\data is pre-processed by splitting trajectories when two consecutive points are too far in space and
time, resulting in more than 45,000 fairly space- and time-continuous trajectories in the final dataset.},
and show that KAM\_CUT and KAM\_REC attain precision and recall that typically are in the range 0.5-0.7,
for $k\in[2,30]$. In these tests, suppression removes between 10\% ($k=2$) and 80\% ($k=16$) of the
trajectories.

Also Gramaglia and Fiore~\cite{gramaglia15} extend GLOVE so as to include suppression. This is realized by
removing points that induce an over-threshold generalization cost during the calculation of the fingerprint
stretch effort. Tests with real-world data\cref{fn:gramaglia15} show that suppressing 5\% of points reduces
the loss of accuracy in space and time by approximately 30-50\%.

\subsubsection{$k$-anonymity via microaggregation and suppression}
\label{sub:kanon_micro}


Microaggregation is a family of two-step perturbative Statistical Disclosure Control (SDC) methods that can be used to implement $k$-anonymity in \data. In the first step (partition), the set of original trajectories is clustered based on similarity, so that each cluster has cardinality at least $k$. In the second step (aggregation), the trajectories in a cluster are replaced by a cluster prototype, computed through an operator over the spatiotemporal points in the cluster. Overall, this effectively $k$-anonymizes the dataset, by making all $k$ or more trajectories in a same cluster identical to the prototype.



A seminal work partially based on microaggregation of spatiotemporal trajectories is that by Domingo-Ferrer and Trujilo-Rasua~\cite{domingo12}. The authors introduce a new pairwise trajectory similarity metric, which we refer to as \emph{synchronized trajectory distance}. The distance is computed in two steps: first, trajectories are synchronized, \ie linearly interpolated and sampled with an identical periodicity; second, the total Euclidean distance between contemporary points is computed. In the case where the two trajectories span different time intervals (\ie the times of their first and last points do not match), all non-overlapping points are suppressed, and the distance metric is divided by the percentage of suppressed points as a similarity penalty. An interesting property of this metric is that it satisfies the triangle inequality, which allows speeding up calculations of all-pair distances.

The SwapLocation algorithm employs a legacy clustering technique based on the synchronized trajectory distance. Then, for each trajectory in a cluster, it swaps all of its spatiotemporal points with points of other trajectories in the same cluster. The exchange of points must respect configurable thresholds in space and time, which can be seen as the maximum allowed distortion of each point in a trajectory; also, a point is suppressed if no switch is possible under the imposed thresholds. We remark that the swap operation is not fully consistent with the standard microaggregation strategy, however the rest of the solution is coherent with such a model.

The authors use both synthetic and real-life datasets%
\footnote{The synthetic dataset is obtained via the Brinkhoff's moving object generator~\cite{brinkhoff03}, and consists of 1,000 trajectories which visit a total of 45,505 locations in Oldenburg, Germany. 
The real-life dataset consists of 4,582 GPS cab traces collected in San Francisco, USA, with a mean of 94 points each. \label{fn:domingo12}} to assess the performance of SwapLocation. The solution imposes significant suppression in the synthetic dataset and with $k=10$: a 1-km spatial threshold in the swap operation leads to removing 50\% of trajectories and 80\% of points. Under a 3-km threshold, suppression is reduced to 5\% for trajectories, but it remains almost unchanged for points; moreover, the average distortion is at 1 km approximately. In the case of real-world data, and $k=2$, 29\% of points are suppressed and the mean spatial distortion is at 2.4 km. All results refer to a 100-minute temporal granularity of the anonymized dataset.


A main limitation of the synchronized trajectory distance is its reduced capability to manage pairs of trajectories that do not perfectly overlap, which leads to substantial suppression of points. Torres and Trujillo-Rasua~\cite{torres16} propose a novel pairwise metric that is based on the \emph{Fr\'echet/Manhattan coupling distance}~\cite{eiter94} and overcomes such an issue. The metric is based on the notion of coupling, \ie a sequence of matching point pairs (one per trajectory) that respects the time ordering of points and ensures that all points are considered. The Fr\'echet/Manhattan coupling distance is then the coupling that minimizes the sum of Euclidean distances between matched points. Notably, the metric does not formally account for temporal distances between points, and its computation has a limited complexity $\mathcal{O}(pq)$, where $p$ and $q$ are the number of points in the two input trajectories.

The Fr\'echet/Manhattan coupling distance is then leveraged in a randomized clustering algorithm, and prototype trajectories are obtained for each cluster by means of an obfuscation process: linear interpolation and downsampling are first used to homogenize trajectories in time, and spatial averaging is then adopted to compute prototype locations at each instant.
In order to evaluate their solution, the authors employ synthetic data%
\footnote{The data were generated with Brinkhoff's moving object generator~\cite{brinkhoff03}, and consist of 5,000 trajectories containing 492,105 locations in Oldenburg, Germany, with 98,421 locations per trajectory on average.} and perform spatiotemporal range queries~\cite{trajcevski04}, which aim at inferring if a specific trajectory has some or all points inside a target region during a given time interval. Distortions of query results are between 0.2 and 0.95 for $k=2$, which are lower than those induced by the approaches proposed by Nergiz \etal~\cite{nergiz09} and Domingo-Ferrer and Trujilo-Rasua~\cite{domingo12}.

A special case of microaggregation is considered by  Naini \etal~\cite{naini16}, who adapt the approach to the case where histograms of popular locations are to be $k$-anonymized, rather than the complete trajectories. The authors then use a normalized information loss metric that is suitable for the probability distributions they target, and run a legacy clustering algorithm based on that metric. Evaluations with real-world datasets%
\footnote{The authors use three different datasets: the two-week CDR of 50,000 Orange customers in Ivory Coast released within the context of the D4D Challenge~\cite{d4d}, the Web browsing history of 472 users from the Web History Repository~\cite{wbh}, and 154 users with an average of 15.4 weeks of data each from the GeoLife experiment~\cite{geolife}.}
show that re-identification is impaired by $k$-anonymity with $k>10$, however that comes at a substantial information loss above 65\%.

\subsubsection{Generalized $k$-anonymity with spatial uncertainty}

\begin{figure}
\centering
\includegraphics[width=0.75\columnwidth]{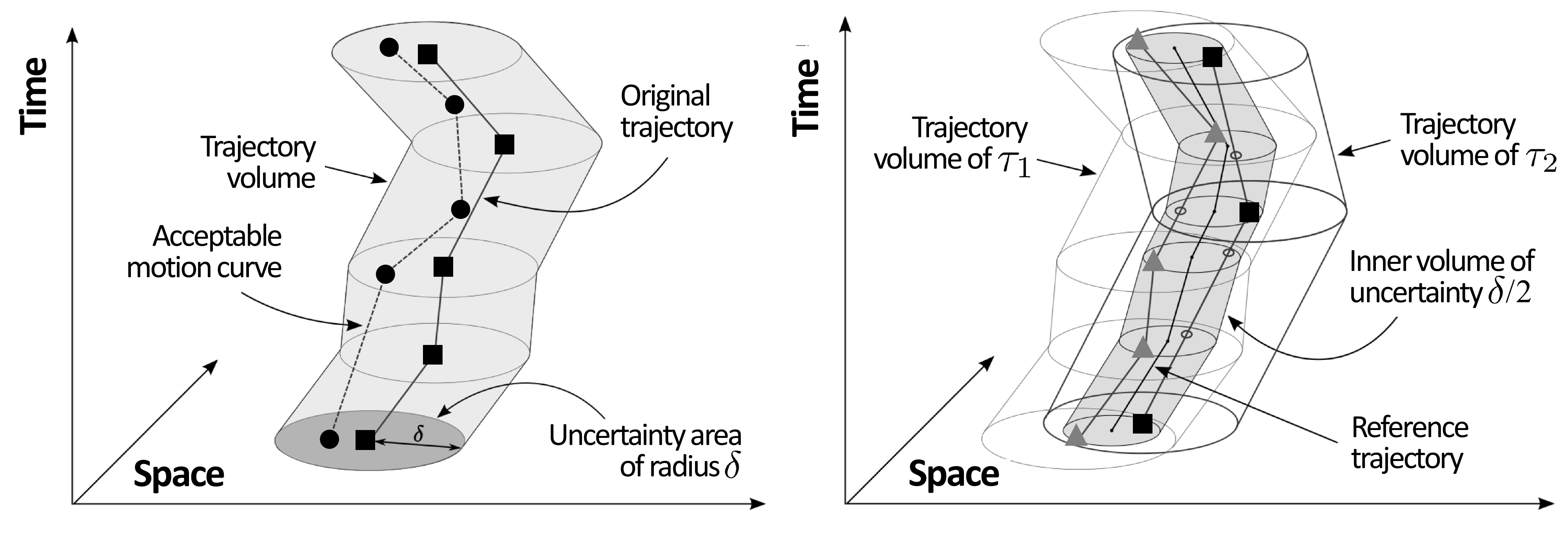}
\vspace*{-8pt}
\caption{Example of $(2,\delta)$-anonymity. Left: an original trajectory (solid line with square markers) is assigned an uncertainty radius $\delta$ around each of its spatiotemporal points, which yields a trajectory volume resulting from joining all uncertainty areas. An acceptable distorted motion curve for the original trajectory (\eg the one denoted with dashed line and circle markers) lies entirely within such a volume. Right: the original spatiotemporal points (black squares and grey triangles, respectively) of two trajectories $\tau_1$ and $\tau_2$ are included into volumes (black for $\tau_1$, grey for $\tau_2$) characterized by uncertainty $\delta$. The inner cylindrical volume is drawn with an uncertainty $\delta/2$ around a reference trajectory corresponding to the arithmetic mean of the points of the two original trajectories ((interpolated if necessary): a minimum translation of points within such inner volume ensures $(2,\delta)$-anonymity. In the example, no translation is needed since only two trajectories are concerned.
Adapted from Abul \etal~\cite{abul08}.}
\label{fig:abul08_fig1-2}
\vspace*{-8pt}
\end{figure}

Abul \etal~\cite{abul08} propose a generalization to the $k$-anonymity criterion, by assuming that published trajectories must be indistinguishable within an \emph{uncertainty threshold} $\delta$. In other words, the strong indistinguishability requirement of $k$-anonymity that each trajectory must be identical to at least $k-1$ others (see Definition 1 in Section\,\ref{sub:k-anon_def}) is relaxed, allowing a spatial distance up to $\delta$ among each point of an anonymized trajectory and the concurrent ones of the $k-1$ other trajectories. The intuition is that uncertainty among points within a geographical distance $\delta$ still provides sufficient protection, while it requires less distortion.
The generalized criterion is named \emph{$(k,\delta)$-anonymity}, and its concept is illustrated in Figure~\ref{fig:abul08_fig1-2}. The criterion offers inherently reduced privacy guarantees with respect to pure $k$-anonymity: as shown by Trujillo-Rasua and Domingo-Ferrer~\cite{trujillo11}, the two criteria match only under perfect spatial identity of the anonymized trajectories, \ie $\delta=0$.

In the original work by Abul \etal~\cite{abul08}, an algorithm named Never Walk Alone (NWA) is proposed to implement $(k,\delta)$-anonymity in uniformly sampled \data while minimizing distortion. NWA is organized in three phases. In the first phase, trajectories are trimmed so that they start and end at a limited set of time instants, and are then divided into groups with identical time spans. In the second phase, trajectories in a same group are separated into clusters of cardinality $k$ based on Euclidean distance, and possibly suppressed if the operation cannot be performed while keeping distances below an adjustable threshold. In the third and final phase, NWA performs, for each time instant and each cluster, a minimum translation of spatial points of all trajectories, so that they are within a distance $\delta/2$ from their arithmetic mean, as exemplified in Figure~\ref{fig:abul08_fig1-2}. It is apparent that NWA performs a partial microaggregation, and can be classified along solutions in Section\,\ref{sub:kanon_micro} when $\delta=0$.

Experiments with both real-world and synthetic trajectory datasets%
\footnote{Real-life \data consists of 273 trajectories of trucks~\cite{frentzos05}. The second one has been generated using Brinkhoff's network-based synthetic generator of moving objects~\cite{brinkhoff03}, and contains 100,000 trajectories during one day in the city of Oldenburg, Germany.}
show that NWA attains $(k,\delta)$-anonymity by inducing spatial displacements from several kilometers to several tens of kilometers, as $k$ grows from 2 to 100, and $\delta$ increases from a few hundreds to several thousand meters.

NWA only operates on trajectories with identical, periodic sampling. To overcome this important limitation, Abul \etal~\cite{abul10} also propose an improved method that can achieve $(k,\delta)$-anonymity with proper spatiotemporal trajectories, named Wait For Me (W4M). W4M largely builds on NWA, but instead of relying on Euclidean distance it uses pairwise trajectory distance measures that account for different temporal samplings of the input trajectories. The considered metrics are: the \emph{Edit Distance on Real sequences (EDR)}~\cite{chen05}, which targets quality preservation but has complexity $\mathcal{O}(pq)$, where $p$ and $q$ are the number of points in the two input trajectories; and the linear spatiotemporal distance (LSTD), which is designed for efficiency, having linear complexity $\mathcal{O}(p+q)$. 
Adopting EDR and LSTD also has the advantage that the first phase of NWA can be skipped, and all trajectories can be processed at once in the clustering phase.
Tests with heterogeneous datasets%
\footnote{The authors employ a real-world dataset of 45,000 GPS trajectories of cars in Milan, Italy, collected during one week by the GeoPKDD project~\cite{giannotti09}, and synthetic data generated with Brinkhoff’s network-based simulator of moving objects~\cite{brinkhoff03} for 100,000 trajectories over one day in Oldenburg, Germany.}
prove that W4M achieves substantially lower distortion of spatiotemporal range queries~\cite{trajcevski04} than NWA, and induces spatial and temporal translations in the order of several kilometers and hours, respectively. Also, the approach suppresses 5\% of points and creates 5\% new points to attain $k=2$; these figures grow up to 20\% when $k>30$.


Further extensions to NWA and W4M are proposed by Kopanaki \etal~\cite{kopanaki16}, who introduce a suite of algorithms under the denomination of Who-Cares-about-Others'-Privacy (WCOP). The variants of WCOP allow accounting for \textit{(i)} personalized values of $k$ and $\delta$ that vary for each trajectory, \textit{(ii)} a temporal segmentation of the trajectories such that each time segment of the dataset can be anonymized in isolation, and \textit{(iii)} bounded distortion of the output trajectories.
%
%
%


\subsubsection{Indistinguishability beyond $k$-anonymity: $l$-diversity and $t$-closeness}

All anonymization solutions presented above implement indistinguishability at the record level, hence are suitable countermeasures in presence of record linkage attacks. However, they do not address the indistinguishability of attributes, leaving the door open to the attribute linkage (or homogeneity) attacks discussed in Section\,\ref{sub:homo} and Section\,\ref{sub:att-tax-obj}. This holds for both cases where sensitive attributes are separated from or embedded in the trajectory data, as per the discussion in Section\,\ref{sub:att-tax-obj}. As an example, let us consider the case where the adversary is interested in finding some sensitive location embedded in the trajectory of the target user: by accessing a $k$-anonymized database with a few known spatiotemporal samples, the attacker would retrieve $k$ movement patterns; if, however, they all include the position of a gentlemen club, a privacy breach occurs.

In fact, it is well known that $k$-anonymity is not a sufficient criterion in the case of attribute linkage, which asks instead for more complex privacy definitions. Specifically, popular criteria designed to counter attribute linkage attacks are \emph{$l$-diversity}~\cite{machanavajjhala2006diversity} and \emph{$t$-closeness}~\cite{li2007t}. The former assumes that a precise set of attributes (either separated from or embedded in the trajectory data, as per the discussion in Section\,\ref{sub:att-tax-obj}) is identified as sensitive: then, any trajectory must be indistinguishable from a set of others whose sensitive attributes are sufficiently different from those of the original trajectory. Formally:

\begin{definition}
Let $\mathcal{D}$ be a database of \data, which includes a set of sensitive attributes; also, let $LBQID$ be the location-based quasi-identifier associated
with $\mathcal{D}$, 
and $\mathcal{D}[LBQID]$ the set of records returned by a query for $LBQID$ on $\mathcal{D}$. Then, $\mathcal{D}$ is said to satisfy $l$-diversity if and only if the records in $\mathcal{D}[LBQID]$ contain at least $l$ ``well-represented'' values for the sensitive attributes. Multiple notions of the concept of ``well-represented'' are possible, the simplest one being that at least $l$ distinct values for the sensitive fields be present in $\mathcal{D}[LBQID]$.
\end{definition}

A step further, $t$-closeness imposes a statistical constraint on the sensitive attributes, rather than the numerical one defined by $l$-diversity. The reason is that in practical cases the semantics of the attribute values are critical: for instance, a query returning a set of records with $l$ different but correlated attribute values (\eg $l$ variants of the same rare illness) satisfy $l$-diversity but still reveals sensitive information about the target user (\eg the fact that she suffers from the rare illness). To avoid these situations, $t$-closeness ensures that there is no substantial statistical difference between the attribute values in every set of indistinguishable users and those in the whole user population. Formally: 

\begin{definition}
Let $\mathcal{D}$ be a database of \data, which includes a set of sensitive attributes; also, let $LBQID$ be the location-based quasi-identifier associated
with $\mathcal{D}$, 
and $\mathcal{D}[LBQID]$ the set of records returned by a query for $LBQID$ on $\mathcal{D}$. Then, $\mathcal{D}$ is said to satisfy $t$-closeness if and only if the records in $\mathcal{D}[LBQID]$ contain sensitive attributes whose distribution has a distance lower than $t$ to the distribution of the attributes in the whole $\mathcal{D}$.
\end{definition}

While the problem has long been addressed in the context of LBS, for instance in the early study by Riboni \etal~\cite{riboni09}, the only works to date that tackles the anonymization of \data in a way to achieve both $l$-diversity and $t$-closeness are those by Tu \etal~\cite{tu2017beyond,tu18}. The authors focus on semantic attacks where the sensitive information is embedded in the spatiotemporal points, and corresponds to the points of interest (PoI) present in a target geographical region. Therefore, they propose an algorithm that builds on GLOVE by Gramaglia and Fiore~~\cite{gramaglia15}; as such, it leverages both generalization and suppression of samples, however these operations are augmented to ensure that each generalized sample fulfills the $t-closeness$ (and, implicitly, $l$-diversity) criterion. Specifically, the difference between the PoI distributions within each sample and in the whole database, measured in terms of Kullback–Leibler (KL) divergence, must be below a threshold $t$. The performance evaluation with measurement data%
\footnote{Two datasets are used for the evaluation. The first dataset is from a Chinese cellular network in Shanghai, and covers one week of data generated by 5,9 million users with an average number of 261 points each. The second dataset is from 15,500 users of a mobile application in Shanghai, and features a mean of 496 points per user.}
shows that the proposed solution can reduce KL divergence by a factor three while sacrificing an additional $30\%$ of the spatial and temporal resolution over the baseline $k$-anonymity granted by GLOVE.

\subsection{Solutions providing uninformativeness}
\label{sub:anon-uninf}

The second important privacy principle explored in the literature is that of uninformativeness. This principle aims to cope with probabilistic attacks and has received substantial attention in recent times. As anticipated in Section\,\ref{sub:dp_def}, the de-facto standard criterion to implement uniformativeness is differential privacy, a popular privacy criterion first introduced by Dwork \etal~\cite{dwork06} for PPDM. Implementing differential privacy is especially elegant and simple in presence of algorithms that execute numeric or categorical queries. In the former case, the output is a vector of scalars, and differential privacy is obtained by the \textit{Laplacian mechanism}, which adds Laplacian noise to such an output as first proposed by Dwork \etal~\cite{dwork06}. In the second case, the output is a probability distribution over a discrete, finite set of outcomes, and differential privacy is obtained by randomizing the probability according to an \textit{exponential mechanism}, as first explained by McSherry and Talwar~\cite{mcsherry07}. In both situations above, the level of noise or randomization is calibrated according to $\epsilon$, as well as to the maximum difference among all possible outputs when a single record is removed.
In addition, under such query models, differentially private algorithms enjoy composition properties that describe how multiple queries consume the budget $\epsilon$; this allows calibrating noise to the type and frequency of queries permitted on the database.


In the context of trajectory data, differential privacy has been successfully used to guarantee location privacy, \ie warranting that queries on single spatiotemporal points satisfy the uninformativeness principle. Criteria like \textit{geo-indistinguishability}, first introduced by Andr\'es \etal~\cite{andres13}, or based on the location-privacy metrics proposed by Shokri \etal~\cite{shokri12} adapt differential privacy to the specific case of location data. A number of works have implemented and possibly enhanced the criteria above, including those by Assam \etal~\cite{assam12}, Chatzikokolakis \etal~\cite{Chatzikokolakis2014}, Bordenabe \etal~\cite{bordenabe14}, Xiao and Xiong~\cite{xiao15}, or Ngo and Kim~\cite{ngo15}. However, as explained in Section~\ref{sub:intro-remarks}, solutions that anonymize queries on instantaneous locations are relevant for LBS, but not for trajectory PPDP.

Closer to our context of data publishing, a fairly large body of works have concentrated on PPDP of aggregate statistics from \data. A commonly studied class of aggregates is that of spatial densities, especially in the form of \textit{quadtrees}, \ie hierarchical spatial structures that allow for efficient querying: solutions such as those proposed by Cormode \etal~\cite{cormode12}, Qardaji \etal~\cite{qardaji13}, Li \etal~\cite{li14} or by Zhang \etal~\cite{zhang16} allow generating differentially private density databases from the actual trajectories, which can then be publicly released and safely queried. Extensions, such as those by Acs and Castelluccia~\cite{Acs2014} or Alaggan \etal~\cite{alaggan15}, consider spatiotemporal densities from \data, developing solutions that account for the temporal dimension of the aggregate statistics in addition to the spatial one. Other classes of trajectory data aggregates that can be transformed to meet differential privacy guarantees include weighted spatial graphs that describe transit counts between locations, such as those considered by Brunet \etal~\cite{Brunet2016}, or histograms, such as those assumed by Hay \etal~\cite{hay16}. Further investigations, \eg by Kellaris \etal~\cite{Kellaris2014} or Cao \etal~\cite{cao15}, adapt the techniques above to the case of streaming data, where privacy-preserving spatial density information needs to be continuously published.
Nevertheless, these works do not allow releasing spatiotemporal trajectories, but only their density or count statistics; hence, they are beyond the focus of our review on PPDP of \data.

When applied to our target milieu, \ie publishing actual trajectory data, differential privacy recommends that the output of an algorithm run on the released database is not affected by any single original trajectory beyond the privacy budget $\epsilon$.
Unfortunately, due the very high dimensionality of each trajectory, there is no current method to achieve such a goal by directly adding noise to the \data with existing mechanisms such as Laplacian or exponential. Therefore, two alternative approaches have been explored: \textit{(i)} considering softened notions of differential privacy; or, \textit{(ii)} generating synthetic trajectories that mimic the properties of true individual user movements yet ensure that the differential privacy criterion is fully met.
	
Below, we review solutions that adopt the first strategy in Section~\ref{sub:weak-dp}, and present works that instead take the second approach in Section~\ref{sub:synth-dp}. We also present a couple of works that adopt other criteria than differential privacy to realize uninformativeness, in Sections~\ref{sub:plausible} and ~\ref{sub:kte}.

\subsubsection{$(\epsilon,\delta)$-differentially private \data}
\label{sub:weak-dp}

A weaker notion of differential privacy that has been successfully adopted for the PPDP of trajectories is \textit{$(\epsilon,\delta)$-differential privacy}. This is a relaxation of the basic notion of differential privacy provided in Section~\ref{sub:dp_def} (which we recall to be also referred to as $\epsilon$-differential privacy), where privacy breaches are allowed to occur with a (small) probability $\delta$.
Formally:
\begin{definition}
A randomized algorithm $\mathcal{A}$ offers $(\epsilon,\delta)$-differential privacy if for all datasets $\mathcal{D}'$ and $\mathcal{D}''$ differing in one element ($|\mathcal{D}'-\mathcal{D}''|=1$), and all subsets $S$ of the output of $\mathcal{A}$, then $Pr[\mathcal{A}(D') \in S] \leq e^\epsilon \times Pr[\mathcal{A}(D'') \in S] + \delta$.
\end{definition}

Shao \etal~\cite{shao13} propose techniques that achieve $(\epsilon,\delta)$-differential privacy by combining trajectory sampling and interpolation, either in this order (a-priori) or in the reverse one (a-posteriori). The sampling phase realizes a $(0,\delta)$ form of differential privacy, by preserving one original point in every $1/\delta$: these points are publicly disclosed, and represent the privacy breach. The interpolation (a classic cubic B\'ezier) instead completes the gaps in between the retained points; by the composition properties of differential privacy, such a deterministic operation preserves the privacy properties of the sampling.
Then, under the important assumption that the initial and final points of each trajectory are publicly known and their disclosure does not represent a privacy breach, both strategies attain $(0,\delta)$-differential privacy.
Experiments%
\footnote{The study uses one-hour GPS data of ships in the Singapore Straits.}
show that the a-posteriori method tends to have better results in terms of average error when querying the privacy-preserving database.

\subsubsection{Differentially private synthetic \data}
\label{sub:synth-dp}

Proper differential privacy can be ensured by a different process where \textit{(i)} some representation of the original \data is randomized so as to meet differential privacy requirements, and \textit{(ii)} synthetic trajectories are derived from such representations. Then, databases of synthetic trajectories can be distributed with provable privacy guarantees.

\begin{figure}
\centering
\includegraphics[width=0.7\columnwidth,trim={0 0 0 0},clip]{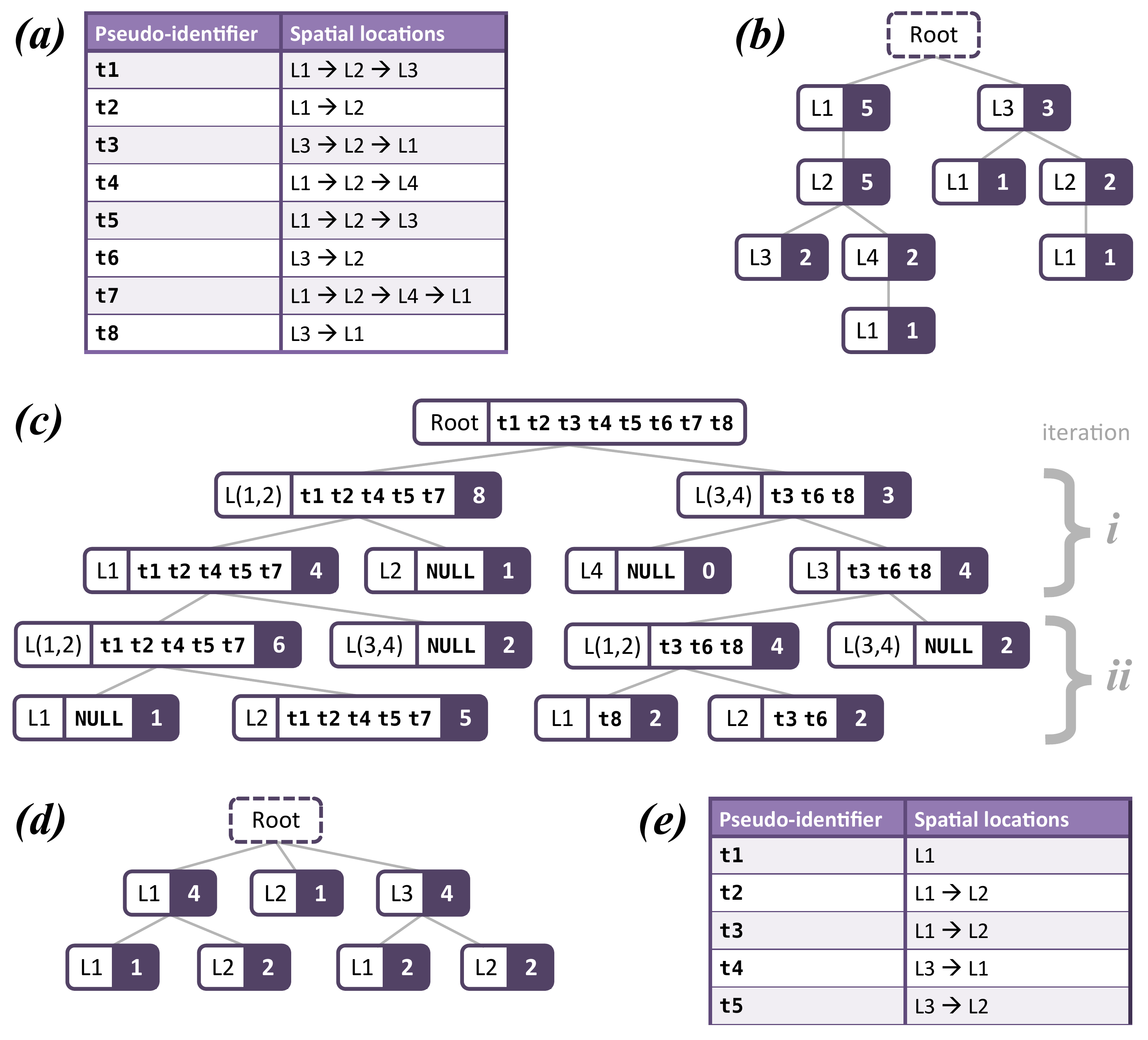}
\vspace*{-10pt}
\caption{Toy example of the approach by Chen \etal~\cite{chen12} to generate differentially private synthetic trajectories. (a) Original spatial trajectory database. (b) Prefix tree structure summarizing the original database. (c) Differentially private prefix tree with height $2$, under two levels of generalization: high-level generalized locations $L(1,2)$ and $L(3,4)$, and the low-level full-granularity locations $L1,L2,L3,L4$. Two iterations $i$ and $ii$ are needed to attain the desired tree height of $2$. Within each iteration, nodes for all possible higher-level generalizations -- $L(1,2)$ and $L(3,4)$ in the example -- are created from each leaf of the previous iteration. Such nodes are then expanded to lower-level full-granularity nodes only if their Laplacian noisy count is above a threshold (set to $3$ here). The model can accommodate more than two levels of generalization, which creates additional layers within each iteration $i$ and $ii$, one for each generalization level.
(d) Differentially private prefix tree upon pruning of all nodes for generalized locations, and with noisy counts made consistent among parent and child nodes. (e) Example of the synthetic trajectories extracted from the final prefix tree structure. Adapted from Chen \etal~\cite{chen12}.}
\vspace*{-10pt}
\label{fig:chen12}
\end{figure}

\textbf{Representing \data as trees.}
The first work to adopt the methodology above is that by Chen \etal~\cite{chen12}. They model the original database as a \textit{prefix tree}, \ie a hierarchical structure where trajectories are grouped based on matching location subsequences whose length grows with tree depth%
\footnote{Although we present it in the context of \data, the approach is general, and can operate on any type of sequential data.}. A privacy-preserving version of the prefix tree is then obtained by considering multiple levels of spatial generalization based on a predetermined location taxonomy, and iterating on the following operations at each prefix tree layer. First, nodes are created for all locations at the highest level of generalization, as children of each leaf from the previous iteration; second, Laplacian noise is added to the count of trajectories associated to each generalized node at the current prefix tree layer; third, nodes with a noisy count below a tunable threshold are not expanded further, while nodes with noisy counts above threshold generate children nodes for all locations at the following level of generalization. The process is repeated from the second step above. Iterations conclude once a user-defined tree height is reached, with Laplacian noises set so that the total privacy budget $\epsilon$ is equally divided across all tree levels and nodes within each level. An example is provided in Figure~\ref{fig:chen12}, plots (a)-(c).

Then, differentially private synthetic trajectories can be derived from the sanitized prefix tree. To this end, the tree is pruned so that only nodes at the lowest level of generalization (\ie retaining the maximum spatial granularity) are preserved. Then, the noisy counts associated to such nodes are made consistent across levels, ensuring that the count of each node is not less than the sum of counts of its children nodes. Finally, the synthetic trajectories are generated by visiting the resulting prefix tree. An example is provided in Figure~\ref{fig:chen12}, plots (d)-(e).

We remark that the solution is introduced for trajectories that are defined on a discrete space, but it can be extended to include discrete time information as well. Tests in a real-world case study%
\footnote{The authors employ information collected by the Soci\'et\'e de Transport de Montr\'eal (STM) about the transit history of passengers in the underground and bus networks of Montreal, Canada. The data contains over $1.5$ million trajectories, with an average of around $5$ locations each, out of a universe of $90$ an $121$ locations. \label{fn:chen12}}
show that the private synthetic trajectories can be mined to count passengers at stations, as well as to identify frequent sequential patterns of public transport usage, with a relatively low error.

\begin{figure}
\centering
\includegraphics[width=0.9\columnwidth]{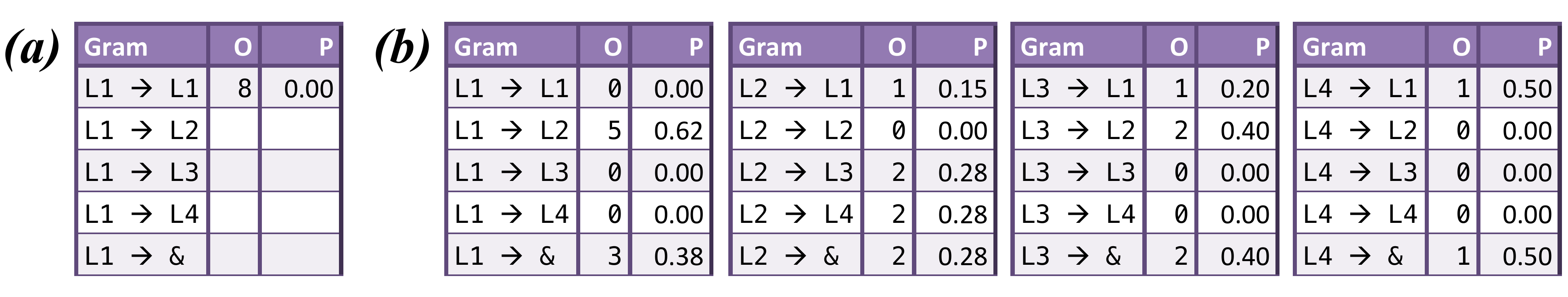}
\vspace*{-8pt}
\caption{Examples of $n$-grams used as trajectory representation by Chen \etal~\cite{chen12_ccs}, for the database of \data in Figure~\ref{fig:chen12}. Columns report the sequence (Gram), number of occurrences (O), and probability (P) for (a) $1$-grams, and (b) $2$-grams. Adapted from Chen \etal~\cite{chen12_ccs}.}
\vspace*{-8pt}
\label{fig:chen12_ccs}
\end{figure}

Chen \etal~\cite{chen12_ccs} propose a variant to the strategy above, where the main difference is that the prefix tree is replaced by an \textit{$n$-gram} representation. This probabilistic model describes trajectories as transition probabilities based on a past history of $(n-1)$ locations, \ie correspond to a Markovian model of order $(n-1)$. Figure~\ref{fig:chen12_ccs} illustrates the concept of $n$-grams.
The rest of the procedure is equivalent, by deriving a private prefix tree from the $n$-grams rather than from the original trajectories, and skipping intermediate generalization. Specifically, properly calibrated Laplacian noise is added to the counts of all $1$-grams, expanding them to $2$-grams only if their noisy count is above a threshold. Then, the procedure is repeated for always longer $n$-grams, descending into the prefix tree structure. The final differentially private variable-length $n$-grams can then be publicly released, or used to generate synthetic trajectories. Experiments with real-life datasets%
\footnote{The authors use the same STM dataset described in footnote~\ref{fn:chen12}, as well as 989,000 sequences of URL categories browsed by users on msnbc.com~\cite{dua17} with a mean length of 5.7 categories in a total set of 17.}
show that also in this case count queries and frequent pattern mining run on the synthetic data return reliable results.

Various refinements of the techniques above are proposed, \eg by Bonomi and Li~\cite{Bonomi13} and Qardaji \etal~\cite{qardaji13}. Note however, that these works aim at developing differentially private tree synopses of trajectory databases, but do not leverage them for the generation of synthetic \data.
Instead, Zhang \etal~\cite{zhang16} propose to extend PrivTree -- a method they originally introduced for privacy-preserving release of spatial density data -- to the case of synthetic sequential data generation. They adopt a \textit{prediction suffix tree} model of \data that is similar in spirit to the prefix tree considered by Chen \etal~\cite{chen12}; however, adapting PrivTree to work on prediction suffix trees allows the authors to remove two limitations of previous techniques. First, it automatically adapts the tree height to the data, which is thus not a fixed parameter anymore; second, the decision of expanding a tree node is not based on a simple count, but adopts a more advanced strategy that also accounts for the entropy of the eventual children nodes.
A comparative evaluation against the solution proposed by Chen \etal~\cite{chen12_ccs} proves that synthetic trajectories generated by PrivTree from real-life sequential data%
\footnote{The study uses 80,362 learners' sequences of activities (among 8 possible states) on a MOOC platform, as well as 989,818 sequences of URL categories browsed by visitors at msnbc.com during a 24-hour period.}
allow for a 10\% or higher improvement in \textit{(i)} top-k frequent string mining, and \textit{(ii)} approximating the distribution of sequence lengths.


He \etal~\cite{he15} demonstrate that the approaches above work well with coarse trajectories defined on small location domains, but fail to scale to realistic database where fine-grained trajectories unfold over moderately large geographical span. The reason is that the representations used by Chen \etal~\cite{chen12_ccs} grow in size as a power law of the number of discrete locations, with an exponent equal to the depth of the prefix tree. Therefore, the authors propose to generate multiple prefix trees, each referring to a different spatial resolution; each transition in a trajectory contributes to one specific tree, based on the travelled distance (\ie low-resolution trees for long distances, and high-resolution trees for short distances). This results in multiple trees with a very small branching factor each, and in a significant reduction of the overall number of counts maintained.
Then, the usual procedure of adding Laplace noise to counts, pruning the prefix trees, and extracting the synthetic trajectories is followed. In this last step, the authors also adopt an original sampling technique that allow preserving the correct directionality in the output trajectories. The proposed solution, named Differentially Private Trajectories (DPT), is evaluated with both real and synthetic datasets%
\footnote{Experiments are run on over 4 million GPS trajectories of 8,600 cabs in Beijing, China, during May 2009. Space is discretized into over 138,000 $100\times100$-m\textsuperscript{2} cells, leading to an average of 20 points per trajectory. Further tests employ Brinkhoff’s network-based generator for moving objects~\cite{brinkhoff03} The data consists of 15 million trajectories of 50,000 synthetic individuals in the region of Oldenburg, Germany, with a spatial resolution of $50\times50$-m\textsuperscript{2}.}
that are queried for distributions of diameters and trips, and for frequent sequential patterns. Results show that DPT largely outperforms the $n$-grams-based approach by Chen \etal~\cite{chen12_ccs} in the considered case studies.


\begin{figure*}
\centering
\includegraphics[width=0.85\columnwidth]{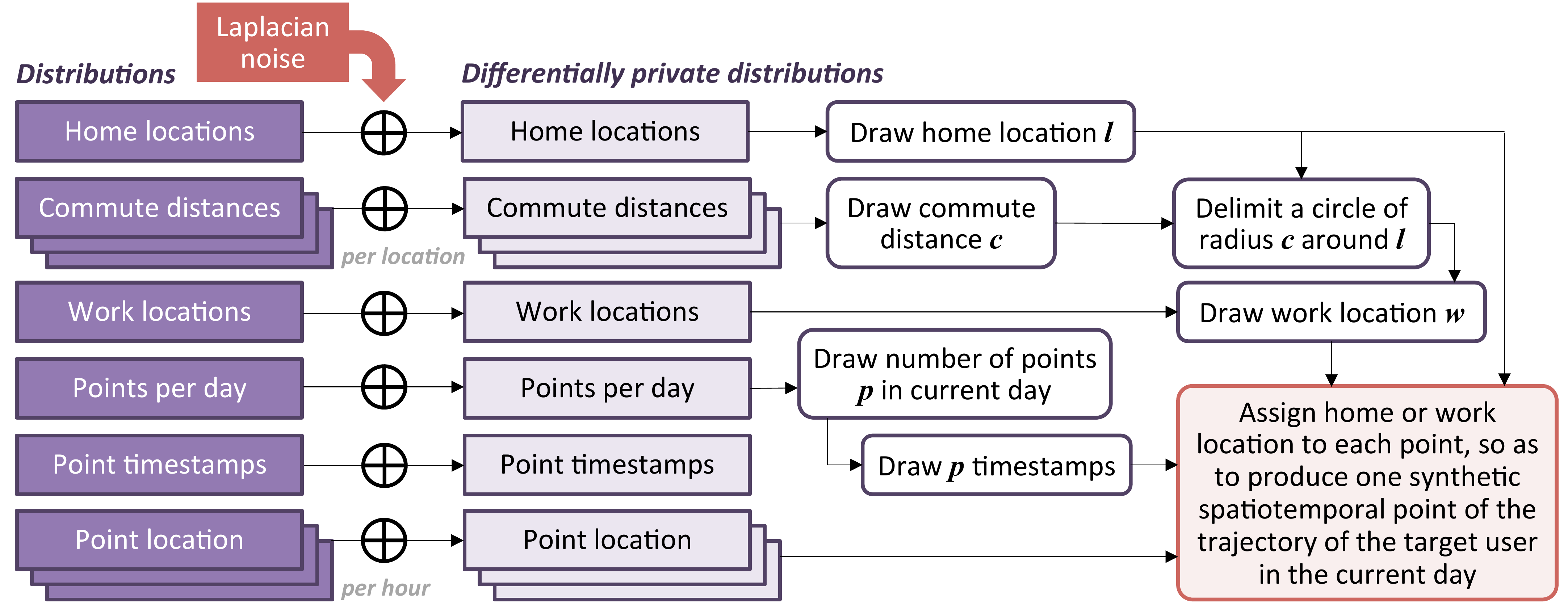}
\vspace*{-4pt}
\caption{Block diagram of the operation of the DP-WHERE solution by Mir \etal~\cite{mir13}. DP-WHERE creates differentially private noisy version of distributions of home and work locations, commute distances (based on the home location), points per day, time at which points are recorded, and locations at which points are recorded (on a hourly basis). Then, values are drawn from such distributions, and combined so as to associate home and work locations to each synthetic user, and generate spatiotemporal points that are then associated with a user. Adapted from Mir \etal~\cite{mir13}.} 
\vspace*{-8pt}
\label{fig:mir13}
\end{figure*}

\begin{figure*}
\centering
\includegraphics[width=0.9\columnwidth]{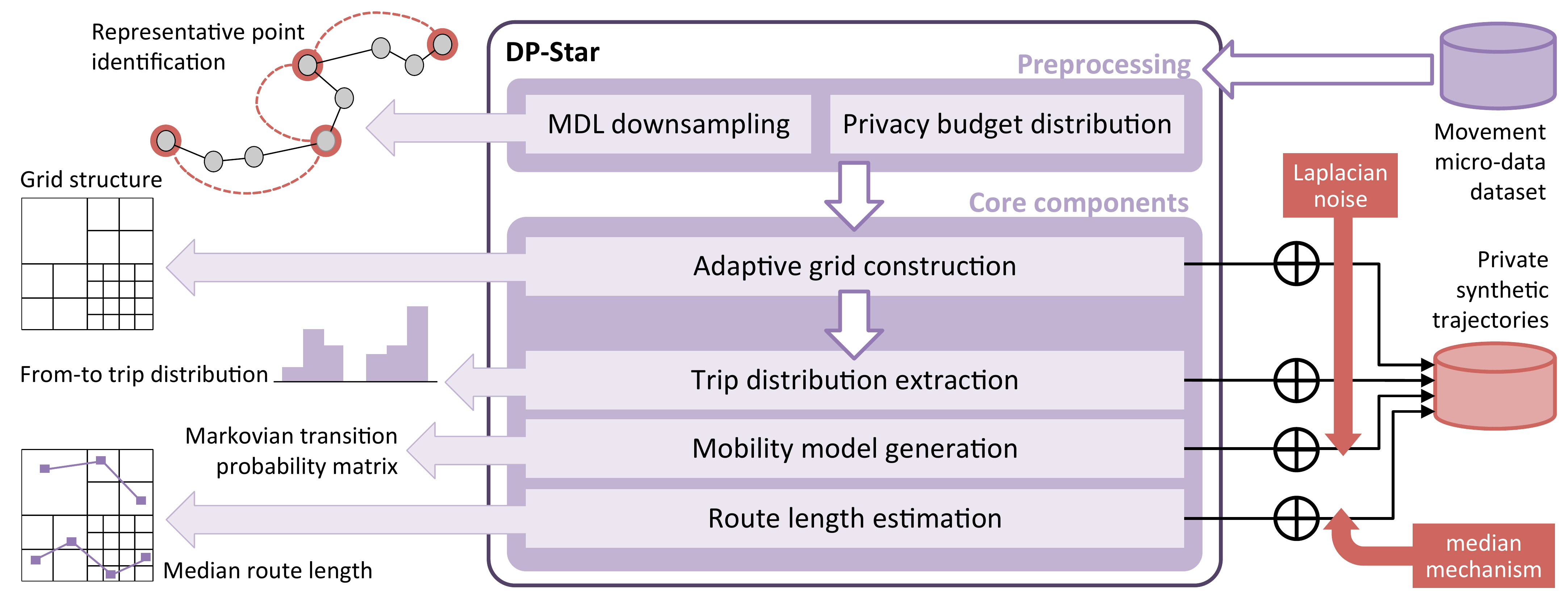}
\vspace*{-4pt}
\caption{Block diagram of the operation of the DP-Star solution by Gursoy \etal~\cite{gursoy18}. The input \data is downsampled so as to only retain utility-preserving movement information. Then, multiple representations are extracted from the downsampled trajectories. The representations are made differentially private my means of Laplacian and median mechanisms. Finally, privacy-preserving synthetic  trajectories are generated from the noisy representations. Adapted from Gursoy \etal~\cite{gursoy18}.}
\vspace*{-8pt}
\label{fig:gursoy18}
\end{figure*}

\textbf{Representing \data as probability distributions.}
A quite different strategy from those above is proposed by Mir \etal~\cite{mir13}. They introduce DP-WHERE, a differentially private synthetic trajectory generator that does not rely on a tree model of the original \data. Instead, DP-WHERE performs the following steps: \textit{(i)} derives a number of distributions that describe different statistical features of the movements in the original trajectory database, such as the spatial distribution of home and work locations, or the number of spatiotemporal points in a trajectory; \textit{(ii)} adds Laplacian noise to such distributions; \textit{(iii)} extracts realizations from the noisy distributions to generate synthetic trajectories. A more detailed view of the considered distributions and their combination is in Figure~\ref{fig:mir13}.
The synthetic movement data%
\footnote{Experiments are carried out on 10,000 synthetic users generated from 1-billion CDR of 250,000 subscribers in the region of New York, NJ, USA, during three months in 2011. The spatial resolution of the synthetic data is of 7 miles, \ie around 11 km.}
produced by DP-WHERE is proven to preserve population density distributions over time, as well as daily ranges of commutes in the reference area.

Roy \etal~\cite{Roy2016} follow a similar approach in their proposed Sanitization Model. First, they remove outlying records from the original dataset by applying the statistical interquartile range rule to all attributes. Second, they run legacy independence and homogeneity tests on attributes, and group attributes with high associativity in non-disjoint sets. Third, they derive synthetic distributions for each attribute group, and add Laplacian noise to them based on the available privacy budget. Fourth, synthetic records are generated by drawing samples from these distributions and aggregating them.
An interesting aspect of the work is the strategy it adopts to assess the quality of the synthetic \data obtained via the Sanitization Model. The authors consider a database published during a data visualization contest%
\footnote{The study leverages data provided by the Hubway bike sharing initiation and the Metropolitan Area Planning Council (MAPC) of Boston within their Hubway Data Visualization Challenge. The database consists of historical data about over one million bike trips in the grater Boston area. For each trip, the data include the start and end spatiotemporal points, as well as non-positioning information about the gender and subscription type of the rider.},
and replicate the competition submissions using both the original and differentially private trajectories. They find that the vast majority of the results are nearly identical, although it should be noted that the reference data is limited to trajectories where only the start and end locations and times are known.

The most recent proposal in probability-distribution-based approaches is DP-Star by Gursoy \etal~\cite{gursoy18}, whose operation is summarized in Figure~\ref{fig:gursoy18}. DP-Star first runs a preprocessing phase, during which raw trajectories are downsampled via Minimum Description Length (MDL), and reduced minimum sequences of representative points; also during preprocessing, the privacy budget $\epsilon$ is automatically split among the different core components by solving an optimization problem. The, DP-Star generates discretized representations of: \textit{(i)} space, as a non-uniform grid whose cell granularity is adapted to the geographical density of trajectory points; \textit{(ii)} trips between start and end locations, as a probability distribution; \textit{(iii)} internal trip structures, as a Markovian model of transition probability among locations; \textit{(iv)} route lengths, as the median distance covered by trajectories starting at each location.
Representations in \textit{(i)}--\textit{(iii)} are perturbed with Laplacian noise, while the noisy median route lengths are obtained with the \textit{median mechanism} proposed by Cormode \etal~\cite{cormode12}.

Differentially private synthetic \data is then extracted by combining the representations above, by selecting start and end grid cells, determining a route length based on the start cell, defining a sequence of cells via the Markovian model, and finally converting cells to actual points. We remark that DP-Star generates spatial trajectories that do not include temporal information, unlike, \eg DP-WHERE.
Evaluations with substantial real-world data%
\footnote{Three different datasets are considered: 14,650 GPS trajectories from the GeoLife project~\cite{geolife}, with an average of over 900 points each; 30,000 taxi traces collected in Porto, Portugal, with 43 points each on average; 50,000 synthetic trajectories created using Brinkhoff's generator~\cite{brinkhoff03}.}
shows that DP-Star retains significantly higher utility than the $n$-gram-based solution by Chen \etal~\cite{chen12_ccs} and DPT by He \etal~\cite{he15}, under several types of queries on \data.

\subsubsection{Plausible deniability}
\label{sub:plausible}

Bindschaedler and Shokri~\cite{bindschaedler16} propose an alternative definition of privacy for synthetic trajectories that also aims at realizing the uninformativeness principle. Their criterion is based on \textit{plausible deniability}: only a subset of \textit{seed} original trajectories is leveraged to generate the synthetic output \data, and the inclusion of a particular real trajectory among the seeds must be plausibly deniable. Such a condition is achieved by requiring that each synthetic trajectory in the released database could have been generated by a sufficiently large number of users in the original database, including those that were not selected as seeds. Formally:
\begin{definition}
A synthetic trajectory $f$ generated from a seed trajectory $s\in\mathcal{S}\subset\mathcal{D}$ satisfies $(k,\delta)$-plausible deniability if there are at least $k\geq1$ alternative trajectories $a\in\mathcal{D}$ such that the similarity $\sigma$ of $f$ and $s$ is not much higher than the same similarity measured between $f$ and any $a$, or $|\sigma(s,f)-\sigma(a,f)|\leq\delta$.
\end{definition}
In the definition above, $k$ denotes the number of trajectories in the original database that is large enough to ensure that the deniability of the presence of $s$ in the released data is actually plausible. Such original trajectories must yield a similarity with the synthetic trajectory, measured by a metric $\sigma$, within a threshold $\delta$ from that between the synthetic and seed trajectories.

\begin{figure}
\centering
\includegraphics[width=0.96\columnwidth]{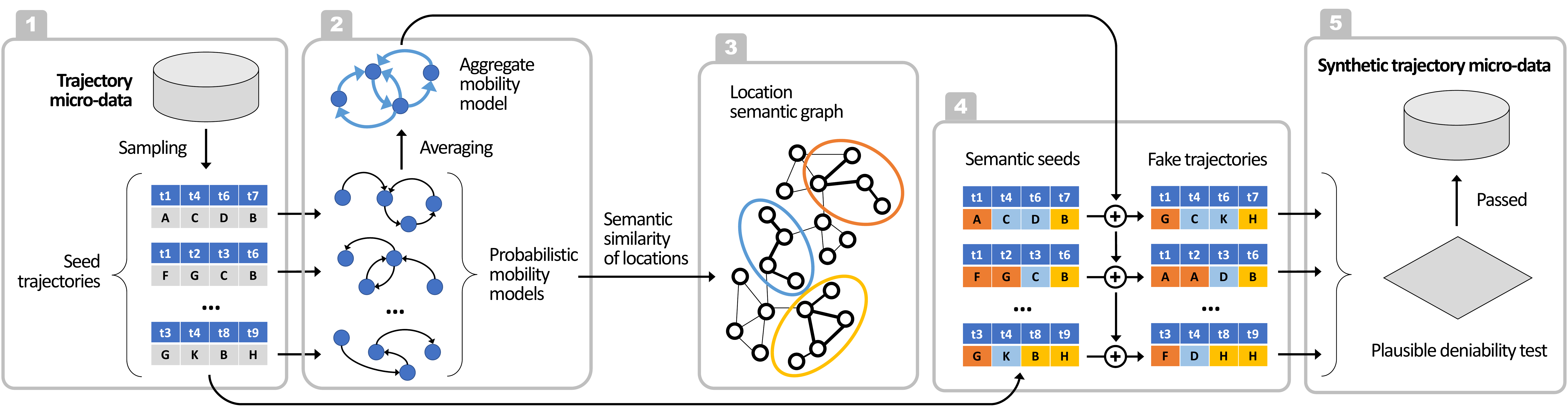}
\vspace*{-8pt}
\caption{Descriptive overview of the technique adopted to implement plausible deniability in synthetic \data. Steps 1 to 4 correspond to the transformation phase \textit{(i)}, and step 5 maps to the verification phase \textit{(ii)}. Adapted from Bindschaedler and Shokri~\cite{bindschaedler16}.}
\label{fig:bindschaedler16_fig1}
\vspace*{-8pt}
\end{figure}

In order to ensure plausibly deniability of synthetic \data, the authors adopt a strategy of: \textit{(i)} transforming each seed input trajectory into a semantic space, and then probabilistically transform it back to the original geographic space; \textit{(ii)} verifying if the proposed re-transformed trace satisfies plausible deniability when confronted to the original database, and only adding it to the output database if the answer is positive.
Figure\,\ref{fig:bindschaedler16_fig1} offers an overview of the proposed scheme.

Phase \textit{(i)} allows generating a synthetic copy of an original trajectory so that spatiotemporal features are preserved, and its main challenge is defining the transformation. To this end, the authors proceed as follows. First, each seed input trajectory is summarized into a mobility model that captures statistical information on the visiting probability to every location and the transition probabilities among such locations. A pairwise semantic similarity of mobility models is then computed as their maximum geographical closeness under all possible mappings of visited locations: the rationale is that two similar trajectories follow equivalent spatiotemporal patterns (\eg the same home $\rightarrow$ work $\rightarrow$ other $\rightarrow$ home repeated sequences) at different locations (\ie the exact home, work and other locations are distinct for the two users, and possibly far apart), hence a suitable mapping of locations (\eg considering the two home locations to be the same, and similarly for work and other) can reveal their resemblance.
Finally, all similarity information is aggregated in a location semantic graph: the nodes are the locations, and edge weights are the average semantic similarity between two locations over all pairs of mobility models. Intuitively, high weights characterize pairs of locations that are visited in similar ways by many users in the original database. The scheme thus performs a clustering on the graph, so that semantically similar (but geographically distinct) locations are grouped together in a same class.

The transformation in \textit{(i)} above consists in replacing locations in the original seed trajectory with their classes, which thus represent the semantic space. The re-transformation occurs according to an aggregate mobility model (\ie an average of all individual mobility models), which is run across the semantic space under the constraint that its locations are a subset of the locations of the target semantic trace. The constraint ensures that each synthetic trajectory shares the same semantic trace with its original version.

The probabilistic nature of the aggregate mobility model allows generating multiple synthetic trajectories for a same original seed user, which is leveraged in phase \textit{(ii)} of the solution. Once one synthetic trajectory is generated, it undergoes a privacy test based on plausible deniability: specifically, it is verified that it does not leak more information about the real mobility of the user than it does for other original trajectories in the input database. If the test fails, then a different synthetic trajectory is generated for the user, and the process is iterated.
In a follow-up work, Bindschaedler \etal~\cite{bindschaedler17} show that a randomized form of the solution above achieves in fact $(\epsilon,\delta)$-differential privacy, under a rigid set of $\epsilon$ and $\delta$ values.

Experiments with measurement data%
\footnote{The authors generate synthetic trajectories from the interpolated GPS data of 30 seed users participating in the Nokia Lausanne Data Collection Campaign~\cite{nokia}. The data is preprocessed so that all trajectories have a fixed sampling interval of 20 minutes, and a duration of one day; also, rarely visited locations are clustered together, so that the total number of locations is reduced by 60\%. A different day of mobility of the same 30 users is leveraged as the alternative database during the test phase that ensures plausible deniability.}
show that the synthetic \data retains, under $k=1$, significant utility in terms of visit frequency distributions, top-$n$ location coverage, user time allocation, spatiotemporal and semantic mobility features.

\subsubsection{$k^{\tau,\epsilon}$-anonymity}
\label{sub:kte}

\begin{figure}
\centering
\includegraphics[width=0.7\columnwidth]{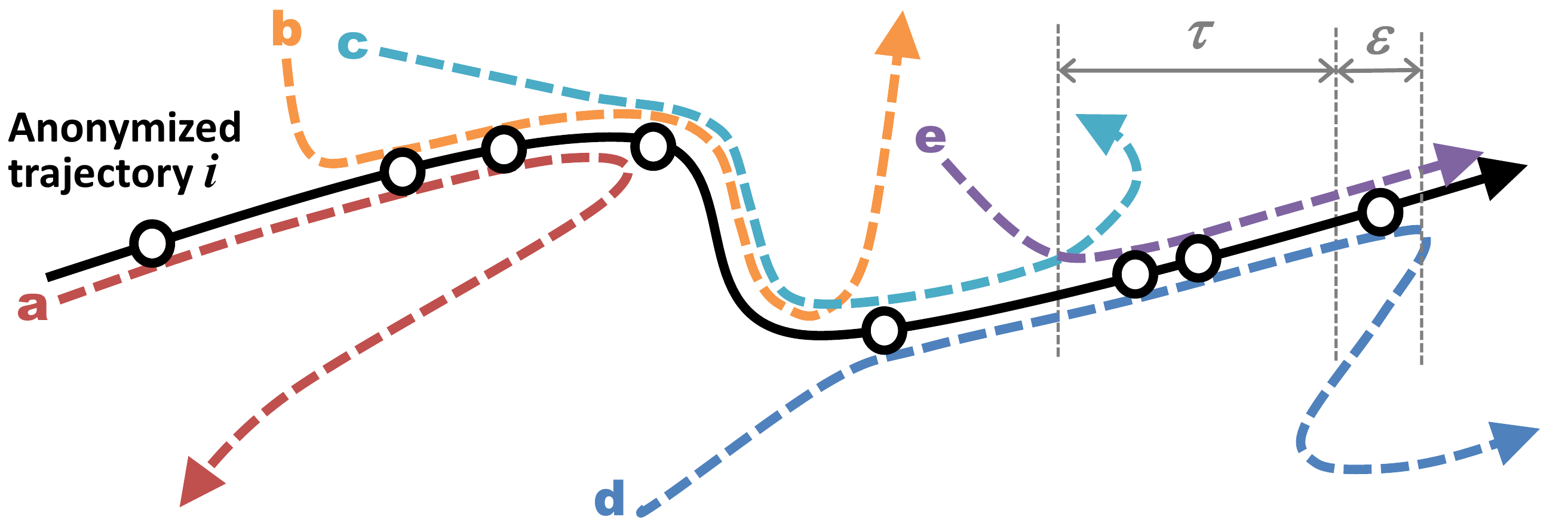}
\vspace*{-8pt}
\caption{Illustrative example of $k^{\tau,\epsilon}$-anonymity of user $i$, for $k=2$, using the movements of five other subscribers $a$, $b$, $c$, $d$, $e$. An adversary knowing a sub-trajectory of $i$ during any time interval of duration $\tau$ always finds at least one (\ie $k-1$) other user with a movement pattern that is identical to that of $i$ during that interval, but different elsewhere. With this knowledge, the adversary cannot tell apart $i$ from the other subscriber, and thus cannot attribute full trajectories to one user or the other. As this holds no matter where the knowledge interval is shifted to, the attacker can never retrieve the complete movement patterns of $i$: this achieves the uninformativeness principle. Still, the adversary can increase its knowledge in some cases. Let us consider the interval $\tau$ indicated in the figure: the trajectories of $i$, $d$ and $e$ are identical for some time after $\tau$, which allows associating to $i$ the movements during $\epsilon$: the attacker learns one additional spatiotemporal point of $i$. Adapted from Gramaglia \etal~\cite{gramaglia17}.}
\label{fig:gramaglia17}
\end{figure}

Gramaglia \etal~\cite{gramaglia17} introduce $k^{\tau,\epsilon}$-anonymity, a privacy criterion that stems from the original $k$-anonymity but has fundamentally different semantics. The idea behind $k^{\tau,\epsilon}$-anonymity is ensuring that any subset of spatiotemporal points known to the adversary matches $k$ trajectories, and that such $k$ trajectories are sufficiently diverse from each other in the rest of their points. Formally:

\begin{definition}
Let $\mathcal{D}$ be a database of \data and $LBQID$ the location-based quasi-identifier associated with it comprising $\tau$ spatiotemporal points. 
Also, let
$\mathcal{D}[LBQID]$ be the set of records returned by a query for $LBQID$ on $\mathcal{D}$. Then, $\mathcal{D}$
is said to satisfy $k^{\tau,\epsilon}$-anonymity if and only if the records in $\mathcal{D}[LBQID]$ are at least $k$, and only share an additional set of $\epsilon$ points other than the $\tau$ in the $LBQID$.
\end{definition}

According to the definition above, given a subset of any $\tau$ points in a target trajectory, the subset is $k$-anonymous, \ie is indistinguishable from points in $k-1$ other trajectories. However, apart from the $\tau$ points, the target trajectory only shares a small number of additional $\epsilon$ points with the $k-1$ trajectories used to anonymize the subset under consideration.
The correct implementation of $k^{\tau,\epsilon}$-anonymity grants that the attacker cannot distinguish among $k$ separated alternative trajectories that stem from his known points, and cannot infer substantial additional positioning information about his target user by accessing the database. Hence, $k^{\tau,\epsilon}$-anonymity realizes the uninformativeness principle.
As an interesting remark, when $\tau$ maps to the whole target trajectory, $k^{\tau,\epsilon}$-anonymity reduces to the original $k$-anonymity; therefore, solutions providing the former can be naively reduced to offer the latter privacy criterion.


The solution proposed by Gramaglia \etal~\cite{gramaglia17}, named \texttt{kte-hide}, satisfies the privacy criterion under the assumption that the points of the LBQID are adjacent in time; in this case, $\tau$ and $\epsilon$ can be mapped to time intervals rather than numbers of points. The algorithm then alternates different and partially overlapping anonymity sets over subsequent intervals of duration $\tau+\epsilon$, realizing the structure illustrated in the toy example of Figure~\ref{fig:gramaglia17}.
The performance evaluation of \texttt{kte-hide} is based on real-world datasets%
\footnote{The authors use nationwide and citywide CDR-based trajectories released in the context of the D4D Challenge~\cite{d4d}, as well as CDR data collected by the University of Minnesota~\cite{zhang2014exploring}. The resulting datasets follow 30,000--300,000 users for 1 day to 2 weeks in Abidjan, Dakar, Ivory Coast, Senegal and Shenzhen.},
and shows that the method is capable of attaining $2^{\tau,\epsilon}$-anonymity under seamless adversary knowledge ranging between 10 minutes and 4 hours, and $\epsilon=\tau$. The solution retains a median accuracy of the anonymized trajectory data of 1-2 km in space and less than 1 hour in time.

%
%
%
%
%

\section{Discussion and perspectives}
\label{sec:disc}

Based on the comprehensive survey of attacks against released databases of \data, and of countermeasures against such threats, our main takeaway message is that PPDP of trajectories is still a largely open problem. There is substantial space for improvement at all levels, and we outline below some directions for future research.

\subsection{Realistic and credible risk assessments}

It is important that the privacy risks associated with the publication of \data are assessed in practical settings. The vast majority of the works in the literature highlight very high re-identification (\ie successful record linkage) rates, announcing dramatic hazard for the privacy of the monitored individuals. However, these results have to be interpreted with a grain of salt. Many assume that the adversary knows some spatiotemporal points of its target user, which happen to be exactly in the target database (\ie a spatiotemporal subset format of the side information, according to our classification): this is very unlikely to happen in real life, as the adversary would have to anticipate when the user's location will be sampled by the positioning system. Similarly, it is simplistic to assume that the adversary is aware of its target's locations sampled with similar temporal frequency and spatial accuracy than those in the target dataset; or, equivalently, it is naive to expect that an attacker can build mobility profiles that are as detailed as those it can infer from the target \data. Indeed, there is a legitimate question on whether an attacker having such a substantial knowledge would be actually interested in making a large effort to retrieve ``more of the same'' data.

Note that we are not downplaying the privacy issues in \data -- which we believe are many and extremely relevant. However, we advocate for more realistic risk assessments that are representative of the actual conditions an attacker could operate in. Practical attacks require identifying and retrieving useful side information, and performing a reliable match with the target data; moreover, in most cases the attacker has to deal with uncertainty about the presence of its target user in the target database, as well as about an eventual match (since it does not possess any ground truth information guaranteeing that the match is correct). In absence of these practical considerations, studies may lead to over-pessimistic claims on privacy risks, which are instead mitigated when attacks are run in the wild.

The recent work by Wang \etal~\cite{wang18}, who show that figures on attack success rates in the literature are largely exaggerated when considering closer-to-reality settings, is a first evidence in this sense. However, it is not a definitive one, as the authors still retain many assumptions that simplify the attacker's work. More realistic and credible assessments of the actual risks associated with record linkage of \data are required.

\subsection{Risks beyond record linkage}

Record linkage absorbs almost the whole literature on attacks against \data, as it is well illustrated in Table\,\ref{tab:attacks}. However, these are, at least in theory, the simplest forms of menace against trajectory databases. Therefore, our considerations above on risk assessment are exacerbated in the case of attacks that are more complex than record linkage.

The privacy risks of, \eg attribute linkage (just to consider the next level of threat) are basically unexplored. Homogeneity, \ie the weakness that paves the way for attribute linkage, is a clearly understood concept, for which toy examples are easily constructed, and for which practical cases have been demonstrated in the context of relational databases. However, whether homogeneity actually exists, and, if so, to which level, remains a fully open question when it comes to the sensitive attributes one could link to \data. To date, we can only imagine that the risk may exist, but we do not even have a rough picture of its practical viability. The situation is similar for probabilistic attacks. Therefore, and even more than in the case of record linkage, realistic risk assessments of attribute linkage or probabilistic attacks represent an opportunity for future investigation.

\subsection{Silver bullet anonymization}

Anonymizing \data is extremely complex, and this is apparent from the number of solutions proposed over the past few years. We have understood that mitigation techniques simply do not work: reducing the spatial or temporal resolution of the data does not help, and also shortened or intertwined trajectories retain re-identifiability risks. Unfortunately, also more complex approaches are far from perfect.

On the one hand, techniques that grant $k$-anonymity are today fairly mature, preserve individual trajectories, and can retain a decent level of precision in the anonymized data (see Table~\ref{tab:anon-feat}). However, they typically scale poorly with $k$. More importantly, they only offer a protection against record linkage, and leave the data prone to more complex attacks, disregarding for the moment the question if these are actually feasible or not (see above).

On the other hand, differential privacy and its extensions for location data are very hard to apply to \data. As of today, all solutions implementing such privacy principle construct some model from the original data, apply noise so as to make the model differentially private, and then generate synthetic trajectories from the noisy model. It is clear that the anonymized dataset only retain global properties, and prevents analyses that require following actual individuals. Moreover, the global properties that can be explored through data mining are the same that are preserved by the noisy model: \ie there is no guarantee that features that are lost during the modelling phase will be reflected in the output database. Again, this poses potential limits to the nature of queries one can safely run on the anonymized data. Finally, most solutions for differential privacy also do not scale well, and are only demonstrated with simplistic databases of trajectories that are either very short, only defined over space, or spanning a small set of total locations.

As a result, the quest for a silver bullet anonymization solution for \data is still open, and it may pass through new privacy principles that go beyond $k$-anonymity or differentialy privacy.


\subsection{Reproducible research and comparative evaluations}

A striking aspect of most works in the literature on \data anonymization is that they provide very little in terms of comparison with previous solutions. This is clearly an issue that hinders our capability of untangling the body of literature and name a clear winner in the contest for the current state-of-the-art. We identify three main reasons for such an undesirable situation.

First, there is a lack of reference dataset of \data. Synthetic datasets (\eg those presented in~\cite{yarovoy09,bettini09,uppoor14} struggle to rise to such a status, due to their artificial nature. Publicly available datasets collected in the real world (\eg those in the CRAWDAD repository) are fairly old and limited in size. Some larger real-life databases have been released, \eg as part of challenges by mobile network operators~\cite{d4d}, yet they are protected by non-disclosure agreements that prevent their open distribution. Many works thus rely on proprietary data that is not made accessible to the research community, again due to agreements with the data providers, which are typically companies, for legal reasons related to the publication of real mobility traces. Such a scenario makes it hard to develop a reference set of trajectory databases like in other communities, hence limits the possibility of verifying the performance of different solutions on the same ground. Overall, we argue that there is a significant need for some large academic initiative to collect and release such open \data.

Second, the approaches adopted to evaluate different anonymization techniques vary wildly across studies. Works in the literature use a plethora of different quality measures, error metrics, queries and data mining analyses, which are however very diverse. Researchers have a tendency to always design new metrics (possibly well suited to their proposed solution), making it impossible to confront the performance evaluations carried out in two different papers. Also in this case, we need a reference set of metrics or tasks for quality assessment of the anonymized data, to be adopted throughout all studies and allowing a direct comparison of the performance figures. Clearly, such a set shall be large enough to cover a vast range of data usages, and avoid favouring a solution over another.


Third, very few researchers release the source code of their solutions. This is a despicable but common practice that curbs not only the reproducibility and comparability, but also the mere verifiability of the results. We argue that, as a community, we should move to a fully verifiable model where all papers proposing anonymization techniques shall be accompanied by their source code, possibly written in a commonly agreed language.

Overcoming the three problems above would make comparison straightforward and unavoidable, and improve the scientific rigour of the process towards solving the problem of anonymization of \data.

\section*{Acknowledgements}
The authors would like to thank Emilie Sirvent-Hien and Marc-Olivier Killijian for the constructive discussions. This research was supported by BPIFrance through the Programme d'Investissement d'Avenir, project n.P128356-2659748 (ADAGE).


\bibliographystyle{plain}
\bibliography{biblio}      


\end{document}